\documentclass[11pt,a4paper]{article}

\usepackage[T1]{fontenc}
\usepackage{eurosym}
\usepackage{booktabs}
\usepackage{multirow}
\usepackage{sectsty}
\usepackage{graphicx}
\usepackage{amsmath}
\usepackage{amssymb}
\usepackage{subcaption}
\usepackage{caption}
\usepackage{todonotes}
\usepackage{hyperref}
\hypersetup{colorlinks=true,linkcolor=blue,urlcolor=blue}

\usepackage{fancyhdr}
\usepackage[left=2cm,top=1.5cm,bottom=1.5cm,right=2cm,includefoot,includehead,headheight=13.6pt]{geometry}
\usepackage{lastpage}
\usepackage{enumitem}
\usepackage{lipsum}
\usepackage{titlesec}
\usepackage[toc]{multitoc}

\numberwithin{equation}{section}

\newcommand{\simas}[1]{\raisebox{-.1ex}{
            $\stackrel{\small{#1}}{\sim}$}}

\newcommand{\ssfill}{\xleaders\hbox to 0.35em{\scriptsize.}\hfill}% d

\makeatletter
\newcommand{\oset}[3][0ex]{%
	\mathrel{\mathop{#3}\limits^{
			\vbox to#1{\kern-2\ex@
				\hbox{$\scriptstyle#2$}\vss}}}}
\makeatother

\newcommand*{\cventry}[7][.25em]{
  \noindent\begin{tabular*}{\textwidth}{l@{\extracolsep{\fill}}r}%
	  {\bfseries #4} & {\bfseries #5} \\%
	  {\itshape #3\ifthenelse{\equal{#6}{}}{}{, #6}} & {\itshape #2}\\%
  \end{tabular*}%
  \ifx&#7&%
    \else{\\\vbox{\small#7}}\fi%
  \par\addvspace{#1}}

\newcommand*{\hintfont}{\bfseries}
\newcommand*{\hintstyle}[1]{{\noindent\hintfont{#1}}}
\newcommand*{\cvitem}[3][.25em]{%
  \ifthenelse{\equal{#2}{}}{}{\hintstyle{#2}: }{#3}%
  \par\addvspace{#1}}

\let\OriginalQuotation\quotation
\renewcommand*{\quotation}{\OriginalQuotation\small\sf}

\begin{document}

\allsectionsfont{\sffamily}

% HEADERS
\fancyhead{}
\fancyfoot{}

\fancyhead[CO]{{}}
\fancyhead[LO]{{}}
\fancyhead[RO]{{}}
\fancyfoot[R]{\thepage\, / {\color[rgb]{0.6,0.,0}\pageref{LastPage}}}
\renewcommand{\headrulewidth}{0pt}

% Title
\newcommand*{\titleGP}{\begingroup % Create the command for including the title page in the document
\thispagestyle{empty}

{\centering % Center all text

\begin{center}
{\LARGE {\sf The gradient flow coupling at high-energy \\[0.75ex] and
    the scale of SU(3) Yang-Mills theory}} \\[0.2\baselineskip]
\end{center}

%\rule{\textwidth}{0.4pt}\vspace*{-\baselineskip}\vspace{3.2pt} % Thin horizontal line
}
\begin{center}
  {\large Mattia~Dalla~Brida$^a$ and Alberto~Ramos$^b$} 
\end{center}
\vspace{0.2cm}
\begin{center}
  $^a${Dipartimento di Fisica, Universit\`a di Milano-Bicocca
                    and INFN, sezione di Milano-Bicocca, \\ Piazza della Scienza 3,
                    I-20126 Milano, Italy}\\
                  \texttt{<mattia.dallabrida@unimib.it>} \\
  $^b${School of Mathematics \& Hamilton Mathematics Institute, Trinity College Dublin, Dublin
  2, Ireland}\\
  \texttt{<alberto.ramos@maths.tcd.ie>}
\end{center}

\vspace{1cm}
\begin{center}
  \large{\sf Abstract}
\end{center}
\rule{\textwidth}{0.4pt}
\noindent
Using finite size scaling techniques and a
renormalization scheme based on the Gradient Flow, we determine
non-perturbatively the $\beta$-function of the $SU(3)$ Yang-Mills theory 
for a range of renormalized couplings $\bar g^2\sim 1-12$. We perform a
detailed study of the matching with the asymptotic NNLO  perturbative  
behavior at high-energy, with our non-perturbative data showing 
a significant deviation from the perturbative prediction down to $\bar{g}^2\sim1$. 
We conclude that schemes based on the Gradient Flow are not
competitive to match with the asymptotic perturbative behavior, even
when the NNLO expansion of the $\beta$-function is known. On the other
hand, we show that matching non-perturbatively the Gradient Flow
to the Schr\"odinger Functional scheme allows us to make safe contact 
with perturbation theory with full control on truncation
errors. This strategy allows us to obtain a precise determination of the
$\Lambda$-parameter of the $SU(3)$ Yang-Mills theory in units of a reference 
hadronic scale ($\sqrt{8t_0}\, \Lambda_{\overline{\rm MS}} = 0.6227(98)$), 
showing that a precision on the QCD coupling below 0.5\% per-cent can 
be achieved using these techniques. 
%%% Local Variables:
%%% mode: latex
%%% TeX-master: "paper"
%%% End:

\\\rule{\textwidth}{0.4pt}\\[\baselineskip] % Thin horizontal line

\tableofcontents

\newpage
\endgroup}

\begingroup % Create the command for including the title page in the document
\thispagestyle{empty}

{\centering % Center all text

\begin{center}
{\LARGE {\sf The gradient flow coupling at high-energy \\[0.75ex] and
    the scale of SU(3) Yang-Mills theory}} \\[0.2\baselineskip]
\end{center}

%\rule{\textwidth}{0.4pt}\vspace*{-\baselineskip}\vspace{3.2pt} % Thin horizontal line
}
\begin{center}
  {\large Mattia~Dalla~Brida$^a$ and Alberto~Ramos$^b$} 
\end{center}
\vspace{0.2cm}
\begin{center}
  $^a${Dipartimento di Fisica, Universit\`a di Milano-Bicocca
                    and INFN, sezione di Milano-Bicocca, \\ Piazza della Scienza 3,
                    I-20126 Milano, Italy}\\
                  \texttt{<mattia.dallabrida@unimib.it>} \\
  $^b${School of Mathematics \& Hamilton Mathematics Institute, Trinity College Dublin, Dublin
  2, Ireland}\\
  \texttt{<alberto.ramos@maths.tcd.ie>}
\end{center}

\vspace{1cm}
\begin{center}
  \large{\sf Abstract}
\end{center}
\rule{\textwidth}{0.4pt}
\noindent
Using finite size scaling techniques and a
renormalization scheme based on the Gradient Flow, we determine
non-perturbatively the $\beta$-function of the $SU(3)$ Yang-Mills theory 
for a range of renormalized couplings $\bar g^2\sim 1-12$. We perform a
detailed study of the matching with the asymptotic NNLO  perturbative  
behavior at high-energy, with our non-perturbative data showing 
a significant deviation from the perturbative prediction down to $\bar{g}^2\sim1$. 
We conclude that schemes based on the Gradient Flow are not
competitive to match with the asymptotic perturbative behavior, even
when the NNLO expansion of the $\beta$-function is known. On the other
hand, we show that matching non-perturbatively the Gradient Flow
to the Schr\"odinger Functional scheme allows us to make safe contact 
with perturbation theory with full control on truncation
errors. This strategy allows us to obtain a precise determination of the
$\Lambda$-parameter of the $SU(3)$ Yang-Mills theory in units of a reference 
hadronic scale ($\sqrt{8t_0}\, \Lambda_{\overline{\rm MS}} = 0.6227(98)$), 
showing that a precision on the QCD coupling below 0.5\% per-cent can 
be achieved using these techniques. 
%%% Local Variables:
%%% mode: latex
%%% TeX-master: "paper"
%%% End:

\\\rule{\textwidth}{0.4pt}\\[\baselineskip] % Thin horizontal line

\tableofcontents

\newpage
\endgroup

\section{Introduction}

Yang-Mills (YM) theories play a central role in our understanding of
natural laws. They lie at the heart of the unification between the
electromagnetic and weak interactions and at the foundations of
Quantum Chromo Dynamics (QCD). The (pure) $SU(3)$ YM theory shares many of the
interesting features of QCD. It is a strongly coupled non-abelian gauge 
theory where the fundamental degrees of freedom, the gluons, are not part 
of the spectrum of the theory. The theory has an intrinsic energy scale given 
by the $\Lambda$-parameter, and at energy scales much larger than 
$\Lambda$ it is well-approximated by perturbation theory
(i.e.~the theory is asymptotically free~\cite{PhysRevLett.30.1343,
  PhysRevLett.30.1346}). Connecting this high energy perturbative
regime of the theory with its low energy spectrum is a difficult
multi-scale problem, very similar to the one faced when one wants to determine the
strong coupling or quark masses in the Standard Model (SM). On the other hand, 
the pure gauge theory is much more tractable from a computational point of 
view when lattice field theory methods are employed. The currently known simulation 
algorithms are far more efficient in the case of the pure gauge theory than in
the case of QCD. In summary, the $SU(3)$ YM theory is very interesting in its own, 
and a perfect laboratory for testing new lattice techniques and ideas before 
applying them to QCD. 

In this work we are interested in the determination of the intrinsic energy scale of  
the $SU(3)$ YM theory, i.e., its $\Lambda$-parameter. In the case of QCD this is equivalent 
to determining the value of the strong coupling. We make use of lattice field theory 
methods~\cite{Wilson:1977nj, Creutz:1980zw}, which allow us to determine non-perturbatively 
the running of the gauge coupling. In particular, thanks to the techniques of finite size
scaling~\cite{Luscher:1991wu, Luscher:1992an}, this running can be
computed over a wide range of renormalization scales. These two
ingredients of our analysis are what allows us to bridge the
3 orders of magnitude that separate the typical energy scales of
the hadronic, strongly coupled regime of the theory, and the high energy 
perturbative regime, \emph{without} making any assumption on the
validity of perturbation theory (see the discussion
in~\cite{DallaBrida:2018rfy, Brida:2016flw}). 

A similar study using the Schr\"odinger Fucntional (SF)
coupling was one of the earliest applications 
of step scaling techniques in this context~\cite{Luscher:1993gh}. More than 20
years later, we are now capable of a much more precise 
computation. This not only because of the increase in computational power over the years, 
but also thanks to the development of new theoretical tools. In particular, the
Gradient Flow (GF)~\cite{Narayanan:2006rf, Luscher:2010iy} allowed us to
introduce new coupling definitions~\cite{Luscher:2010iy,Fodor:2012td,
  Fritzsch:2013je, Ramos:2014kla} which are very compelling for step
scaling studies.  These couplings are constructed from observables with very small
variance and very special properties under renormalization
(see~\cite{Ramos:2015dla} for a comparison among different coupling
definitions). As a result, GF couplings permit to achieve an excellent
statistical precision, especially in the low energy regime of the
theory, where the traditional SF coupling struggles to produce precise
results.  

GF-based couplings, however, have their issues, too. One 
of the main issues are the relatively large cutoff 
effects that have been observed in several studies (see~\cite{Ramos:2015dla} 
for a discussion). Despite a solid theoretical understanding on the anatomy
of these discretization effects~\cite{Ramos:2015baa}, these remain 
the main source of concern in most applications. Thus, using GF couplings one 
can achieve very high statistical precision, but accurate continuum results
require the simulation of large lattices. There are also other challenges if 
one wants to use these couplings for a precise determination of the
strong coupling. First, the known perturbative coefficients of the
$\beta$-function show bad convergence~\cite{Harlander:2016vzb,
  DallaBrida:2017tru}. This implies that even simple estimates of the 
theoretical perturbative uncertainties in the extraction of $\alpha_s$ based on 
GF couplings, even at energies as high as the electroweak scale, are about half
a percent. Second, the relative statistical precision on the GF couplings is typically
$\delta\alpha_{\rm GF}/\alpha_{\rm GF}\propto {\rm const.}$, while the traditional
SF coupling has $\delta\alpha_{\rm SF}/\alpha_{\rm SF}\propto \alpha_{\rm SF}$.
This means that eventually the numerical cost of the GF couplings will be larger 
than for the SF couplings when $\alpha \ll 1$. These issues explain the strategy 
followed by the ALPHA collaboration in the determination of $\alpha_s$~\cite{Brida:2016flw, 
DallaBrida:2016kgh,Bruno:2017gxd, DallaBrida:2018rfy}, where the SF coupling is still
the coupling of choice at high energies (see also refs.~\cite{Korzec:2017ypb,DallaBrida:2018cmc} 
for recent reviews).

In this work we will explore in detail the high energy sector of the
$SU(3)$ gauge theory using the Gradient Flow. One of our aims is
to assess whether precise results for $\alpha_s$ can be achieved using GF couplings. As
mentioned earlier we will have to face the large truncation errors in
these schemes due to the bad behaviour of their perturbative series. 
We will show that, in fact, due to the large truncation effects, it is very
challenging to obtain precise results for the $\Lambda$-parameter using only these
schemes. We will also pay special attention to the linear $\mathcal O(a)$
effects that typically arise in finite volume renormalization
schemes that break translational invariance. They are the main source
of systematic effects in computations based on the SF
couplings~\cite{DallaBrida:2018rfy}, and we will see that GF schemes
allow us to substantially reduce these effects. 

The paper is organized as
follows. Section~\ref{sec:gener-strat-runn} introduces the GF and SF
schemes in our preferred finite volume
setup. Section~\ref{sec:grad-flow-coupl} introduces our numerical
setup and our determination of the running of the GF coupling at high
energies. In this section we shall discuss in detail the matching with 
the asymptotic perturbative behaviour. In section~\ref{sec:hadronic} we 
complete our determination of the running coupling in the non-perturbative 
domain and match our finite volume renormalization schemes with the infinite
volume reference scales $t_0$, and $r_0$. Section~\ref{sec:lambda-parameter}
presents our final results and offers a comparison with the data 
available in the literature, while section~\ref{sec:conclusions} presents the
conclusions of our work. A few appendices are moreover included to 
address some more technical details. In appendix~\ref{sec:UbyU} we provide
evidence that our results for the GF coupling, that span a 
factor 3 in lattice spacing, are suitable to produce accurate 
continuum limit extrapolations. Appendix~\ref{sec:oa} details our
procedure to estimate the boundary $\mathcal O(a)$ effects. Appendix~\ref{sec:ss} 
presents a more traditional analysis of our high energy data, and the related
extraction of the $\Lambda$-parameter. Appendix~\ref{sec:SFPT} discusses the perturbative 
improvement we applied to the SF coupling. Finally, appendix~\ref{sec:raw-data-meas}
contains all raw data measurements of our simulations.

%%% Local Variables:
%%% mode: latex
%%% TeX-master: "paper"
%%% End:

\section{General strategy: running couplings, schemes, and discretizations}
\label{sec:gener-strat-runn}

\subsection{Running couplings and the $\Lambda$-parameter}
\label{sec:general-strategy}

The $SU(3)$ Yang-Mills theory is a fairly simple theory. Formally,
it is defined in terms of a gauge potential $A_\mu$ which lives in 
the Lie algebra $\mathfrak{su}(3)$ of $SU(3)$, and by the action:
\begin{equation}
\label{eq:action}
  S[A_\mu] = \frac{1}{2g^2}\int {\rm d}^4x\, {\rm tr}\{F_{\mu\nu}(x)F_{\mu\nu}(x)\}\,,
\end{equation}
where $F_{\mu\nu}$ is the field strength tensor,
\begin{equation}
  F_{\mu\nu} = \partial_\mu A_\nu - \partial_\nu A_\mu + [A_\mu,A_\nu],
\end{equation}
and $g$ is a dimensionless free parameter: the gauge coupling. 
The renormalization of the theory requires us to introduce a \emph{renormalized}
coupling, $\bar{g}\equiv\bar{g}(\mu)$, through some suitable renormalization 
condition; different conditions define what we refer to as different renormalization
schemes for the coupling. Renormalized couplings depend explicitly on the energy scale, 
$\mu$, at which their defining conditions are imposed. The dependence on this scale
is encoded in their $\beta$-functions, $\beta(\bar{g})$, which are defined by the 
renormalization group (RG) equations:
\begin{equation}
  \label{eq:rg}
  \mu \frac{{\rm d}\bar g(\mu)}{{\rm d}\mu} = \beta(\bar g)\,.
\end{equation}
The $\beta$-functions have an asymptotic perturbative expansion:
\begin{equation}
  \label{eq:betaPT}
  \beta(\bar g) \simas{\bar g\to 0}  -\bar g^3
  \sum_{k=0}b_k\bar g^{2k} \,,
\end{equation}
where the first two coefficients,
\begin{equation}
  \label{eq:beta_univ}
    b_0 = \frac{11}{(4\pi)^2}
    \quad
    \text{and}
    \quad
    b_1 = \frac{102}{(4\pi)^4}\,,
\end{equation}
are universal, i.e., they are independent of the specific renormalization scheme chosen
for the coupling. The scheme dependence only enters through the higher-order coefficients 
$b_k$, with $k>2$.%
\footnote{We note while passing that, at present, the perturbative $\beta$-function is 
	most accurately known in the $\overline{\rm MS}$ scheme of dimensional regularization, 
	where the $b_k$-coefficients have been computed up to $k=4$~\cite{vanRitbergen:1997va,Czakon:2004bu,Baikov:2016tgj,Luthe:2016ima,Herzog:2017ohr}. Other specific 
	cases will be presented in detail below.} 
This implies that although at "low" energy different coupling definitions can behave very 
differently as a function of $\mu$, at high-energy these differences must eventually 
disappear, as all definitions share the property of \emph{asymptotic freedom}: 
$\bar{g}(\mu) \overset{\mu\to\infty}{\to}0$. 

The first-order RG equation (\ref{eq:rg}) has as an implicit solution given by, 
\begin{equation}
  \label{eq:lamdef}
  \frac{\Lambda}{\mu} = 
  \left[b_0\bar g^2(\mu)\right]^{-\frac{b_1}{2b_0^2}}\,
  e^{-\frac{1}{2b_0\bar g^2(\mu)}}\,
  \exp\{-I_g(\bar{g}(\mu),0)\},
  \quad
  I_g(g_2,g_1)= \int_{g_1}^{g_2}{\rm d}x\, \left[\frac{1}{\beta(x)} + 
  \frac{1}{b_0x^3} - \frac{b_1}{b_0^2x}\right],  
\end{equation}
where $\Lambda$ is a constant of mass dimension one. Its value depends on the
exact renormalization scheme chosen for the coupling. From eq.~(\ref{eq:lamdef}) 
it is clear that, given the knowledge of the $\beta$-function, the $\Lambda$-parameter 
is all that is needed to infer the value of the coupling $\bar g(\mu)$ at any renormalization 
scale $\mu$. In particular, the coupling must be a function of $\mu/\Lambda$, which means 
that $\Lambda$ defines what is to be considered "low" or "high" energy.

The $\Lambda$-parameter is a compelling quantity to determine. First of all, its definition
does not rely on perturbation theory: it is non-perturbatively defined once the corresponding
coupling and $\beta$-function are. Second, it is a renormalization group invariant. As 
such it does not depend on any renormalization scale, i.e. $d\Lambda/d\mu=0$. In addition, 
even though its value depends on the scheme, this dependence can be
computed analytically. Given two renormalized couplings 
$\bar{g}_{\rm X}$ and $\bar{g}_{\rm Y}$ and the one-loop perturbative
relation 
\begin{equation}
  \label{eq:CouplingsMatching}
  \bar{g}^2_{\rm Y}(\mu)\oset[.5ex]{\mu\to\infty}{=}
  \bar{g}^2_{\rm X}(\mu)+c_1\,\bar{g}^4_{\rm X}(\mu)+
  \mathcal{O}(\bar{g}^6_{\rm X}(\mu)),
\end{equation} 
with $c_1$ a pure number, one can easily show using eqs.~(\ref{eq:betaPT})-(\ref{eq:lamdef})
that the corresponding $\Lambda$-parameters are \emph{exactly} related by:
\begin{equation}
  \label{eq:LambdaRatio}
  {\Lambda_{\rm Y}\over\Lambda_{\rm X}}= \exp\bigg\{{c_1\over 2b_0}\bigg\}.
\end{equation}
These properties make any $\Lambda$-parameter a natural reference scale for both the low 
and high energy regimes of the theory. In particular, any dimensionfull renormalization group 
invariant quantity, like for instance any "hadronic" quantity, must be proportional to 
$\Lambda$ (or some power of it).%
\footnote{In the following we shall loosely refer to (RG invariant) low-energy scales 
		 of	the $SU(3)$ Yang-Mills theory as "hadronic" scales/quantities, although strictly 
	     speaking there are no hadrons in this theory. Popular examples of low-energy scales
	     are, for instance, the energies composing the spectrum of the theory, the distance  
	     $r_0$ obtained from the potential between two static quarks~\cite{Sommer:1993ce}, and 
	     the gradient flow time $t_0$~\cite{Luscher:2010iy}. We will come back to some of these
	     in later sections.}
The proportionality constants that relate these quantities to $\Lambda$ are fundamental numbers 
that characterize the pure Yang-Mills theory, as they are given solely by the dynamics.

Lattice field theory is the only known framework that allows us to extract low-energy 
physics from first principles. In particular, the value of $\Lambda$ in units of a typical 
hadronic scale $\mu_{\rm had}$, for which $\bar{g}(\mu_{\rm had})=\bar{g}_{\rm had}$, can 
in principle be obtained by employing these techniques and eq.~(\ref{eq:lamdef}). The determination 
requires of course the knowledge of the corresponding $\beta$-function for all energies larger than 
$\mu_{\rm had}$. The basic strategy for this computation is to divide this energy range into two parts. 
While the non-perturbative $\beta$-function is computed from $\mu_{\rm had}$ up to some 
energy $\mu_{\rm PT}$, for which $\bar{g}_{\rm PT}=\bar{g}(\mu_{\rm PT})\ll \bar{g}_{\rm had}$, perturbation 
theory is used for energies larger than $\mu_{\rm PT}$. For large enough $\mu_{\rm PT}$, the relevant 
integral entering the definition of $\Lambda/\mu_{\rm had}$ can be approximated as (cf.~eq.~(\ref{eq:lamdef})):
\begin{equation}
  \label{eq:IntegralPT}
  I_g(\bar{g}_{\rm had},\bar{g}_{\rm PT}) +
  I_g^N(\bar{g}_{\rm PT},0) 
  \oset[1.5ex]{\mu_{\rm PT}\to 0}{\sim}
  I_g(\bar{g}_{\rm had},0) + \mathcal O(\bar  g^{2N-2}_{\rm PT})\,,
\end{equation}
where $I_g^N$ is analogously defined as $I_g$ by replacing the 
$\beta$-function with its perturbative expression up to some order $N$, i.e.,
\begin{equation}
  \label{eq:IntegralPT2}
  \beta(\bar{g})\to\beta_{\rm PT}^{(N)}(\bar{g}) = \sum_{k=0}^{N-1} b_k \bar{g}^{2k}
  \quad
  \Rightarrow
  \quad
  I_g\to I_g^N.
\end{equation}
The ratio of interest, $\Lambda/\mu_{\rm ref}$, is thus approximated by:
\begin{equation}
   \label{eq:lam_extraction}
	\varphi^{(N)}(\bar{g}_{\rm had},\bar{g}_{\rm PT})=
	\left[b_0\bar g^2_{\rm had}\right]^{-\frac{b_1}{2b_0^2}}\,
	e^{-\frac{1}{2b_0\bar g^2_{\rm had}}}\,
	\exp\{-I_g(\bar{g}_{\rm had},\bar{g}_{\rm PT})-I^N_g(\bar{g}_{\rm PT},0)\}
    \oset[0.75ex]{\mu_{\rm PT}\to\infty}{\sim}
	\frac{\Lambda}{\mu_{\rm had}}+ \mathcal O(\bar  g^{2N-2}_{\rm PT})\,.
\end{equation}
There are two important points we must stress here. First, in the extraction of 
the $\Lambda$-parameter the truncation errors are formally of $\mathcal O(\bar{g}^{2N-2}_{\rm PT})$ 
(cf.~eq.~(\ref{eq:lam_extraction})). Since the running of the coupling 
at high energies is logarithmic in $\mu/\Lambda$, reducing the size of these truncation errors of 
a given factor, requires a significantly larger change in the energy scale. Second, 
one must always remember that eq.~(\ref{eq:lam_extraction}) is only an asymptotic statement. 
In principle, non-perturbative corrections (e.g. power corrections) are also present when approximating 
the integral as in eq.~(\ref{eq:IntegralPT}). As we shall clearly see in the following, these  
issues imply that in order to accurately estimate the systematic uncertainties coming from the 
use of perturbation theory at high-energy, as well as to keep these uncertainties small, we must
study the applicability of perturbation theory over a wide range of energies, reaching up to very 
large scales. Therefore, a precise determination of the $\Lambda$-parameter in terms of low-energy 
scales is no easy task.

\subsection{Finite size scaling}
\label{sec:step-scaling}

It is certainly not obvious how in the strategy presented above one can reach large
energy scales via lattice field theory simulations. Indeed, the ultra-violet cutoff of the lattice
theory set by $1/a$, where $a$ is the lattice spacing, has to be much larger than the largest
energy scale one wants to reach; large discretization errors will otherwise affect the results.
At the same time, in order to have finite-volume effects both in the coupling and in
the hadronic quantities well under control, the infra-red cutoff set by the finite extent of the 
lattice, $L$, has to be at least a few femto-meters long. Given the fact that current computational 
resources allow us to simulate lattices with lattice sizes $L/a\sim \mathcal{O}(10^2)$, this 
significantly limits the range of energy scales that can actually be covered in a single lattice 
simulation.

Finite-size scaling techniques overcome these problems by integrating the RG equations non-perturbatively 
in a finite-volume renormalization scheme~\cite{Luscher:1991wu}. Within this strategy the coupling is defined 
through a finite volume observable, and the renormalization scale is identified with the infra-red cutoff, 
i.e., $\mu = 1/L$. Said it differently, we define a renormalized coupling through a finite volume effect. 
In this way, high renormalization scales can be reached by splitting the computation into several 
lattices of smaller and smaller physical size. A quantity of primary role in finite-size scaling 
studies is the step scaling function:
\begin{equation}
  \label{eq:sigma}
  \sigma_s(u) = \bar g^2(\mu/s)\big|_{\bar g^2(\mu) = u}\,.
\end{equation}
It is a discrete version of the $\beta$-function as it measures the change
in the coupling when the renormalization scale is varied by a finite factor
$s$. Note that once the step scaling function is known, the $\beta$-function can
also be determined, by simply noticing that:
\begin{equation}
  \label{eq:beta-sigma}
  \ln \frac{\mu_2}{\mu_1} = \int_{\bar{g}(\mu_1)}^{\bar{g}(\mu_2)} \frac{{\rm d}x}{\beta(x)}
  \qquad
  \Rightarrow
  \qquad
  \ln s = -\int_{\sqrt{u}}^{\sqrt{\sigma_s(u)}}\, \frac{{\rm d}x}{\beta(x)}\,.
\end{equation}

\subsection{Renormalization schemes}
\label{sec:RenormSchemes}

Analytic calculations in gauge theories are simplified by considering for the coupling
the so called modified minimal subtraction ($\overline{\rm MS}$) scheme of dimensional 
regularization. Although this is a convenient choice for perturbative calculations, the 
$\overline{\rm MS}$ scheme is only  defined within perturbation theory and therefore only
applicable at high energies. In this work we are going to employ different renormalization
schemes, which are non-perturbatively defined in a finite Euclidean space-time volume. 
In order to apply the idea of finite-size scaling, the renormalization scale of these couplings
must be linked to the physical size of the system, i.e. $\mu \propto 1/L$.
At high energies, perturbation theory can then be used to relate any of these schemes to the 
more conventional $\overline{\rm MS}$ scheme (cf.~eq.~(\ref{eq:CouplingsMatching})). Note that 
thanks to eq.~(\ref{eq:LambdaRatio}), $\Lambda_{\overline{\rm MS}}$ can be implicitly
defined \emph{non-perturbatively} through any non-perturbative scheme, even though the 
$\overline{\rm MS}$ scheme itself is intrinsically perturbative. 

\subsubsection{Boundary conditions and coupling definitions}

When studying the pure Yang-Mills theory in a finite space-time volume 
the choice of boundary conditions for the fields matters. In this work we consider
the $SU(3)$ Yang-Mills theory with Schr\"odinger functional (SF) boundary 
conditions~\cite{Luscher:1992an}. In this set-up, the gauge field is periodic
in the three spatial directions with period $L$, while its spatial components 
satisfy Dirichlet boundary conditions at Euclidean times $x_0=0,L$, i.e.,
\begin{eqnarray}
  A_k(x)\big|_{x_0=0} = C_k(\mathbf x)\,,\qquad
  A_k(x)\big|_{x_0=L} = C_k'(\mathbf x)\,,
\end{eqnarray}
where $C_k(\mathbf x),C'_k(\mathbf x)\in\mathfrak{su}(3)$ are  given
external fields. A first compelling feature of this type of boundary conditions is
that, for a proper choice of fields $C_k,C'_k$, the action has a unique
global minimum (up to gauge transformations). This avoids several complications when
doing perturbation theory in a finite volume with respect to the general case~\cite{GonzalezArroyo:1981vw}. 
Another advantage of using these boundary conditions is that the system can be probed by 
considering different boundary fields $C_k,C_k'$. In the following we focus on a 
particularly convenient family of fields, given by the Abelian, and spatially constant 
fields of the form~\cite{Luscher:1993gh}:
\begin{eqnarray}
  \label{eq:CCp}
  C_k &=& 
  \frac{i}{L}{\rm diag}
  \bigg\{\eta-{\pi\over 3},\,
  \eta\Big(\nu-{1\over2}\Big),\,
  -\eta\Big(\nu+{1\over 2}\Big)+{\pi\over 3}\bigg\}\,,\\
  C_k' &=& \frac{i}{L}{\rm diag}\bigg\{
  -\eta-\pi,\,
  \eta\Big(\nu+{1\over 2}\Big)+{\pi\over 3},\,
  -\eta\Big(\nu-{1\over 2}\Big)+{2\pi\over 3}\bigg\}\,,
\end{eqnarray}
where $\eta,\nu$ are some (dimensionless) real parameters. Derivatives of the effective 
action of the pure Yang-Mills theory with SF boundary conditions with respect to the 
parameter $\eta$ are renormalized quantities~\cite{Luscher:1992an}. 
They can hence be used to define renormalized couplings. In particular, we can 
define a family of SF couplings as~\cite{Luscher:1992an,Luscher:1993gh,Sint:1995ch,
Sint:2012ae,Brida:2016flw,DallaBrida:2018rfy}: 
\begin{equation}
  \frac{k}{\bar g^2_{{\rm SF},\nu}(\mu)} =  
  \left\langle \frac{\partial S}{\partial \eta}\right\rangle\bigg|_{\eta=0},
  \qquad 
  \mu= L^{-1},
  \qquad
  k=12\pi\,,
\end{equation}
where different values for the parameter $\nu$ define different renormalization schemes.
In fact, this family of couplings can be obtained from a linear combination 
of two distinct observables both defined for $\nu=0$, specifically,
\begin{equation}
  \label{eq:SFnu}
  \frac{1}{\bar g^2_{{\rm SF},\nu}(\mu)} =\frac{1}{\bar g^2_{{\rm SF}}(\mu)} -\nu\bar{v}(\mu),
\end{equation}
where, in terms of expectation values:
\begin{equation}
  \frac{k}{\bar g^2_{{\rm SF}}(\mu)} =  
  \left\langle \frac{\partial S}{\partial \eta}\right\rangle\bigg|_{\eta=\nu=0},
  \qquad
  \bar{v}(\mu) = -\frac{1}{k}  
  \left\langle \frac{\partial^2S}{\partial\eta\partial\nu}\right\rangle\Bigg|_{\eta=\nu=0}.
\end{equation}
The perturbative relation between the SF couplings defined above and the coupling 
in the $\overline{\rm MS}$ scheme, is known to two-loop order, and it is given 
by~\cite{Luscher:1993gh,Bode:1998hd,Bode:1999sm,DellaMorte:2004bc} ($\alpha_{\rm X}\equiv \bar{g}^2_{\rm X}/4\pi$):
\begin{equation}
  \label{eq:SFMS}
  \alpha_{{\rm SF},\nu}(r\mu) = \alpha_{\overline{\rm MS}}(\mu)+
  a^\nu_1(r)\alpha_{\overline{\rm MS}}^2(\mu) + a^\nu_2(r)
  \alpha_{\overline{\rm MS}}^3(\mu)+ \mathcal{O}(\alpha_{\overline{\rm MS}}^4(\mu))\,,
\end{equation}
where $r>0$ and 
\begin{equation}
  a^\nu_1(r) = a_1(r) + 4\pi v_1\nu\,, 
  \qquad
  a^\nu_2(r) - (a_1^\nu(r))^2= a_2(r)-(a_1(r))^2+(4\pi)^2v_2\nu\,,
\end{equation}
with
\begin{gather}
  a_1(r) = -8\pi b_0\log(r) - 1.255621(2), 
  \qquad
  a_2(r) -(a_1(r))^2 = -32\pi^2b_1\log(r) - 1.197(10)\,,
\end{gather} 
and
\begin{equation}
  v_1= 0.0694603(1),
  \qquad
  v_2 =-0.001364(14).
\end{equation}
The knowledge of the two-loop relation (\ref{eq:SFMS}) allows us to determine 
from the 3-loop $\beta$-function in the $\overline{\rm MS}$ scheme~\cite{Tarasov:1980au,Larin:1993tp}, 
the 3-loop $\beta$-function of the SF couplings,
which means to determine the first non-universal coefficient
(cf.~eq.~(\ref{eq:betaPT})):
\begin{equation}
 \label{eq:b2SFnu}
 (4\pi)^3 b^{{\rm SF},\nu}_2= 0.482(7) +(4\pi)^3 \nu(b_0v_2-b_1v_1)
 =  0.482(7) - \nu\times 0.7523(1)\,.
\end{equation}
Using eq.~(\ref{eq:LambdaRatio}) in conjunction with eq.~(\ref{eq:SFMS}), 
we can also obtain the ratio of the corresponding $\Lambda$-parameters:
\begin{equation}
\frac{\Lambda_{\overline{\rm MS}}}{\Lambda_{{\rm SF},\,\nu}}=\exp\bigg\{-\frac{a_1^\nu(1)}{8\pi b_0}\bigg\}.
\end{equation}
The result of eq.~(\ref{eq:b2SFnu}) shows that for values of $|\nu|=\mathcal{O}(1)$, 
the $b^{{\rm SF},\nu}_2$ coefficients are "naturally" small compared to the 
lowest-order results of eq.~(\ref{eq:beta_univ}). Within perturbation theory, one is
hence keen to expect that for these ${\rm SF}_\nu$ schemes truncation
errors in the $\beta$-function are small at small values of the
coupling $\alpha\equiv \bar g^2/(4\pi) \ll 1$.

Other compelling coupling definitions are possible within the SF framework. 
Of particular interest for our study are couplings defined through the Yang-Mills 
gradient flow (GF). The Gradient Flow evolves the gauge field according to a 
diffusion-like equation:
\begin{equation}
 \partial_t B_\mu(t,x) = D_\nu G_{\nu\mu}(t,x),\qquad
 B_\mu(0,x) = A_\mu(x), 
 \label{eq:YMflow}
\end{equation}
where $D_\mu = \partial_\mu + [B_\mu,\cdot]$ denotes the gauge
covariant derivative of the field $B_\mu$, and 
\begin{equation}
  G_{\mu\nu}= \partial_\mu B_\nu - \partial_\nu B_\mu + [B_\mu,B_\nu],
\end{equation}
is the corresponding field strength tensor. The flow time
$t$ has units of length squared, and $B_\mu(t,x)$ can be
seen as a smoothed version of the original gauge field $A_\mu(x)$ over
a length scale $\sim\sqrt{8t}$. The remarkable property of the flow fields 
is that gauge invariant operators made out of these fields are 
renormalized quantities for positive flow times, $t>0$~\cite{Luscher:2011bx}. 
This suggests that, for instance, the dimensionless quantity%
\footnote{We use the convention that ${\rm tr}\{T^aT^b\}=-\frac{1}{2}\delta_{ab}$, 
		  where $T^a$, $a=1,\ldots,8$, are generators of $\mathfrak{su}(3)$.}
\begin{equation}
  \label{eq:Eoft}
  t^2\langle E(t,x) \rangle\,,
  \qquad
  E(t,x)=-\frac{1}{2}{\rm tr}\{G_{\mu\nu}(t,x)G_{\mu\nu}(t,x)\},
\end{equation}
can be used to define renormalized couplings at a scale given by the (inverse)
flow time, e.g. $\mu = 1/\sqrt{8t}$. Clearly, this coupling definition does not 
rely on having specific SF boundary conditions, and for what matters in having
a finite-volume either. Hence, one can take for eq.~(\ref{eq:CCp}) the 
convenient choice: $C_k = C'_k = 0$. Due to the explicit breaking of rotational 
symmetry by the boundary conditions, one can in fact obtain two independent coupling
definitions by considering either the \emph{magnetic} or the \emph{electric}
components of the energy density, eq.~(\ref{eq:Eoft})~\cite{Fritzsch:2013je},
i.e.,
\begin{alignat}{2}
  \label{eq:GFCouplings}
  \bar g^2_{\rm GF,\,m}(\mu) &= 
  \mathcal N_{\rm m}^{-1}\, t^2\langle E_{\rm m}(t,x)\rangle\big|_{\mu=1/\sqrt{8t},\sqrt{8t}=cL,\,x_0=L/2}\,,
  \quad
  E_{\rm m}(t,x)&=-\frac{1}{2}{\rm tr}\{G_{ij}(t,x)G_{ij}(t,x)\},  \\
  \label{eq:GFCouplings2}
  \bar g^2_{\rm GF,\,e}(\mu) &= 
  \mathcal N_{\rm e}^{-1}\,t^2\langle E_{\rm e}(t,x) \rangle\big|_{\mu=1/\sqrt{8t},\sqrt{8t}=cL,\,x_0=L/2}\,,
  \quad
  E_{\rm e}(t,x)&= -\frac{1}{2}{\rm tr}\{G_{0i}(t,x)G_{0i}(t,x)\},
\end{alignat}
where $\mathcal{N}_m$, $\mathcal{N}_e$ are some constants that guarantee the correct 
normalization of the coupling~\cite{Fritzsch:2013je}. In order to properly define a 
coupling suitable for finite-size scaling, we must relate the renormalization scale 
at which the coupling is defined with the size of the finite-volume, we thus 
set: $\mu = 1/\sqrt{8t} = 1/(cL)$, where the constant $c$ is part of the 
scheme definition. In this work we will exclusively take $c=0.3$; the merits 
of this choice have been discussed in ref.~\cite{Fritzsch:2013je}. We note while passing 
that, at fixed flow time, the limit $c\to0$ corresponds to the analogous GF coupling 
definition in infinite space-time volume~\cite{Luscher:2010iy}. In addition,
we choose to measure the flow energy densities for $x_0=L/2$ in order to maximize the 
distance of the observables from the space-time boundaries. When looking at the couplings 
on the lattice, this minimizes the $\mathcal{O}(a)$ contaminations coming from the lattice 
action (cf.~Sect.~\ref{sec:LatticeSetup}).
		
Perturbative computations for flow quantities are usually
involved. Already the 1-loop relation of the GF coupling with the
$\overline{\rm MS} $ coupling in infinite
volume~\cite{Luscher:2010iy} is a challenging computation. On a finite
volume the computation is even more involved
(see~\cite{Bribian:2019ybc}). The two-loop relation requires
substantial effort~\cite{Harlander:2016vzb,DallaBrida:2017tru}, and for our 
particular choice of boundary conditions the result relies on novel 
methods~\cite{Brida:2013mva,DallaBrida:2016dai,DallaBrida:2017pex,DallaBrida:2017tru}
within the framework of Numerical Stochastic Perturbation Theory (NSPT). 
The results are~\cite{DallaBrida:2017tru}:
\begin{equation}
  \label{eq:GFMS}
  \alpha_{{\rm GF},m/e}(\mu) = 
  \alpha_{\overline{\rm MS}}(\mu) + 
  k_1^{m/e} \alpha_{\overline{\rm MS}}^2(\mu) +
  k_2^{m/e}\alpha_{\overline{\rm MS}}^3(\mu) + \mathcal{O}(\alpha_{\overline{\rm MS}}^4(\mu))\,,
\end{equation}
where the coefficients $k_{1,2}^{m/e}$ are collected in table~\ref{tab:gfpt}.
As for the case of the SF couplings, the 2-loop relations (\ref{eq:GFMS}) 
allow us to infer the 3-loop coefficients of the $\beta$-functions of the 
GF couplings using the known results in the ${\overline{\rm MS}}$ scheme,
this yields:
\begin{equation}
  \label{eq:b2GF}
  (4\pi)^3 b_2^{{\rm GF},m} = -3.271(47),
  \qquad
  (4\pi)^3 b_2^{{\rm GF},e} = -2.004(55).
\end{equation}
For the ratios of $\Lambda$-parameters we obtain instead:
\begin{equation}
  \label{eq:LambdaGF2MS}
  \frac{\Lambda_{\overline{\rm MS}}}{\Lambda_{{\rm GF},m}} =  0.4981(17),
\qquad
  \frac{\Lambda_{\overline{\rm MS}}}{\Lambda_{{\rm GF},e}} =  0.5632(23)\,.
\end{equation}
It is interesting to compare these results with those of the analogous GF coupling 
definition in infinite space-time volume ($c=0$)~\cite{Luscher:2010iy,Harlander:2016vzb}. 
For this scheme, the electric and magnetic definitions coincide, and we have~\cite{Harlander:2016vzb}:
 \begin{equation}
   (4\pi)^3 b_{2}^{{\rm GF}} =  -1.90395(4),
   \qquad
   \frac{\Lambda_{\overline{\rm MS}}}{\Lambda_{{\rm GF}}} =  0.534162960405763.
 \end{equation}
Given the above results, it is clear that while the magnetic results for $b_2^{{\rm GF},m}$
are significantly larger than for the infinite volume case, those for the electric
components are similar.\footnote{In fact it seems that the $c$-dependence of $b_2^{{\rm GF},e}$
is quite mild, and its value is close to the corresponding infinite volume one also for 
values of $c$ as large as 0.4 (cf.~ref.~\cite{DallaBrida:2017tru}).} 
In all these cases, however, the 3-loop coefficient is "unnaturally" large and of
opposite sign if compared with the lower-order ones (cf.~eq.~(\ref{eq:beta_univ})).
It is also significantly larger that the ${\rm SF}_\nu$ schemes previously discussed 
(cf.~eq.~(\ref{eq:b2SFnu})). Already from a purely perturbative point of view, 
one is thus worried that higher-order corrections to the $\beta$-function
may be large, even if the coupling is relatively small. These concerns 
will be in fact confirmed by our non-perturbative investigation.

To conclude, for the following it is also useful to work out the perturbative two-loop 
relation between the SF and the GF couplings. Combining eq.~(\ref{eq:SFMS}) with (\ref{eq:GFMS}),
we have:
\begin{equation}
  \label{eq:SF2GF}
  \alpha_{{\rm SF},\nu}(r\mu) = 
  \alpha_{{\rm GF},e/m}(\mu) + 
  d^{\nu,\,e/m}_1(r) \alpha^2_{{\rm GF},e/m}(\mu) + 
  d^{\nu,\,e/m}_2(r)\alpha_{{\rm GF},e/m}^3(\mu)(\mu)+
  \mathcal{O}(\alpha_{{\rm GF},e/m}^4(\mu))\,,
\end{equation}
where
\begin{equation}
	d^{\nu,\,e/m}_1(r) = a^\nu_1(r) - k_1^{e/m},
	\qquad
	d^{\nu,\,e/m}_2(r) -(d^{\nu,\,e/m}_1(r))^2= a_2^\nu(r)-
	(a_1^\nu(r))^2-(k_2^{e/m}-(k_1^{e/m})^2).
\end{equation}

\begin{table}
	\centering
	\begin{tabular}{lllll}
		\toprule
		$c$&$k_1^m$&$k_1^e$&$k_2^m$&$k_2^e$ \\ 
		\midrule
		0.3 &1.220(6) &1.005(7) &-2.17(5) &-1.36(6) \\
		0 & 1.097786736 & 1.097786736 &-0.98225(5) & -0.98225(5)\\
		\bottomrule
	\end{tabular}
	\caption{Coefficients for the perturbative expansion of the magnetic
		and electric GF couplings for two values of $c$ (see eq.~(\ref{eq:GFMS})). 
		The finite-volume SF definition corresponds to $c=0.3$, while $c=0$ refers 
		to the infinite volume definition of the coupling.}
	\label{tab:gfpt}
\end{table}

%%% Local Variables:
%%% mode: latex
%%% TeX-master: "paper"
%%% End:

\section{The gradient flow coupling at high-energy}
\label{sec:grad-flow-coupl}

\subsection{Lattice set-up}
\label{sec:LatticeSetup}

We regularize the $SU(3)$ Yang-Mills theory in terms of the link
variables $U_\mu(x)\in SU(3)$, on a lattice of size $L/a$ in all four
space-time dimensions, where $L$ is the physical size of the lattice
and $a$ is its spacing. For the lattice action we take the standard
Wilson (plaquette) gauge action:
\begin{equation}
  \label{eq:Waction}
  S_{\rm W}[U] = \frac{\beta}{6}\sum_{p}w(p){\rm tr}(1-U_p)\,,
\end{equation}
where the sum is over all the plaquettes of the lattice, and $U_p $
denotes the product of the gauge links around the plaquette
$p$. $\beta=6/g^2_0$, with $g_0$ the \emph{bare} gauge coupling. Due
to our choice of SF boundary conditions we included in
eq.~(\ref{eq:Waction}) the weight factor $w(p)$:
\begin{equation}
  w(p) = \left\{
    \begin{array}{ll}
      c_t(g_0)& \text{if $p$ has one spatial-link on the time-slices $x_0=0,L$,}\\
      1& \text{otherwise.}
    \end{array}
  \right.
\end{equation}
The coefficient $c_t$ can in principle be tuned to cancel the
$\mathcal O(a)$ discretization errors stemming from the boundary of
the lattice~\cite{Luscher:1992an}.  Unfortunately, however, only
perturbative estimates are currently
available~\cite{Luscher:1993gh,Bode:1998hd,Bode:1999sm}. In the
following, we employ the two-loop result~\cite{Bode:1999sm}:
\begin{equation}
  \label{eq:ctstar}
  c_t(g_0) = 1 - 0.08900\times g_0^2 - 0.0294\times g_0^4\,,
\end{equation}
and estimate through dedicated simulations the effect of having the
coefficient truncated at this order (cf.~Appendix~\ref{sec:oa}).

On the lattice, the SF boundary conditions are imposed by setting~\cite{Luscher:1992an}:
\begin{equation}
  U_k(x)\big|_{x_0=0} = \exp(C_k(\mathbf x))\,,
  \qquad
  U_k(x)\big|_{x_0=L} = \exp(C_k'(\mathbf x))\,,
\end{equation}
where $C_k(\mathbf x),C_k'(\mathbf x)$ are either equal to
(\ref{eq:CCp}) with $\eta=\nu=0$, or to $C_k = C_k' = 0$, depending on
whether we are interested in measuring the SF or the GF couplings.

Starting from these definitions a lattice regularization of the SF
couplings naturally follows~\cite{Luscher:1992an,Luscher:1992zx,Luscher:1993gh}. We
refer the reader to the original references for the details. For the
case of the GF couplings, instead, there is quite more freedom in
their lattice definition. First of all, we need to specify a
discretization for the flow equations (\ref{eq:YMflow}). A popular
choice is the \emph{Wilson flow} (no summation over
$\mu$)~\cite{Luscher:2010iy}:
\begin{equation}
  \label{eq:wflow}
  a^2\left(\partial_t V_\mu(t,x)\right) V_\mu(t,x)^\dagger  = 
  -g_0^2 \partial_{x,\mu} S_{\rm W}[V],\qquad V_\mu(0,x) = U_\mu(x) \,,  
\end{equation}
where $V_\mu$ is the lattice flow field and
$\partial_{x,\mu} S_{\rm W}[V]$ is the force deriving from the Wilson
action, eq.~(\ref{eq:Waction}).  The Wilson flow describes the
continuum flow equations up to $\mathcal O(a^2)$ errors.%
\footnote{Note that in order to avoid $\mathcal O(a)$ discretization
  effects with SF boundary conditions one must set $c_t=1$ in
  $S_{\rm W}$ entering the flow equations~\cite{Luscher:2014kea}.}
This can be improved to $\mathcal{O}(a^4)$ by considering the
\emph{Zeuthen flow} (again no summation over
$\mu$)~\cite{Ramos:2015baa}:
\begin{equation}
  \label{eq:zflow}
  a^2\left(\partial_t V_\mu(t,x)\right) V_\mu(t,x)^\dagger  = 
  -g_0^2 \left(1 + 
    \frac{a^2}{12}\Delta_\mu  
  \right) \partial_{x,\mu} S_{\rm LW}[V],\qquad V_\mu(0,x) = U_\mu(x) \,,  
\end{equation}
where $\partial_{x,\mu} S_{\rm LW}[V]$ is now the force deriving from
the Symanzik tree-level $\mathcal{O}(a^2)$ improved (L\"uscher-Weisz)
gauge action, $S_{\rm LW}$~\cite{Luscher:1984xn}.%
\footnote{In the case of SF boundary conditions the exact definition
  of the L\"uscher-Weisz gauge action~\cite{Luscher:1984xn} near the
  time boundaries is not unique. Here we consider the definition of
  ref.~\cite{Luscher:2014kea}.}  Further details can be found in
ref.~\cite{Ramos:2015baa}. Here we just want to comment that the term
proportional to $\Delta_\mu=\nabla_\mu^\ast\nabla_\mu^{}$ is included
for all links, except for those temporal links touching the SF
boundaries at $x_0=0,L$. In this case we set $\Delta_0=0$.

In order to define the GF couplings on the lattice we also need to
specify a discretization for the $E_{\rm e/m}$-fields entering the
definitions, eqs.~(\ref{eq:GFCouplings}). We consider the following
two options (cf.~ref.~\cite{Ramos:2015baa}).  In the case where the
lattice flow is given by Wilson flow (\ref{eq:wflow}), we choose to
discretize $E_{\rm e/m}$ in terms of the clover definition of
$G_{\mu\nu}$ (see also ref.~\cite{Luscher:2010iy}). On the other hand,
in the case where the Zeuthen flow (\ref{eq:zflow}) is used, we take
the $\mathcal O(a^2)$ improved combination:
$E_{\rm e/m}=\frac{4}{3}\, E^{\rm pl}_{\rm e/m} - \frac{1}{3}\,E^{\rm
  cl}_{\rm e/m}$, where $E^{\rm pl}$ and $E^{\rm cl}$, are the energy
densities discretized in terms of the plaquette action density and the
clover definition of the flow field strength tensor,
respectively~\cite{Ramos:2015baa}. Based on an analysis in terms of
Symanzik effective theory the Zeuthen flow/improved observable
combination is preferable, as this choice does not introduce
$\mathcal O(a^2)$ effects when integrating the flow equations or when
evaluating the operators at positive flow times.  In this case,
indeed, $\mathcal O(a^2)$ discretization effects come only from the
action, eq.~(\ref{eq:Waction}) (which also introduces $\mathcal{O}(a)$
effects), and from the incomplete knowledge of a flow improvement
coefficient, $c_b(g_0)$, at $t=0$ (see~\cite{Ramos:2015baa} for a
complete discussion). The numerical experience gained so far seems to
indicate that this results in a better $\mathcal{O}(a^2)$-scaling for
the Zeuthen/improved observable combination than the Wilson
flow/clover one (see e.g.~ref.~\cite{DallaBrida:2016kgh}).

Before giving the final expression for our lattice definition of the
GF coupling we must address one last important point. It is well-known
that numerical simulations of the SF at lattice spacings
$a\lesssim 0.05\,{\rm fm}$ and with $L\gtrsim 0.5\,{\rm fm}$, tend to
show large autocorrelation times due to the infamous problem of
topology freezing~\cite{DelDebbio:2004xh,
  Fritzsch:2013yxa,Luscher:2014kea}. As suggested
in~\cite{Fritzsch:2013yxa}, this problem can be circumvent by defining
the coupling within the sector of topologically trivial gauge
fields. The continuum definitions (\ref{eq:GFCouplings})-(\ref{eq:GFCouplings2}) 
are replaced in this case by:
\begin{equation}
  \bar g^2_{\rm GF,\,m,e}(\mu)\to 
  \bar g^2_{\rm GF,\,m,e}(\mu)=\mathcal N_{\rm m,e}^{-1}\, 
  \frac{t^2\langle E_{\rm m,e}(t,x)\delta_Q \rangle\,}
  {\langle \delta_Q\rangle}\bigg|_{\mu=1/\sqrt{8t},\sqrt{8t}=cL,\,x_0=L/2}\,,
\end{equation}
where $\delta_Q$ is a Dirac $\delta$-function that enforces the
topological charge $Q$ of the gauge fields integrated over in the
functional integral to be zero. We note that this modification
actually defines different renormalization schemes than
eqs.~(\ref{eq:GFCouplings}).  On the other hand, these two set of
couplings are indistinguishable from a perturbative point of view and
thus share the very same perturbative results given in
Sect.~\ref{sec:RenormSchemes}.  On the lattice, we can finally define
the new GF couplings through the expression:
\begin{equation}
  \label{eq:lattgbar}
  \bar g^2_{\rm m,e}(\mu) = t^2\hat{\mathcal N}^{-1}_{\rm e,m}(c,a/L)
  \frac{\langle E_{\rm m,e}(t,x)  
    \hat\delta_Q\rangle}{\langle \,\hat\delta_Q \rangle}
  \bigg|_{\mu=1/\sqrt{8t},\, \sqrt{8t} = cL,\, x_0=T/2} 
  \quad
  (c=0.3)\,.
\end{equation}
To define the topological charge on the lattice we use the clover
discretization of the flow strength tensor~\cite{Luscher:2010iy}:
\begin{equation}
  \label{eq:qtop}
  Q = -\frac{1}{16\pi^2}\sum_x  \epsilon_{\mu\nu\rho\sigma}\,
  {\rm tr}\big\{G_{\mu\nu}^{\rm cl}(t,x)G_{\rho\sigma}^{\rm cl}(t,x)\big\}\,, 
\end{equation}
measured at flow time $\sqrt{8t}=cL$. For the discretization of the
flow equations used for $Q$ we will always use the one employed for 
the discretization of $E_{\rm e/m}$ entering the coupling definition. In addition,
since on the lattice $Q$ is not integer-valued, we replace the Dirac
$\delta$-function with:
\begin{equation}
  \label{eq:deltaq}
  \hat\delta_Q = 
  \begin{cases} 
    1\,, & \text{if } |Q|<0.5,\\
    0\,, & \text{otherwise}\,.
  \end{cases}
\end{equation}
We conclude by noticing that, as proposed in~\cite{Fritzsch:2013je},
the normalization factors $\hat{\mathcal N}_{\rm e,m}(c,a/L)$ are
better computed in lattice rather than continuum perturbation theory,
using the very same lattice discretization employed in the simulations
(which includes the definition of the lattice action, flow, and
observable).%
\footnote{We note that, in fact, we computed the norms $\mathcal{N}_{\rm m,e}$ in 
  tree-level lattice perturbation theory using the set-up described in ref.~\cite{Rubeo:2016szq}. 
  This set-up differs from ours by the way the the temporal links touching 
  the SF boundaries are treated in the Zeuthen flow equation 
  (see ref.~\cite{Rubeo:2016szq} for the details). This difference, 
  however, is an $\mathcal O(a^2)$ effect, which in practice is 
  well below the statistical precision of our non-perturbative data.}
This guarantees that the exact relation:
$\bar g^2_{\rm GF,\,e/m}=g_0^2+\mathcal O(g_0^4)$, holds. All
discretization effects are hence removed at tree-level in perturbation
theory.  For completeness, we collect in table~\ref{tab:norms} the
relevant values of the coupling norms used in this study.

\begin{table}
  \centering
  \begin{tabular}{lllllllllll}
    \toprule
    $L/a$ & 8 & 10 & 12 & 16 & 20 & 24 & 32 & 48\\
    \midrule
    $10^3\times \hat {\mathcal N}_{\rm Z,m}$& 9.73196 & 9.18466 & 8.95746 & 8.76640 & 8.68812 & 8.64814 & 8.61011 & 8.58400\\
    $10^3\times \hat {\mathcal N}_{\rm Z,e}$& 9.90101 & 9.35738 & 9.13180 & 8.94203 & 8.86425 & 8.82451 & 8.78671 & 8.76076\\
    $10^3\times \hat {\mathcal N}_{\rm W,m}$& 7.88614 & 8.14101 & 8.27366 & 8.40243 & 8.46103 & 8.49260 & 8.52383 & 8.54603\\
    $10^3\times \hat {\mathcal N}_{\rm W,e}$& 8.08018 & 8.32976 & 8.45911 & 8.58431 & 8.64116 & 8.67177 & 8.70201 & 8.72349\\
    \bottomrule
  \end{tabular}
  \caption{Lattice norms, eq.~(\ref{eq:lattgbar}). In all cases we use
    $c=0.3$. The labels ${\rm W/Z}$ refers to the Wilson/Zeuthen
    discretization and the labels ${\rm m/e}$ to the magnetic/electric
    GF coupling. See text for more details.}
  \label{tab:norms}
\end{table}

\subsection{Lattice step-scaling function}

As discussed in Sect.~\ref{sec:step-scaling}, the strategy to
determine the non-perturbative running of a renormalized coupling
using finite-size scaling techniques relies on the computation of the
step-scaling function (SSF),
\begin{equation}
  \sigma_s(u) = \bar g^2(\mu/s)\big|_{u=\bar g^2(\mu)}\,,
\end{equation}
where for a finite-volume renormalization scheme, $\mu\propto
1/L$. The corresponding $\beta$-function can then be determined from
$\sigma_s(u)$ using relation (\ref{eq:beta-sigma}). On the lattice, it
is actually straightforward to measure a lattice approximation of the
step scaling function. Indeed, we define the latter as:
\begin{equation}
  \label{eq:Sigma}
  \Sigma_s(u,a/L) = \bar g^2(\mu/s)\big|_{\bar g^2(\mu) = u}\,.
\end{equation}
It is computed by measuring the renormalized coupling on lattices of
size $L/a$ and $sL/a$, at the same value of the bare coupling
$g_0$. The continuum step scaling function eq.~(\ref{eq:sigma}) is
then obtained by taking the continuum limit at a fixed value of the
renormalized coupling $\bar g(\mu)$, i.e.,
\begin{equation}
  \lim_{a/L \to 0}\Sigma_s(u,a/L) = \sigma_s(u)\,.
\end{equation}
In the following section will apply this strategy to the GF couplings
and discuss the determination of their $\beta$-functions at relatively
high-energy scales. As we shall see, the approach to the perturbative
asymptotic regime is dramatically slow for these schemes.  This poses
some severe limitations if one aims at extracting the
$\Lambda$-parameter to high-precision using these coupling
definitions.

\subsection{Datasets}
\label{sec:DataSetGF}

We measured the GF couplings on lattices with
$L/a=8,10,12,16,20,24,32,48$, and for values of the bare coupling,
$\beta\in[6,11]$. In total we collected between $1,000-40,000$
measurements depending on the exact ensemble.  The complete list of
simulation parameters and corresponding results is given in
Appendix~\ref{sec:raw-data-meas}. This choice of parameters covers a
range of renormalized couplings:
$u=\bar g^2_{\rm GF}(\mu)\sim [1,12]$, and it allows us to
determine the lattice step-scaling function $\Sigma_2(u,a/L)$ for
$L/a=8,10,12,16,24$, and $\Sigma_{3/2}(u,a/L)$ for $L/a = 8, 16, 32$.%
\footnote{For ease of notation we shall omit in general the subscripts
  e/m for the electric and magnetic components of the coupling when
  we generically refer to both.}

Concerning the simulation algorithm, we used a combination of
heatbath~\cite{Creutz:1980zw, Fabricius:1984wp,Kennedy:1985nu} and
over-relaxation~\cite{Creutz:1987xi} as suggested in
ref.~\cite{Wolff:1992nq}. In particular, we chose to alternate 1
heat-bath sweep with $L/a$ over-relaxation sweeps. Since measuring the
coupling (i.e.~integrating the flow equations) is numerically more
expensive than performing a Monte Carlo update, we repeated this
process $L/a$ times between subsequent measurements.  In this way we
took into account the expected $\mathcal{O}(a^2)$ scaling of the
integrated auto-correlations of our observables. As a result, for
basically all values of the simulation parameters, we obtained
coupling measurements which are completely uncorrelated.  The only
exceptions are a few simulations for our largest lattices,
$L/a=24-48$, with $\bar g^2_{\rm GF}\sim 10$; here the measured
integrated autocorrelation times are $\tau_{\rm int}\sim 1$. In any
case, we always take autocorrelations into account in our analysis
through the $\Gamma$-method~\cite{Madras1988,
  Wolff:2003sm,Schaefer:2010hu}, implemented along the lines described
in ref.~\cite{Ramos:2018vgu}. We moreover note that in order to integrate 
the flow equations we use the adaptive step size integrator described in
ref.~\cite{Fritzsch:2013je}. This results in an improvement in computer 
time close to a factor 10 on our largest lattices compared to a fixed
step-size integration of the flow equations.

\subsection{The non-perturbative $\beta$-function at high-energy}
\label{sec:beta-function-he}

In this section we study the viability of using the GF couplings to
extract the $\Lambda$-parameter at high-energy. To this end, we first
introduce a convenient high-energy scale, $\mu_{\rm ref}$, by
specifying a relatively small value for the GF
couplings. Specifically, we define this scale in terms of the magnetic
component of the coupling, and set:
\begin{equation}
  \label{eq:gmuref_m}
  \bar g^2_{\rm GF,\,m,\,{\rm ref}} \equiv
  \bar g^2_{\rm GF,\,m}(\mu_{\rm ref}) \equiv \frac{4\pi}{5}
  \sim 2.5132\ldots\,,
\end{equation}
which corresponds to have, exactly,
$\alpha_{\rm GF,\,m}(\mu_{\rm ref})=0.2$.  In the following we also
need the corresponding value of the coupling in the electric scheme:
$\bar g^2_{\rm GF,\,e,\,{\rm ref}}\equiv\bar g^2_{\rm GF,\,e}(\mu_{\rm
  ref})$.  This is given in eq.~(\ref{eq:gmuref_e}), and we assume it
known for the time being; we shall come back shortly to its
determination. With these definitions at hand, the quantity we are
interested to compute is:
\begin{equation}
  \label{eq:LambdaFromGF}
  \frac{\Lambda_{\overline{\rm MS}}}{\mu_{\rm ref}} = 
  \frac{\Lambda_{\overline{\rm MS}}}{\Lambda_{\rm GF}}
  ( b_0 \bar g_{\rm GF,\,ref}^2)^{-\frac{b_1}{2b_0^2}}\,{\rm e}^{-\frac{1}{2b_0 \bar{g}_{\rm GF,\,ref}^2}}  \times
  \exp\{-I^{\rm GF}_g(\bar{g}_{\rm GF,\,ref},0)\}\,,
\end{equation}
where GF may stand for either the magnetic or electric coupling
scheme; clearly, $I^{\rm GF}_g$ is defined in terms of the proper
$\beta$-function (cf.~eq.~(\ref{eq:lamdef})). Note that the results
in the GF schemes are expressed in the $\overline{\rm MS}$ scheme
using the known relations between $\Lambda$-parameters, eq.~(\ref{eq:LambdaGF2MS}).

To evaluate eq.~(\ref{eq:LambdaFromGF}) the necessary ingredient is
the $\beta$-function in the range:
$\bar{g}^2_{\rm GF} \in[0,\bar{g}^2_{\rm GF,\,ref}]$.  Our preferred
strategy to obtain this is to consider a parametrization of the
$\beta$-function of the form:
\begin{equation}
  \label{eq:fit_beta_b2fix}
  \beta(x) = -x^3\left(b_0+b_1x^2+b_2x^4+\sum_{k=3}^{n_b}p_kx^{2k}\right)\,,  
\end{equation}
where the coefficients $b_0,b_1,b_2$ are fixed to their perturbative
values of eqs.~(\ref{eq:beta_univ}),(\ref{eq:b2GF}). This enforces the
correct asymptotic behaviour of $\beta(g)$ for $g\to0$. The
coefficients $p_k$ are then determined by fitting our set of
non-perturbative data. More precisely, we introduce the function:
\begin{equation}
  \label{eq:F}
  F(a,b) = -\int_{\sqrt{a}}^{\sqrt{b}}\, \frac{{\rm d}x}{\beta(x)}\,,
\end{equation}
and define the $\chi^2$-function:
\begin{equation}
  \label{eq:chi2GF}
  \chi^2 = \sum_{i=1}^{N_{\rm data}} \left[\frac{\log(s) + \rho^{(s)}(u_i)(a/L)^2 -
      F_i^{(s)}}{\delta F_i^{(s)}}\right]^2\,,
\end{equation}
where $F_i^{(s)} = F(u_i,\Sigma_s^i)$ is computed from the measured
value of the GF coupling, $u_i$, at a given $i=(L/a,g_0)$, and from
the corresponding result for the lattice step-scaling function
$\Sigma_s^i=\Sigma_s(u_i,a/L)$. In the above expression:
\begin{equation}
  \big(\delta F_i^{(s)}\big)^2 = \frac{1}{4u_i}\left[\beta\left(\sqrt{u_i}\right)
  \right]^{-2}(\delta u_i)^2 +
  \frac{1}{4\Sigma_s^i}\left[\beta\left(\sqrt{\Sigma_s^i}\right)
  \right]^{-2}(\delta \Sigma_s^i)^2\,,
\end{equation}
while
\begin{equation}
  \label{eq:rhoGF}
  \rho^{(s)}(u) = \sum_{k=0}^{n_c}\rho_k^{(s)} u^k,
\end{equation}
parametrizes the cutoff effects in the data. Note that the data
corresponding to different scale factors $s$ can be combined into a
single fit as long as we consider different parametrizations for the
cutoff effects i.e., the coefficients $\rho^{(s)}_k$ are independent
parameters for different values of $s$. Concerning the nature of the
discretization errors, as discussed in Sect.~\ref{sec:LatticeSetup}, 
our data is in principle affected by $\mathcal{O}(a)$ errors. Rather 
than including an explicit $\mathcal{O}(a)$ term in eq.~(\ref{eq:chi2GF}), 
however, we decided to proceed in the following way. The data strongly supports 
the conclusion that within our statistical precision, the dominant
discretization errors for our lattice resolutions and coupling values
are of $\mathcal{O}(a^2)$ rather than $\mathcal{O}(a)$ (cf.~Appendix
\ref{sec:UbyU}).  We thus estimated through dedicated simulations the
systematic effect on the value of the coupling induced by the deviation of 
the 2-loop result for $c_t(g_0)$, used in the simulations, from an
educated guess for its actual, non-perturbative, value. We then added
this systematic uncertainty to the measured values of the coupling, and 
performed the fits including this systematic effect. Appendix~\ref{sec:oa} 
contains a detailed discussion about this point.  

Restricting to values of $u\leq \bar{g}^2_{\rm GF,\,ref}$ the dataset
described in Sect.~\ref{sec:DataSetGF}, the fits defined by
eqs.~(\ref{eq:fit_beta_b2fix})-(\ref{eq:rhoGF}) give in general
excellent $\chi^2$'s.  More precisely, we consider fits with $n_b=4,5$
in the parametrization of the $\beta$-function, and take $n_c=2,3$
($n_c=2$) to describe the cutoff effects of the data with $s=2$
($s=3/2$).%
\footnote{In fact, the results for
  $\Lambda_{\overline{\rm MS}}/\mu_{\rm ref}$ given below show very
  little dependence on the number of terms used to parametrize the
  cutoff effects.  We have explicitly checked that using values as
  large as $n_c=10$ does not significantly change the result.}  The
resulting fits all have a $\chi^2/{\rm dof}\sim0.5-0.9$, where the
fits to the electric components of the GF coupling always have smaller
$\chi^2$'s than those to the magnetic ones.  The main reason for this
is that the estimated $\mathcal O(a)$ uncertainties are significantly
larger in the electric case, resulting in larger errors
for the couplings and SSFs (cf.~Appendix \ref{sec:oa}).

In the case of the electric components of the GF coupling, in addition
to the determination of the $\beta$-function, the evaluation of
eq.~(\ref{eq:LambdaFromGF}) requires also the determination of
$ \bar g^2_{{\rm GF,\,e,\,ref}} = \bar g^2_{\rm GF,\,e}(\mu_{\rm
  ref})$. We can obtain the latter by performing a linear fit to our
19 data points for $\bar g^2_{\rm GF}\in [2,3]$, on lattices with
$L/a=8,12,16,24,32,48$. Specifically, we fit the quantity,
\begin{equation}
  \frac{1}{\bar g_{\rm GF,\,e}^2} - \frac{1}{\bar g_{\rm GF,\,m}^2} = a_0 + a_1\,\bar
  g_{\rm GF,\,m}^2 + \left[ \tilde \rho_0 + \tilde \rho_1\,\bar g_{\rm GF,\,m}^2\right]
  \left(\frac{a}{L} \right)^2\,,
\end{equation}
where $a_0,a_1,\rho_0,\rho_1$ are fit parameters. The fit has a very
good $\chi^2\sim 0.9$, and allows us to obtain the precise result:
\begin{equation}
  \label{eq:gmuref_e}
  \bar g^2_{{\rm GF,\,e,\,ref}} = \bar g^2_{\rm GF,\,e}(\mu_{\rm ref}) =
  2.46508(95) \qquad [0.04\%]\,.
\end{equation}
This result is very stable under a change of the number of fit
parameters, or of the data included in the fit. (Note that the data
spans a factor 6 in the lattice spacing.)  In practice, the tiny
uncertainty that we obtain for $\bar g^2_{{\rm GF,\,e,\,ref}}$ could
be completely neglected in the following analysis, but we include it anyway.

\begin{table}
  \centering
  \begin{tabular}{lllllllll}
    \toprule
    &&&\multicolumn{2}{c}{All
        $L/a$}&\multicolumn{2}{c}{$L/a>8$}&\multicolumn{2}{c}{$L/a>10$}
    \\
    \cmidrule(lr){4-5}     \cmidrule(lr){6-7}     \cmidrule(lr){8-9} 
    Scheme & $n_b$ & $n_c^{(s=2)}$ & Wilson & Zeuthen& Wilson & Zeuthen& Wilson & Zeuthen\\
  \midrule
ele & 4 & 2 & 0.0830(11)  & 0.0790(10) & 0.0812(14)  & 0.0798(13) & 0.0805(17) & 0.0798(17)\\
ele & 4 & 3 & 0.0832(12)  & 0.0789(11) & 0.0813(14)  & 0.0798(14) & 0.0815(19) & 0.0807(18)\\
ele & 5 & 2 & 0.0851(16)  & 0.0787(14) & 0.0824(18)  & 0.0802(18) & 0.0835(26) & 0.0821(26)\\
ele & 5 & 3 & 0.0858(17)  & 0.0788(16) & 0.0835(22)  & 0.0811(21) & 0.0826(28) & 0.0813(27)\\
  \midrule
mag & 4 & 2 & 0.08274(92) & 0.07910(86) & 0.0814(11)  & 0.0803(11) & 0.0819(15) & 0.0813(15)\\
mag & 4 & 3 & 0.08295(95) & 0.07902(89) & 0.0814(12)  & 0.0801(12) & 0.0823(16) & 0.0816(16)\\
mag & 5 & 2 & 0.0834(13)  & 0.0775(12) & 0.0810(15)  & 0.0791(15) & 0.0819(22) & 0.0809(21)\\
mag & 5 & 3 & 0.0831(14)  & 0.0771(13) & 0.0809(18)  & 0.0789(17) & 0.0810(23) & 0.0800(23)\\
    \bottomrule
  \end{tabular}
  \caption{Results for $\Lambda_{\overline{\rm MS}}/\mu_{\rm ref}$ from
  different analysis. We use our data for the step-scaling function (cf.~Sect.~\ref{sec:DataSetGF}),
  to determine the $\beta$-function both in the electric and magnetic schemes. We choose 
  different parametrizations for the $\beta$-function and the cutoff effects in our data 
  (cf.~eqs.~(\ref{eq:fit_beta_b2fix}) and (\ref{eq:rhoGF})), as well as different cuts 
  in the lattice resolution. }
  \label{tab:res_he}
\end{table}

%%% Local Variables:
%%% mode: latex
%%% TeX-master: "../paper"
%%% End:

\begin{figure}
  \centering \includegraphics[width=0.8\textwidth]{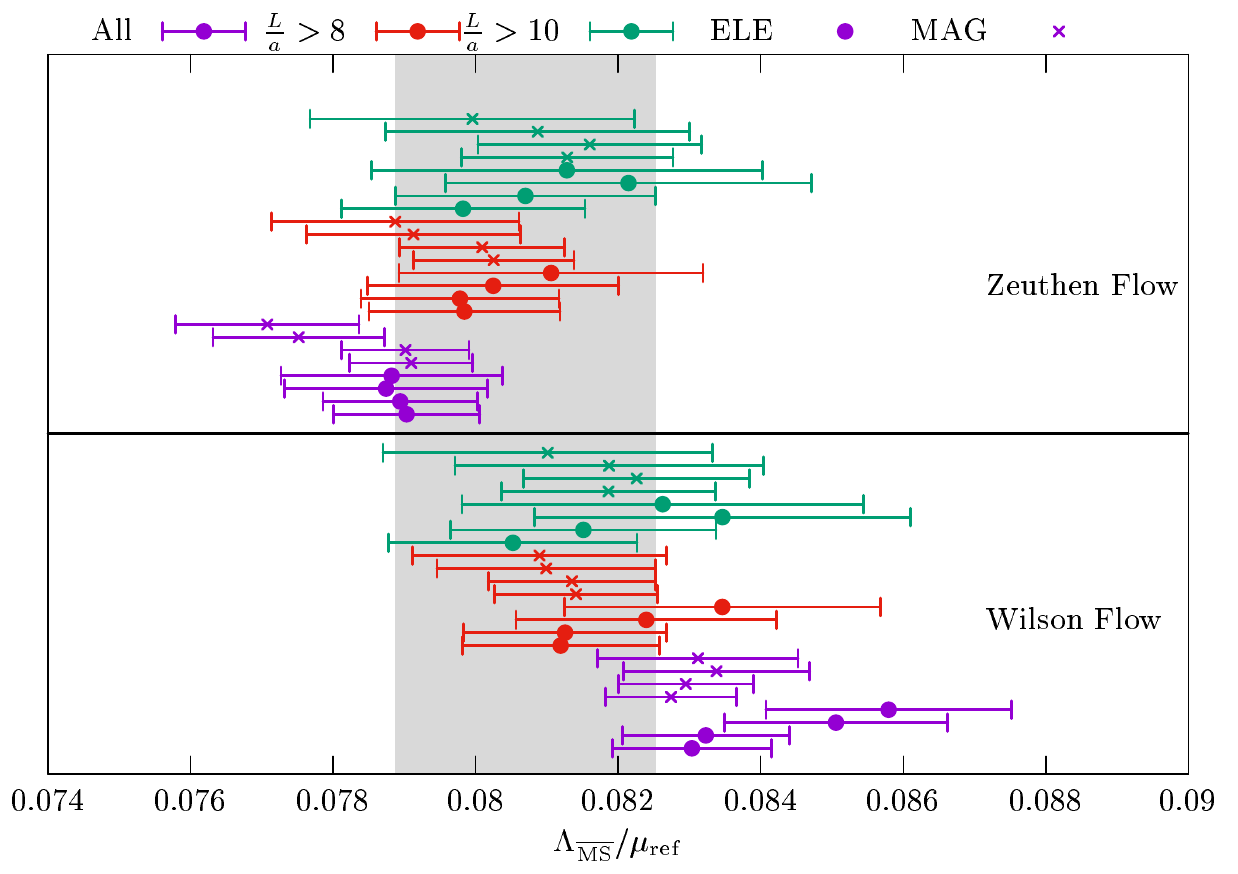}
  \caption{Landscape of the results for $\Lambda_{\overline{\rm MS}}/\mu_{\rm ref}$ 
  	given in table~\ref{tab:res_he}. The shaded region is our final result
    eq.~(\ref{eq:res_he}). Once the coarser lattice $L/a=8$ is
    discarded, different analysis techniques give compatible results.}
  \label{fig:res_he}
\end{figure}

The results for ${\Lambda_{\overline{\rm MS}}}/{\mu_{\rm ref}}$
obtained according to the above strategy are summarized in
table~\ref{tab:res_he}, and also presented in
figure~\ref{fig:res_he}. Different parametrizations of the
$\beta$-function and different data sets corresponding to different
lattice discretizations of the GF couplings all produce consistent
results. There is also perfect agreement between the determinations
from the electric and magnetic components. This is a highly
non-trivial test, since the ratio of $\Lambda$-parameters in these
schemes is $\sim 1.15$. Only when converted to the common scheme
$\Lambda_{\overline{\rm MS} }$ our results agree. As our final result for
${\Lambda_{\overline{\rm MS}}}/{\mu_{\rm ref}}$ we quote the analysis
based on the Zeuthen flow data for the electric components of the GF
coupling with $L/a>10$, and with $n_b=5,n_c^{(s=2)}=3$ in the
parametrization of the $\beta$-function. This gives:
\begin{equation}
  \label{eq:res_he}
  \frac{\Lambda_{\overline{\rm MS}}}{\mu_{\rm ref}} =  0.0807(18)
  \qquad  [2.26 \%]\,.
\end{equation}
The central value of any other fit that uses the Zeuthen flow
discretization with $L/a>8$ is well included in this error band. The
quoted uncertainty is rather conservative as we discard all data with
$L/a=8,10$. For completeness, we also give the corresponding fit
parameters for the $\beta$-function:
\begin{equation}
  p_3 =   0.00022135\,, \qquad
  p_4 =  -0.00000173\,,
\end{equation}
with their covariance,
\begin{equation}
  {\rm cov}(p_3,p_4) = \left(
    \begin{array}{ll}
      5.00835383\times 10^{-8}  & -1.98372733\times 10^{-8}\\
      -1.98372733\times 10^{-8} &   8.01483182\times 10^{-9}\\
    \end{array}
  \right)\,.
\end{equation}
Before moving to the next section, we want to mention that in order to
gain further confidence in our determination of
${\Lambda_{\overline{\rm MS}}}/{\mu_{\rm ref}}$ we also considered
some alternative fitting strategy. One example is to determine the
$\beta$-function entering eq.~(\ref{eq:LambdaFromGF}) by fitting our
dataset according to the $\chi^2$-function
(cf.~eq.~(\ref{eq:chi2GF})):
\begin{equation}
  \label{eq:chi2GF2}
  \chi^2 = 
  \sum_{i=1}^{N_{\rm data}} 
  \left[\frac{\Sigma_s^i - G_s(u_i)- \rho^{(s)}(u_i)(a/L)^2}{\delta \Sigma_s^i}\right]^2\,,
\end{equation}
where the function $G_s(u_i)$ is the solution of the equation,
\begin{equation}
  \label{eq:ImplicitBetaGF}
  F(u_i,G_s(u_i)) + \log s = 0\,,
\end{equation}
with $F$ defined in eq.~(\ref{eq:F}). As before, in the expression for
$F$ we can simply consider a parametrization of the $\beta$-function
as eq.~(\ref{eq:fit_beta_b2fix}).  Alternatively, we can take:
\begin{equation}
  \label{eq:inverse_beta}
  \beta(x) = -\frac{x^3}{\sum_{k=0}^{n_b}p_kx^{2k}}\,,
  \qquad
  p_0=\frac{1}{b_0}\,,
  \quad
  p_1=-\frac{b_1}{b_0^2},
  \quad
  p_2=-\frac{b_0b_2-b_1^2}{b_0^3}\,.
\end{equation}
It is clear that in the limit $x\to0$ this parametrization reproduces
the correct perturbative expression for the $\beta$-function.  The
nice feature of this choice is that we can now evaluate explicitly the
function $F(a,b)$ entering the $\chi^2$ functions. Indeed,
\begin{equation}
  \label{eq:ScaleRatio}
  F(a,b) = H(a) - H(b),
  \qquad
  H(x)= 
  \frac{1}{2 b_0 x^2} + \frac{b_1}{b_0^2} \log x + \frac{(b_0 b_2-b_1^2)x^2}{2b^3_0}  
  -\sum_{k=2}^{n_b-1} p_{k+1} \frac{x^{2k}}{2k}.
\end{equation}
Once the function $F$ is determined, it is straightforward
to compute (\ref{eq:LambdaFromGF}). Without entering into more
details, very similar conclusions apply for these alternative fits as
for our preferred strategy.  The fits describe our dataset well, and
different discretizations and schemes all give results for
$\Lambda_{\overline{\rm MS}}/\mu_{\rm ref}$ which are perfectly
compatible with eq.~(\ref{eq:res_he}).

\subsubsection{Comparison with perturbation theory}

\begin{figure}
  \centering
  
\begin{subfigure}[t]{0.49\textwidth}
  \includegraphics[width=\textwidth]{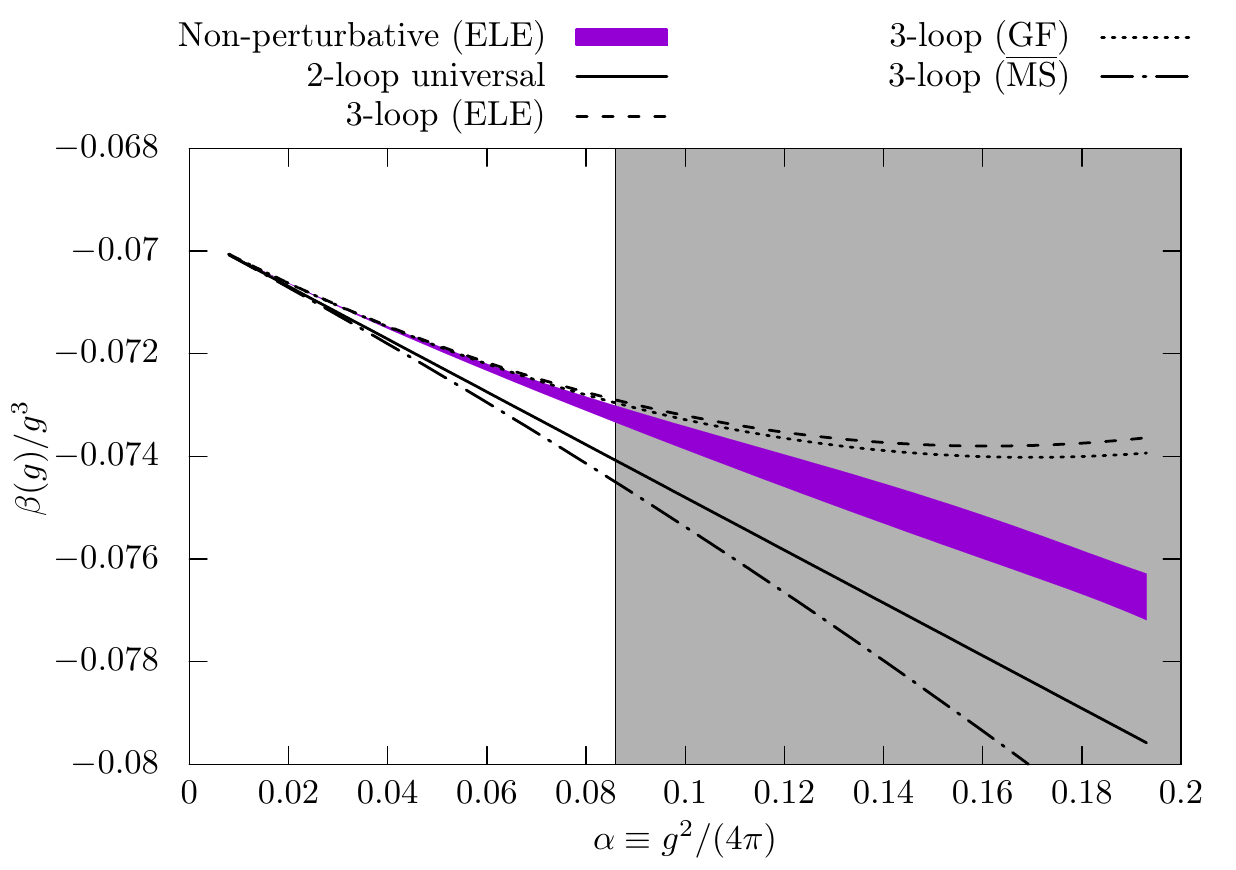}
  \caption{Electric scheme.}
  \label{fig:beta_ele}
\end{subfigure}
\begin{subfigure}[t]{0.49\textwidth}
  \includegraphics[width=\textwidth]{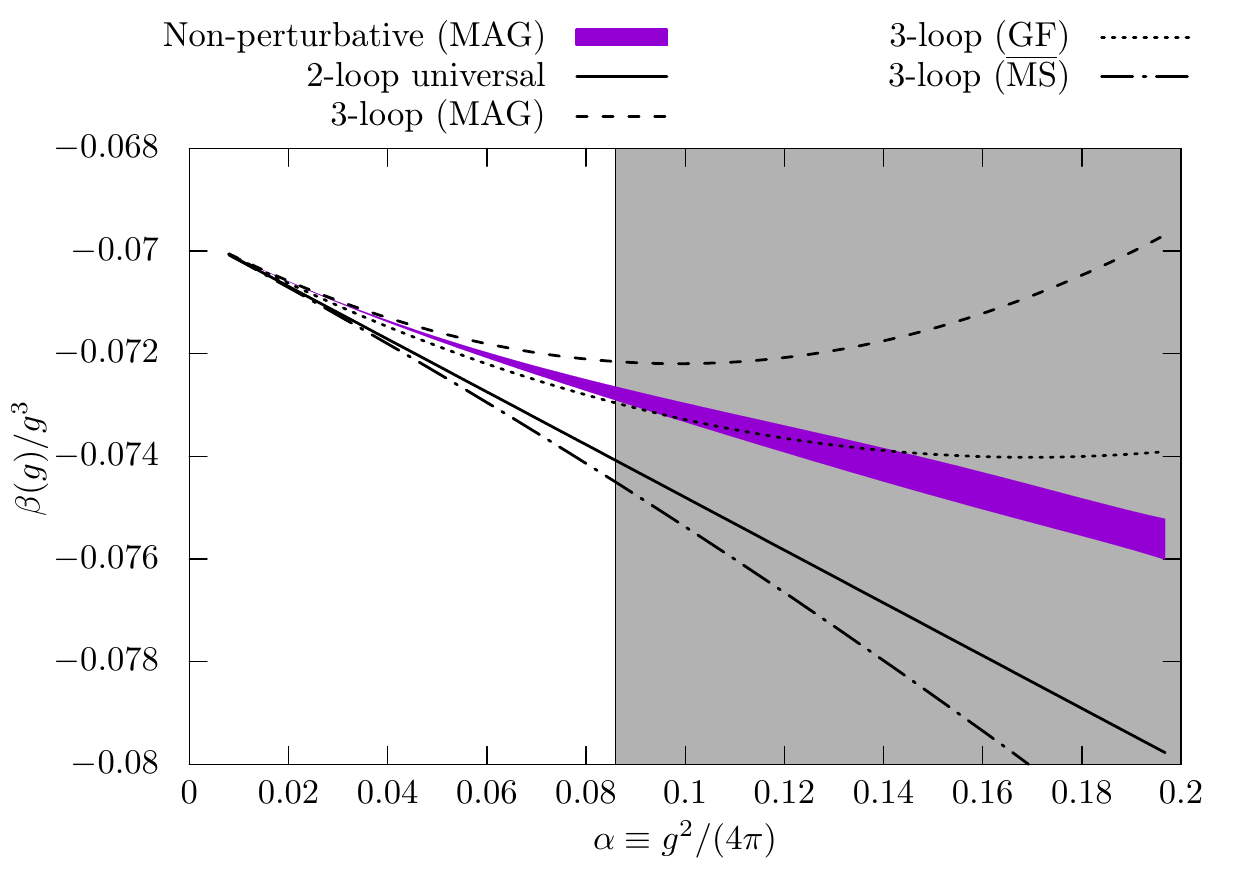}
  \caption{Magnetic scheme}
  \label{fig:beta_mag}
\end{subfigure}
\caption{Comparison of the $\beta$-function in the electric/magnetic
  schemes with their perturbative predictions. We also show the
  perturbative predictions in the GF infinite volume and
  $\overline{\rm MS}$ schemes. The shaded region
  corresponds to the range where the data is available
  $\bar g^2\sim [1,2.5]$. See text for a discussion.}
\label{fig:beta}
\end{figure}

It is instructive to plot our results for the $\beta$-functions and
compare them with their 3-loop perturbative predictions.  This is
illustrated in figure \ref{fig:beta}. It is clear from the figure that
the GF based schemes show a very poor convergence to their expected
perturbative behaviour. For the electric scheme, the non-perturbative
data is barely compatible within errors with the 3-loop
$\beta$-function at the most perturbative point, where
$\alpha\sim 0.08$. The magnetic scheme shows even poorer
convergence, with a clear deviation at $\alpha \sim 0.08$ between
our data and the 3-loop prediction. In this case,
looking even beyond the range covered by the data, our parametrization
for the $\beta$-function appears to deviate from its 3-loop
approximation down to couplings as small as $\alpha\sim 0.05$.
With hindsight, these conclusions are not too surprising, given the
large value of the $b_2$-coefficients in these schemes
(cf.~eq.~(\ref{eq:b2GF})).  The slightly better convergence of the
electric scheme with respect to the magnetic one made us favour as our
final result, eq.~(\ref{eq:res_he}), a determination based on the
electric rather than the magnetic coupling.  Either way, however, a
determination of the $\Lambda$-parameter based on the GF schemes is
very much limited in the attainable precision due to these issues.

\begin{figure}
  \centering \includegraphics[width=0.8\textwidth]{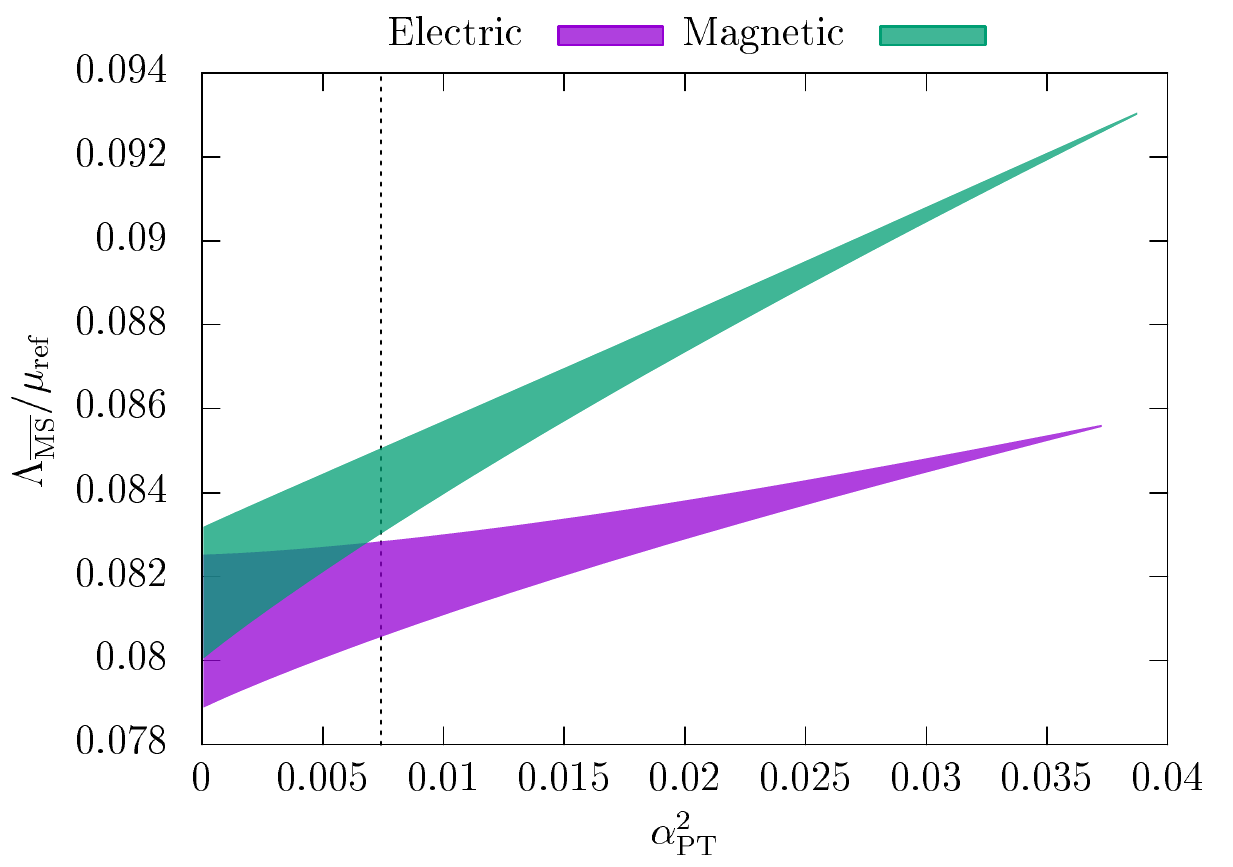}
  \caption{The ratio $\Lambda_{\overline{\rm MS}}/\mu_{\rm ref}$
    estimated from the function $\phi_{\rm GF}(\alpha_{\rm PT})$ (see
    eq.~(\ref{eq:lam_truncated})) as a function of the matching scale
    with perturbation theory (represented by
    $\alpha_{\rm PT} = \bar g^2_{\rm PT}/4\pi$). At
    $\alpha_{\rm PT}\sim 0.08$ (our most non-perturbative data point)
    the NNLO predictions using the electric and magnetic schemes
    disagree by $\sim2.5\%$. }
  \label{fig:lovm}
\end{figure}

The effect of the poor convergence to perturbation theory of these
schemes on the determination of
$\Lambda_{\overline{\rm MS}}/\mu_{\rm ref}$ can be quantitatively
studied by examining the dependence of
$\Lambda_{\overline{\rm MS}}/\mu_{\rm ref}$ on the scale
$\mu_{\rm PT}$ at which perturbation theory is used to extract
it. More precisely, we can look at:
\begin{eqnarray}
  \label{eq:lam_truncated}
  \phi_{\rm GF}(\alpha_{\rm PT}) = \frac{\Lambda_{\overline{\rm MS}}}{\Lambda_{\rm GF}}\, 
  \varphi_{\rm GF}^{(3)}(\bar{g}_{\rm GF,\,ref},\bar{g}_{\rm GF,\,PT})
  \oset[1ex]{\mu_{\rm PT}\to\infty}{=} 
  \frac{\Lambda_{\overline{\rm MS}}}{\mu_{\rm ref}} + \mathcal O(\alpha_{\rm PT}^2)\,,
  \qquad
  \alpha_{\rm PT}\equiv\frac{\bar{g}^2_{\rm GF,\,PT}}{4\pi}\,,
\end{eqnarray}
where, as usual,
$\bar{g}_{\rm GF,\,X}\equiv \bar{g}_{\rm GF}(\mu_{\rm X})$, and GF may
refer to either the electric or the magnetic components of the GF
coupling.  The function $\varphi^{(3)}_{\rm GF}$ is defined by
eq.~(\ref{eq:lam_extraction}) in terms of the GF $\beta$-function and
its 3-loop perturbative approximation.  As anticipated in the above
equation, in the limit $\alpha_{\rm PT}\to 0$ the function
$\phi_{\rm GF}$ approaches $\Lambda_{\overline{\rm MS}}/\mu_{\rm ref}$
with $\mathcal{O}(\alpha^2_{\rm PT})$ corrections. The latter are
expected to be small if and only if 3-loop perturbation theory is a
good approximation for the $\beta$-function at scales of
$\mathcal O(\mu_{\rm PT})$. As stated above, however, this is not 
the case for the GF schemes, even at the largest energy scales we explored. 
At values of the couplings $\alpha_{\rm PT}=0.2$, for instance, where
3-loop perturbation theory is typically expected to be accurate, we
find:
\begin{equation}
  \frac{\phi_{\rm GF,\,e}(0.2)-\phi_{\rm GF,\,e}(0)}{\phi_{\rm GF,\,e}(0)} = \phantom{1}6.1(2.1)\% \,,
  \qquad
  \frac{\phi_{\rm GF,\,m}(0.2)-\phi_{\rm GF,\,m}(0)}{\phi_{\rm GF,\,m}(0)}  = 12.6(1.7)\% \,.
\end{equation}
The approximations $\phi_{\rm GF,\,e}$, $\phi_{\rm GF,\,m}$ thus
deviate from our final result eq.~(\ref{eq:res_he}) by $\sim 6$\%
and 13\%, respectively. These numbers are between 3 to 6 times larger
than our uncertainty on eq.~(\ref{eq:res_he}).  Even at the largest
energy scale reached by our simulations, where
$\alpha_{\rm PT}\sim 0.08$, both schemes show a significant
deviation from eq.~(\ref{eq:res_he}) (cf.~figure \ref{fig:lovm}).

As anticipated, the issue of poor convergence has a dramatic effect on
the precision that can be attained for $\Lambda_{\overline{\rm MS}}/\mu_{\rm ref}$ 
using GF based schemes. Due to the large corrections in eq.~(\ref{eq:lam_truncated})
one is forced to take as final estimate for $\Lambda_{\overline{\rm MS}}/\mu_{\rm ref}$ 
the value of $\phi_{\rm GF}(0)$, which requires an extrapolation. A competitive
scheme, on the other hand, should allow us to quote as final result the estimate for 
$\Lambda_{\overline{\rm MS}}/\mu_{\rm ref}$ at, say, $\alpha\sim0.1$, i.e. at the 
smallest couplings reached by our simulations (cf.~Sect.~\ref{subsec:GF2SF}). This clearly 
permits to reach a significantly higher precision. The result for, 
\begin{equation}
  \phi_{\rm GF,\,m}(0.1) = 0.08492(85)\,,
\end{equation}
for instance, has an error 50\% smaller than eq.~(\ref{eq:res_he}). Unfortunately,
however, at this value of $\alpha$ the results in different schemes differ by
a few standard deviations.  These values are also significantly
different from $\phi_{\rm GF}(0)$, indicating that a safe contact with
perturbation theory has not been reached. In conclusion, using GF
based schemes, 3-loop perturbation theory is not accurate enough to
extract the $\Lambda$-parameter with $\sim1\%$ error at
$\alpha\sim 0.1$.
The conclusions of this section are further corroborated by a more
traditional step-scaling strategy to extract
$\Lambda_{\overline{\rm MS}}/\mu_{\rm ref}$.  We refer the interested
reader to Appendix \ref{sec:ss} for more details.

\subsection{Non-perturbative matching to the SF scheme}
\label{subsec:GF2SF}

Over the years the SF couplings have been widely employed in
step-scaling
studies~\cite{Luscher:1992zx,Luscher:1993gh,DellaMorte:2004bc,Sommer:2010ui,Brida:2016flw},
and recently their perturbative behaviour has been carefully
investigated in the context of determining the $\Lambda$-parameter of
QCD to a few per-cent accuracy. From these studies, we expect the
matching of the SF schemes with perturbation theory to be
quantitatively much better than what we observed above for the GF
schemes. In particular, 3-loop perturbation theory may be precise
enough for our purposes at scales where $\alpha\sim 0.1$. This in
fact will be confirmed below.

Given these observations, in this section we consider a different
strategy for determining $\Lambda_{\overline{\rm MS}}/\mu_{\rm
  ref}$. We will first match non-perturbatively the SF and the GF
schemes, and then extract the $\Lambda$-parameter using perturbation
theory in the SF schemes. To achieve this, we have measured the SF
couplings on lattices of size $L/a=6,8,10,12,16$, and at values of the
bare coupling $g_0$ that match our measurements of the GF couplings on
$2L/a$ lattices, i.e. $L/a=12, 16, 20, 24, 32$, respectively. We thus
combine a change of scheme with a change in the renormalization scale
by a factor $s=2c$, where $c=0.3$ in our study. We collected from 
about $5\times10^5$ measurements on the smaller lattices with $L/a=6$,
up to $2\times 10^6$ measurements for $L/a=16$ (see
table \ref{tab:SF} for the exact figures). We measured both
the SF coupling $\bar{g}^2_{\rm SF}$ and the observable $\bar{v}$,
which gives us access to $\bar{g}^2_{{\rm SF},\nu}$ for any value of
$\nu$ (cf.~eq.~(\ref{eq:SFnu})). In this section we focus exclusively
on $\nu=0$.  The consistency between determinations for different
$\nu$-values is discussed in Appendix \ref{sec:ss}.
 
With these results at hand, we fit the data for the GF and SF
couplings as:
\begin{equation}
  \label{eq:gf2sf}
  \frac{1}{\bar g^2_{\rm SF}(\mu)} - \frac{1}{\bar g^2_{\rm GF}(\mu/(2c))} =
  f(u) + \tilde \rho(u)\left( \frac{a}{L} \right)^2 \,,\qquad (u = \bar g^2_{\rm GF}(\mu/(2c)))\,.
\end{equation}
The function $\tilde \rho(u)$ parametrizes the cutoff effects in 
this relation by a simple polynomial:
\begin{equation}
  \tilde \rho(u) = \sum_{k=0}^{n_{\tilde \rho}} \tilde \rho_n u^n\,.
\end{equation}
We also use a polynomial for $f(u)$:
\begin{equation}
  \label{eq:fexp}
  f(u) = \sum_{k=0}^{n_f} f_n u^n\,.
\end{equation}
Using the perturbative relation, eq.~(\ref{eq:SF2GF}), we could in
principle fix the coefficients $f_0,f_1$, to their perturbative
values. Here, however, we refrain to do so. Clearly,
this implies that the non-perturbative matching between the SF and GF
schemes is only trustworthy within the available range of couplings
i.e., for $\bar g_{\rm GF,\,SF}^2\sim 1-2$.
We obtain good fits by taking, e.g., $n_f=2,3$ and
$n_{\tilde \rho}=2$.  We note however that our final results for
$\Lambda_{\overline{\rm MS}}/\mu_{\rm ref}$ do not depend
significantly on this particular choice. We only see some dependence
if we include or not our coarser lattices (i.e. $L/a=6$ for the SF
coupling and $L/a=12$ for the GF couplings). Any disagreement 
disappears once we perturbatively improve the SF data to 2-loop order
in lattice perturbation theory (see appendix~\ref{sec:SFPT}). Having
observed this, in order to be conservative, we prefer to discard the
coarser lattice spacing for computing the matching, and use the 2-loop 
improved data.  All our fits are then good with a
$\chi^2/\text{dof}\sim 0.5-1$.

\begin{figure}[t!]
  \centering
  \begin{subfigure}[t]{0.9\textwidth}
    \centering \includegraphics[width=0.8\textwidth]{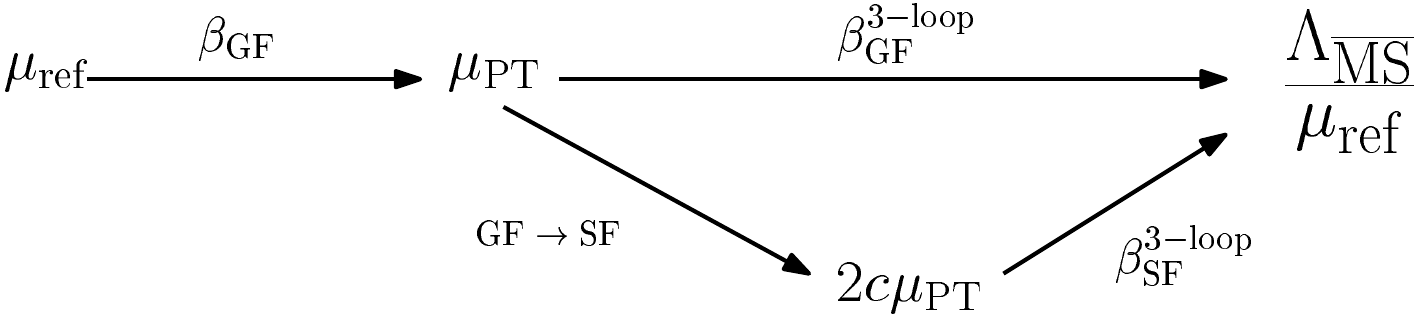}
    \caption{By using the non-perturbative matching between the SF and
      the GF schemes (eq.~(\ref{eq:gf2sf})), we can determine the
      $\Lambda_{\overline{\rm MS}}$ making contact with perturbation
      theory in the SF scheme. Due to the much better perturbative
      behavior (compare figure~\ref{fig:beta} and
      figure~\ref{fig:betaGF2SF}), this strategy allows a more acurate
      and precise determination of $\Lambda_{\overline{\rm MS}}$.}
    \label{fig:diag}
  \end{subfigure}
  
  \begin{subfigure}[t]{0.45\textwidth}
    \includegraphics[width=\textwidth]{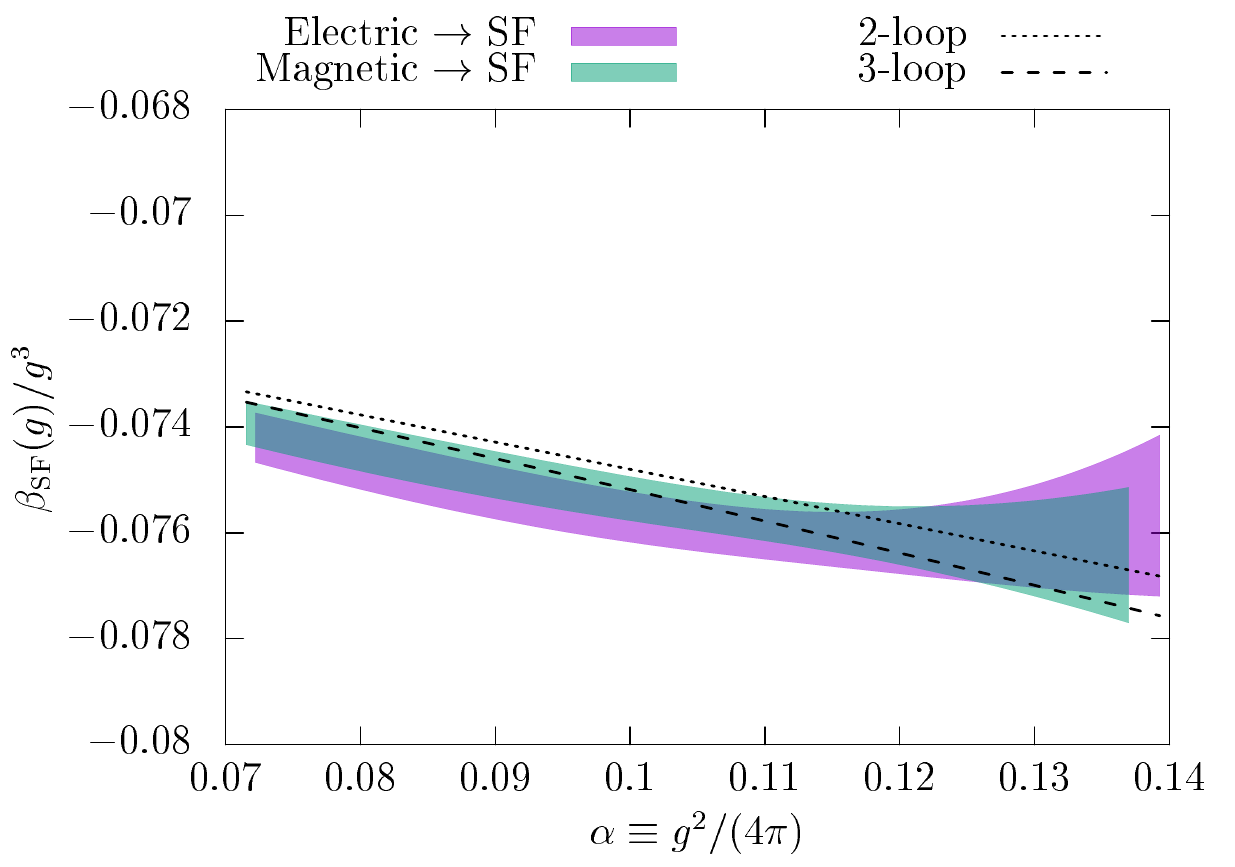}
    \caption{The $\beta$-function in the SF scheme determined by
      matching non-perturbatively the SF and the GF schemes (see
      eq.~(\ref{eq:betaSF})). In all the range of coupling our results
      agree with the perturbative 3-loop prediction. This contrast
      with the behaviour in the GF schemes (see
      figure~\ref{fig:beta}).  }
    \label{fig:betaGF2SF}
  \end{subfigure}
  \begin{subfigure}[t]{0.45\textwidth}
    \includegraphics[width=\textwidth]{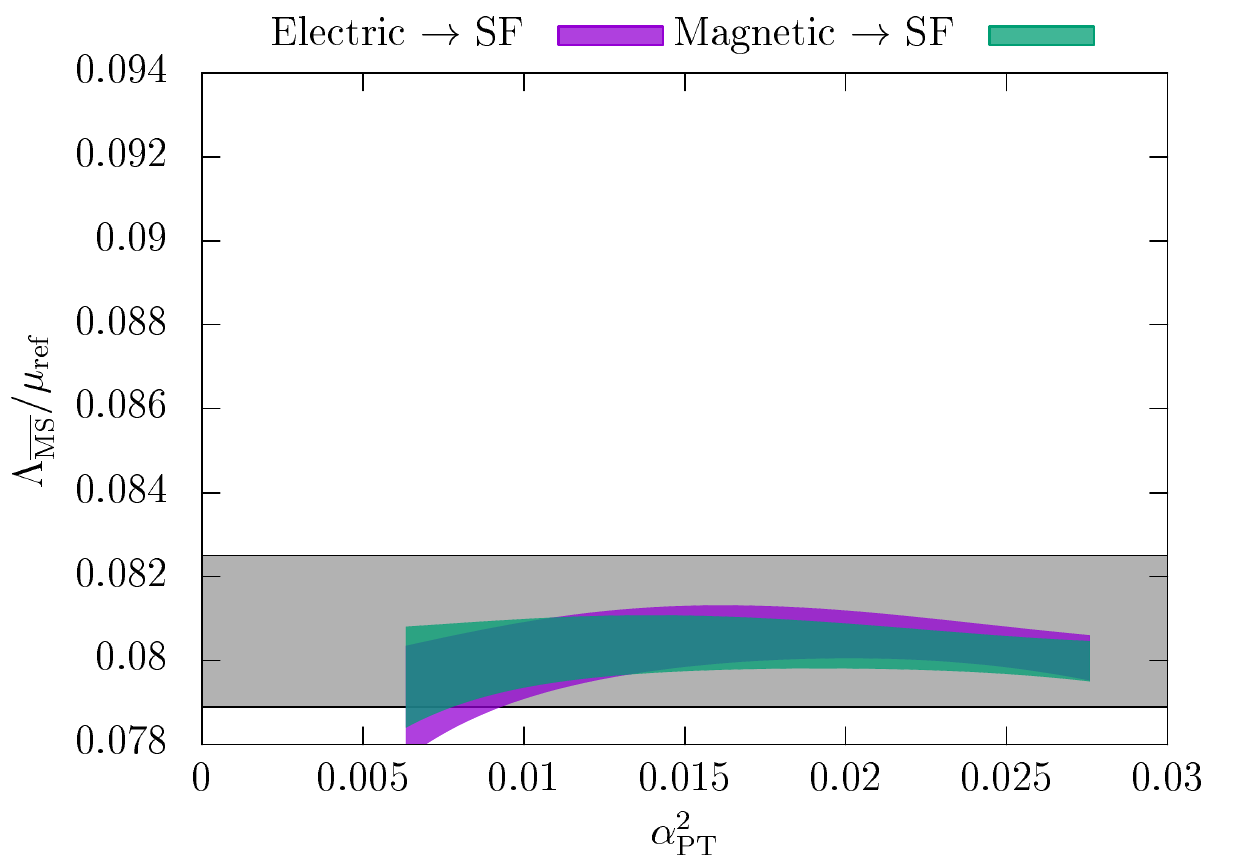}
    \caption{The determination of
      $\Lambda_{\overline{\rm MS}}/\mu_{\rm ref}$ from the function
      $\phi_{\rm SF}(\alpha_{\rm PT})$ (see
      eq.~(\ref{eq:varphi_SF})). The figure shows that the value of
      $\phi_{\rm PT}(\alpha_{\rm PT})$ is independent on
      $\alpha_{\rm PT}$, showing that the corrections
      $\mathcal O(\alpha_{\rm PT}^2)$ are negligible within our
      statistical precision. This is in contrast with the behaviour 
      in the GF schemes (see figure~\ref{fig:lovm}). The error band 
      is the result of eq.~(\ref{eq:res_he}).}
    \label{fig:lovmSF}
  \end{subfigure}
  \caption{The determination of
    $\Lambda_{\overline{\rm MS}}/\mu_{\rm ref}$ from a
    non-perturbative matching between the GF and SF schemes.}
\end{figure}

Once the function $f(u)$ has been determined, the $\beta$-function in
the SF scheme can be inferred from the one in the GF scheme using the
relation:
\begin{equation}
  \label{eq:betaSF}
  \mu \frac{{\rm d}{\bar g}_{\rm SF}(\mu)}{{\rm d}\mu} =  \beta_{\rm SF}(\bar
  g_{\rm SF}(\mu)) = \sqrt{1 + uf(u)}\left[- \frac{f'(u) - 1/u^2}{(f(u) + 1/u)^2} \right]
  \beta_{\rm GF}(\sqrt{u})
  \qquad (u = \bar{g}_{\rm GF}^2(\mu/(2c)))\,.
\end{equation}
Note that since we are not interested in matching the GF schemes with
their perturbative expressions, we can consider also fits for
$\beta_{\rm GF}(g)$ where the $b_2$ coefficient is treated as a fit
parameter, rather than being fixed to its perturbative value
(cf.~eq.~(\ref{eq:fit_beta_b2fix})). In this case, the results for
$\beta_{\rm SF}(g)$ only use as perturbative information the universal
coefficients of the $\beta$-function, eq.~(\ref{eq:beta_univ}). In
particular the known value for $b_2^{\rm SF}$ is not used.  
We have determined $\beta_{\rm SF}$ using the data for
$\beta_{\rm GF}$ both in the electric and magnetic GF scheme. The SF
$\beta$-function determined this way is shown in
figure~\ref{fig:betaGF2SF}. As one can see from the plot, as expected
there is agreement between the determination from the electric 
and magnetic GF schemes. The non-perturbative data then match the 
perturbative prediction in all the range of couplings we covered.

\begin{figure}[t!]
  \centering
  \includegraphics[width=0.8\textwidth]{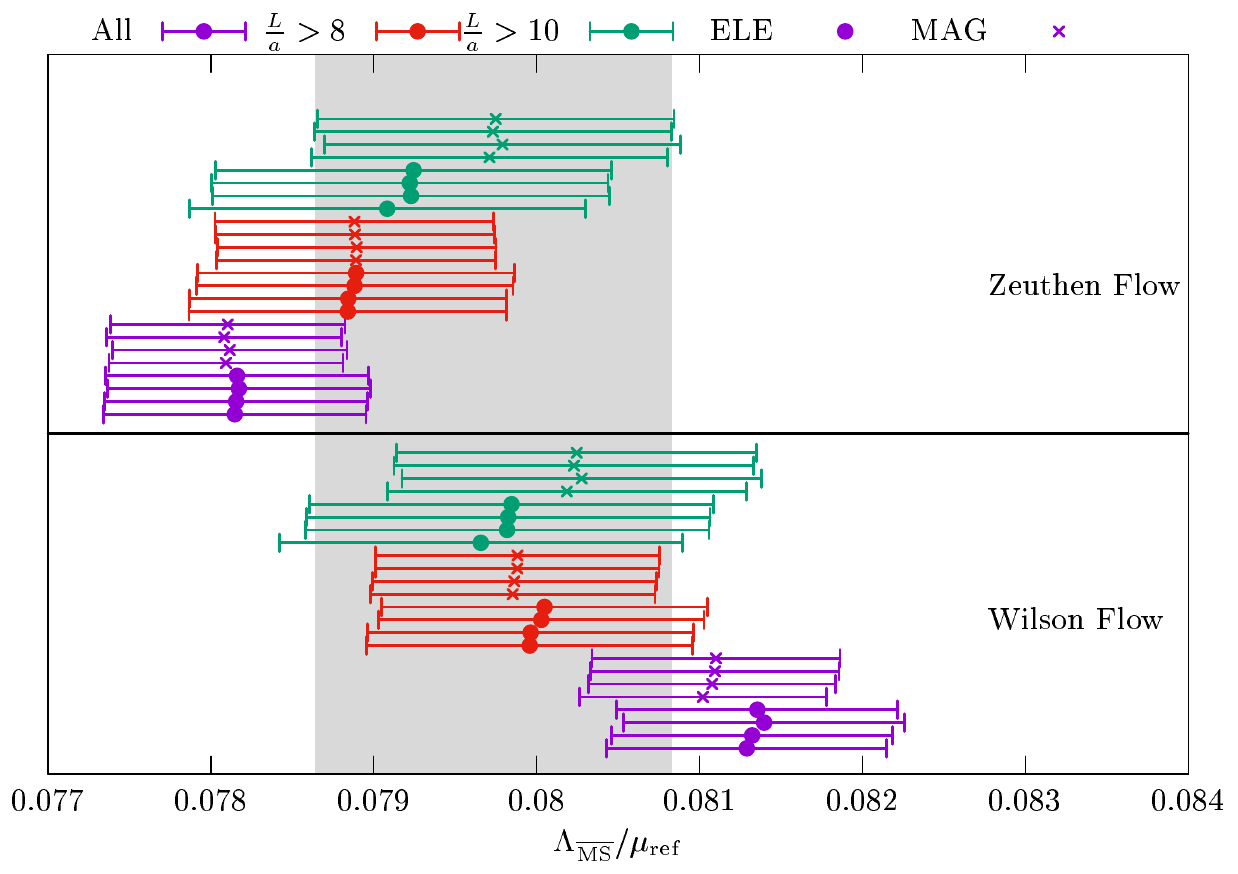}
  \caption{Results for the ratio
    $\Lambda_{\overline{\rm MS}}/\mu_{\rm ref}$ by quoting the value
    of $\phi_{\rm SF}(\alpha_{\rm PT})$ at the smaller value of the
    coupling $\alpha_{\rm PT} = 1.05/(4\pi)\sim 0.0836...$. 
    Different analysis correspond to different fits to the GF 
    $\beta$-function as discussed in section~\ref{sec:beta-function-he}.}
  \label{fig:results_SF}
\end{figure}

Similarly to what we did in the previous section for the GF coupling,
we can now explore in the SF scheme the effect on
$\Lambda_{\overline{\rm MS}}/\mu_{\rm ref}$ of matching with
perturbation theory at different energy scales, $\mu_{\rm PT}$.  To
this end, we introduce the function
($\bar{g}_{\rm X,\,Y}\equiv \bar{g}_{\rm X}(\mu_{\rm
  Y})$):

\begin{multline}
  \label{eq:varphi_SF}
  \phi_{\rm SF}(\alpha_{\rm PT})= 2c\frac{\Lambda_{\overline{\rm
        MS}}}{\Lambda_{\rm SF}}
  ( b_0 \bar g^2_{\rm GF,\,ref})^{\frac{-b_1}{2b_0^2}} {\rm e}^{-\frac{1}{2b_0 \bar g^2_{\rm GF,\,ref}}} \times\\
  \times\exp\big\{-I_g^{\rm GF}(\bar{g}_{\rm GF,\,ref},\bar{g}_{\rm
    GF,\,PT}) -I_{g}^{\rm SF, 3}(\bar{g}_{\rm SF, 2c\mu_{\rm
      PT}},0)\big\}, \quad \alpha_{\rm PT}\equiv\frac{\bar{g}^2_{\rm
      GF,\,PT}}{4\pi},
\end{multline}
where $I_g^{\rm GF}$ is defined in terms of the non-perturbative
$\beta$-function in either the electric or magnetic GF scheme
(cf.~eq.~(\ref{eq:lamdef})), while $I_{g}^{\rm SF,3}$ integrates the
perturbative 3-loop $\beta$-function in the SF scheme
(cf.~eq.~(\ref{eq:IntegralPT2})). This quantity is entirely analogous
to eq.~(\ref{eq:lam_truncated}), the only difference being that once
we arrive at the scale $\mu_{\rm PT}$ with the running in the GF
scheme, we change to the SF scheme at the scale $2c\mu_{\rm PT}$, and
we then use perturbation theory in the SF scheme (see figure
\ref{fig:diag} for a cartoon). Note that the factor $s=2c$ appearing
on the r.h.s. of eq.~(\ref{eq:varphi_SF}) compensates the scale
difference in matching the SF and GF schemes. The function
$\phi_{\rm SF}$ has, of course, an asymptotic expansion of the form:
\begin{equation}
  \label{eq:varphi_to_zero}
  \phi_{\rm SF}(\alpha_{\rm PT}) = \frac{\Lambda_{\overline{\rm
        MS}}}{\mu_{\rm ref}} + \mathcal O(\alpha_{\rm PT}^2)\,,
\end{equation}
in complete analogy with $\phi_{\rm GF}$. The crucial difference lies
in the size of the $\mathcal O(\alpha_{\rm PT}^2)$ corrections. As
clearly shown in figure~\ref{fig:lovmSF}, in this case the corrections
are negligible within our statistical precision. In contrast
to the case of the GF schemes, the value of
$\phi_{\rm SF}(\alpha_{\rm PT})$ is essentially independent on
$\alpha_{\rm PT}$ in the range $\alpha_{\rm PT}\sim 0.1-0.2$. This
allows us to take as our estimate for
$\Lambda_{\overline{\rm MS}}/\mu_{\rm ref}$ the value of
$\phi_{\rm SF}(\alpha_{\rm PT})$ at the smallest coupling reached in
our simulations, which is:
$\alpha_{\rm PT} = 1.05/(4\pi)\sim 0.0836$.  It is important to
stress that this result does not rely on the 3-loop $\beta$-function
coefficient $b_2$ of the GF schemes. Different fits to the GF
$\beta$-function then give compatible results (see
figure~\ref{fig:results_SF}).  In particular, the determinations show
very good agreement once the coarser lattice, $L/a=8$, is dropped from
the analysis of the GF $\beta$-function. Moreover, as already
mentioned, changing the parametrization of the matching function
$f(u)$ in eq.~(\ref{eq:gf2sf}) results in insignificant
changes in the final numbers.

As the final result of this strategy that combines the GF and SF
couplings, we choose to quote the one based on the Zeuthen flow data
for the magnetic GF scheme with $L/a>10$:
\begin{equation}
  \label{eq:res_he_sf}
  \frac{\Lambda_{\overline{\rm MS}}}{\mu_{\rm ref}} =  0.0797(11)
  \qquad
  [1.37\%]\,.
\end{equation}
This number has one of the largest uncertainties of all the analysis
we considered, and its error covers all central values of the
determinations which use only lattices with $L/a>8$.  In conclusion,
thanks to the better perturbative behaviour of the SF schemes, by
non-perturbatively matching the GF and SF schemes we can reliably
quote an uncertainty significantly smaller than in
eq.~(\ref{eq:res_he}).  Similar conclusions are obtained using a more
conventional step-scaling strategy based on the GF couplings, and
considering different values of $\nu$ for the SF scheme. We refer the
interested reader to Appendix \ref{sec:ss}.

%%% Local Variables:
%%% mode: latex
%%% TeX-master: "paper"
%%% End:

\section{Connection to an hadronic scale}
\label{sec:hadronic}

To compute the $\Lambda$-parameter in units of an hadronic quantity,
like $t_0$~\cite{Luscher:2010iy} or $r_0$~\cite{Sommer:1993ce}, we
must relate this quantity with the technical scale $\mu_{\rm ref}$.
In this section we consider two different strategies to 
compute this relation. Both rely on the determination of the
$\beta$-function of the GF coupling at relatively low-energy
scales. In the first strategy, we first use the $\beta$-function
to relate $\mu_{\rm ref}$ with a convenient low-energy finite-volume 
scale, $\mu_{\rm had}$.  We then relate, in a second step, this  scale
with the infinite volume scales $t_0$ and $r_0$. We shall 
refer to this strategy as "fixed scale determination". 
In the second strategy, instead, we use the knowledge of the 
$\beta$-function to match directly $\mu_{\rm ref}$ with $t_0$ and $r_0$;
we shall refer to this as "global analysis".

\subsection{The $\beta$-function at low-energy}
\label{sec:beta-function-le}

We begin by defining two convenient low-energy finite-volume scales,
one for each GF scheme. We do so through the conditions:
\begin{equation}
\label{eq:muhad}
\bar g^2_{\rm GF,\,e}(\mu_{\rm had, e}) = 10.95
=
\bar g^2_{\rm GF,\,m}(\mu_{\rm had, m})\,;
\end{equation}
note that $\mu_{\rm had, e}\neq\mu_{\rm had, m}$.
The desired ratios of scales $\mu_{\rm ref}/\mu_{\rm had}$ can now be
inferred once the $\beta$-function in the two schemes is known in the
range of couplings:
$\bar{g}^2\in[\bar{g}^2_{\rm GF,\,ref},\bar{g}^2_{\rm GF,\,had}]$. In
this range, we find convenient to employ a parametrization of the
$\beta$-function of the form:
\begin{equation}
\label{eq:beta_low}
\beta(x) = -\frac{x^3}{\sum_{k=0}^{n_b}p_kx^{2k}}\,.
\end{equation}
This is completely analogous to eq.~(\ref{eq:inverse_beta}), the
difference being that here we do not fix the coefficients
$p_0,p_1,p_2$, to their perturbative values. As we saw for
eq.~(\ref{eq:ScaleRatio}), this parametrization allows us to express
in a straightforward way the function $F$ of eq.~(\ref{eq:F}) in terms
of the parameters of the $\beta$-function, i.e.:
\begin{equation}
\label{eq:F2}
F(a,b) = H(a) - H(b),
\qquad
H(x)=  \frac{p_0}{2 x^2} - p_1 \log x - \sum_{k=1}^{n_b-1} p_{k+1}\frac{x^{2k}}{2k}.
\end{equation}
As done previously for the $\beta$-function at high-energy, the fit
coefficients $p_k$ can then be determined by minimizing the
$\chi^2$-function, eq.~(\ref{eq:chi2GF}) (or eq.~(\ref{eq:chi2GF2}));
we again consider the parametrization (\ref{eq:rhoGF}) for the cutoff
effects. Once $F$ is determined, the desired ratios of scales are
given by:
\begin{equation}
\log\bigg(\frac{\mu_{\rm ref}}{\mu_{\rm had}}\bigg) = 
F(\bar{g}^2_{\rm GF,\,ref}, \bar{g}^2_{\rm GF,\,had}) \,.
\end{equation}
We consider fits to data with $\bar g^2_{\rm GF}\in[2-4.7]$, and their
corresponding SSFs. We obtain good fits choosing $n_b=2,3$, and we
typically take $n_{c}=2,3$, although the final results are pretty much
independent on the number of parameters used to parametrize the cutoff
effects. Figure~\ref{fig:low_energy} collects a landscape of
results. As one can see from the figure, we have excellent agreement
between different analysis as long as we discard the coarser lattices
with $L/a=8$. As our final results we quote:
\begin{equation}
	\label{eq:murat}
	\frac{\mu_{\rm ref}}{\mu_{\rm had,m}} = 6.528(32)\quad [0.48\%]\,,
	\qquad
	\frac{\mu_{\rm ref}}{\mu_{\rm had,e}} = 6.516(35)\quad [0.54\%]\,.
\end{equation}
For completeness, we give below the fit coefficients and their
covariance matrix, which describe the $\beta$-function in the electric
scheme:

\begin{figure}
	\begin{subfigure}[t]{0.45\textwidth}
		\includegraphics[width=\textwidth]{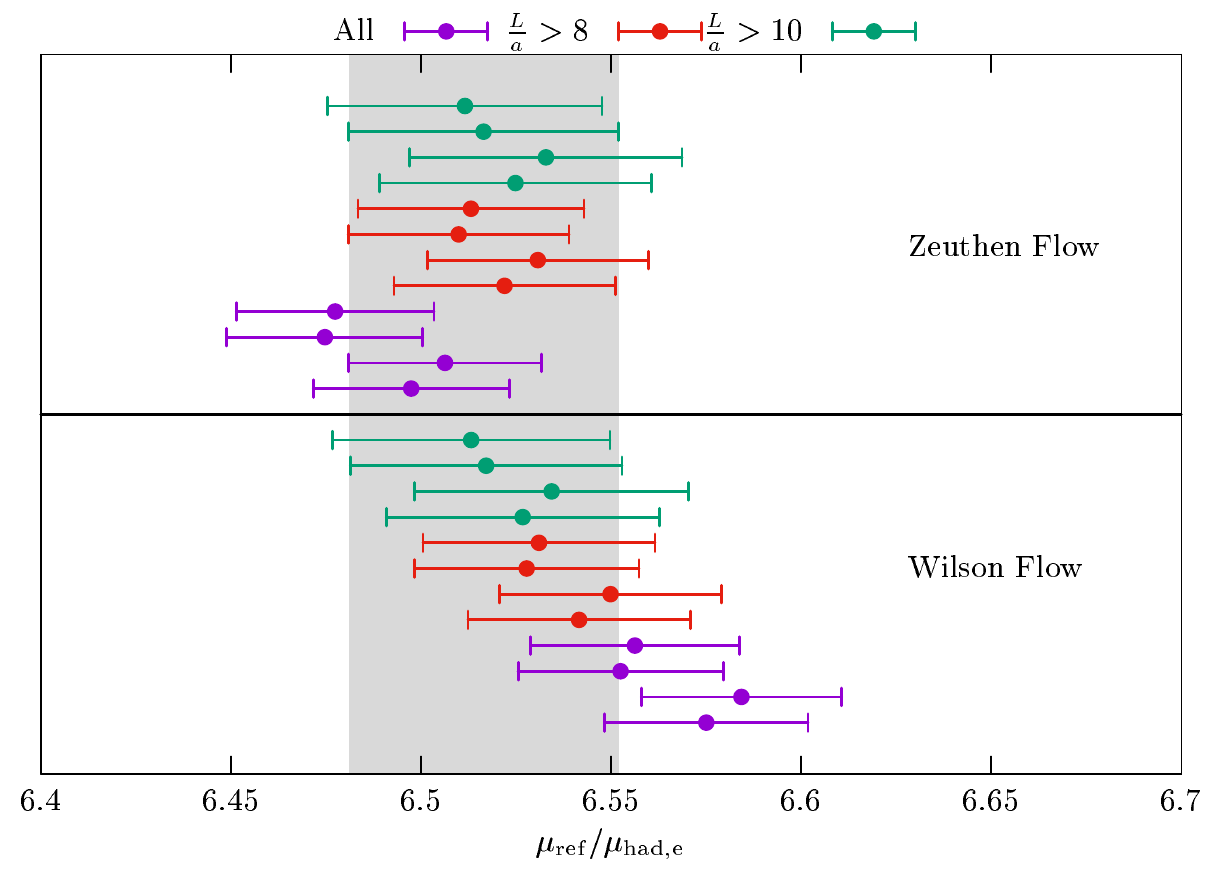}
		\caption{$\mu_{\rm ref}/\mu_{\rm had,e}$}
	\end{subfigure}
	\begin{subfigure}[t]{0.45\textwidth}
		\includegraphics[width=\textwidth]{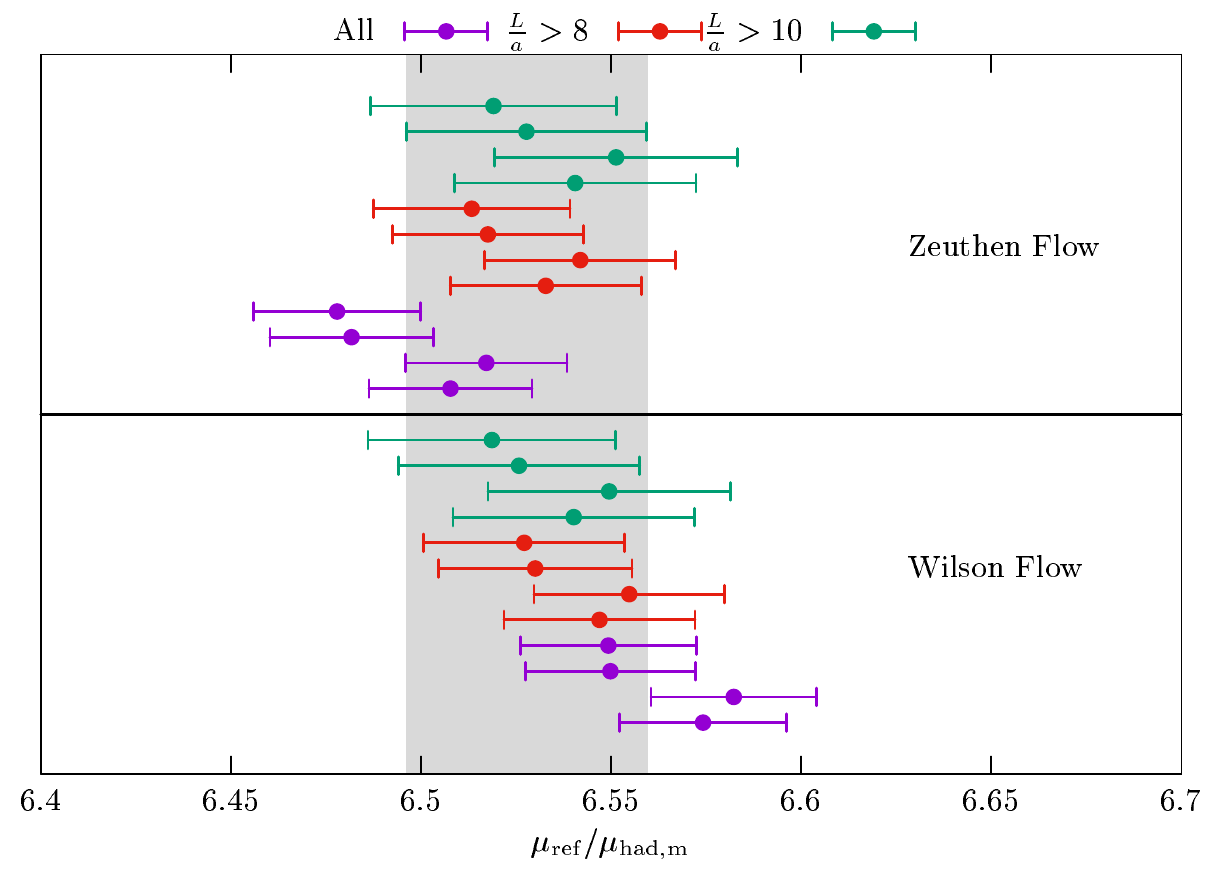}
		\caption{$\mu_{\rm ref}/\mu_{\rm had,m}$}
	\end{subfigure}
	\caption{Results for the ratios $\mu_{\rm ref}/\mu_{\rm had,e}$ and
		$\mu_{\rm ref}/\mu_{\rm had,m}$. The different analysis correspond
		to different choices of discretization (labelled Zeuthen and
		Wilson), parametrizations of the $\beta$-function and cuts on the
		lattice resolutions (see text for more details).}
	\label{fig:low_energy}
\end{figure}

\begin{equation}
p_0 = 14.86156402\,,\ 
p_1 =  1.03507289\,,\ 
p_2 = -0.15702286\,,\ 
p_3 =  0.01250783\,,
\end{equation}

\begin{displaymath}
{\rm cov}(p_i,p_j) = \left(
\begin{array}{cccc}
7.19145589\times 10^{\text{-1}} &   4.50192418\times 10^{\text{-1}} &  -8.34709691\times 10^{\text{-2}} &   4.60384717\times 10^{\text{-3}}\\
4.50192418\times 10^{\text{-1}} &   2.87728456\times 10^{\text{-1}} &  -5.41797562\times 10^{\text{-2}} &   3.02449220\times 10^{\text{-3}}\\
-8.34709691\times 10^{\text{-2}} &  -5.41797562\times 10^{\text{-2}} &   1.03377600\times 10^{\text{-2}} &  -5.83491819\times 10^{\text{-4}}\\
4.60384717\times 10^{\text{-3}} &   3.02449220\times 10^{\text{-3}} &  -5.83491819\times 10^{\text{-4}} &   3.32558532\times 10^{\text{-5}}\\
\end{array}
\right)\,.
\end{displaymath}

\begin{table}
  \centering
  \begin{tabular}{ll|ll|ll}
    \toprule
    $\beta$&$t_0/a^2$&$\beta$&$t_0/a^2$&$\beta$&$r_0/a$ \\
    \midrule
5.9600 & 2.7854(62)  \(^\dagger\)      & 6.4200 & 11.241(23)  \(^\dagger\) & 5.7000 & 2.9220(90)  \(^{\dagger\dagger}\)\\  
5.9600 & 2.7875(53)  \(^\ddagger\)     &  6.4500 & 12.196(21) \(^\ddagger\)        & 5.8000 & 3.6730(50)  \(^{\dagger\dagger}\)\\  
6.0500 & 3.7834(47)  \(^\ddagger\)     &  6.5300 & 15.156(28) \(^\ddagger\)        & 5.9500 & 4.898(12)  \(^{\dagger\dagger}\)\\   
6.1000 & 4.4329(32)  \(^{\star}\)  &  6.6100 & 18.714(30)  \(^\ddagger\)     &  6.0700 & 6.033(17)  \(^{\dagger\dagger}\)\\   
6.1300 & 4.8641(85)  \(^\ddagger\)     &  6.6720 & 21.924(81) \(^{\star}\)   & 6.1000 & 6.345(13)  \(^{\star}\)\\       
6.1700 & 5.489(14) \(^\dagger\)       &   6.6900 & 23.089(48) \(^\ddagger\)        & 6.2000 & 7.380(26)  \(^{\dagger\dagger}\)\\   
6.2100 & 6.219(13)  \(^\ddagger\)     &   6.7700 & 28.494(66) \(^\ddagger\)        & 6.3400 & 9.029(77)  \(^{\star}\)\\       
6.2900 & 7.785(14)  \(^\ddagger\)     &   6.8500 & 34.819(84)  \(^\ddagger\)       & 6.4000 & 9.740(50)  \(^{\dagger\dagger}\)\\   
6.3400 & 9.002(31)  \(^{\star}\)  &  6.9000 & 39.41(15) \(^{\star}\)      & 6.5700 & 12.380(70)  \(^{\dagger\dagger}\)\\  
6.3400 & 9.034(29)  \(^{\star}\)  &   6.9300 & 42.82(11) \(^\ddagger\)           &6.5700 & 12.176(97)\(^{\mathsection}\) \\        
6.3700 & 9.755(19)  \(^\ddagger\)     &    7.0100 & 52.25(13)  \(^\ddagger\)      & 6.6720 & 14.103(92)  \(^{\star}\)\\        
6.4200 & 11.202(21)  \(^\ddagger\)    &           &                       &     6.6900   &     14.20(12)\(^{\mathsection}\) \\  
           & & & &  6.8100   &     16.54(12)\(^{\mathsection}\) \\
           & & & &  6.9000   & 18.93(15)  \(^{\star}\)  \\
           & & & &  6.9200   &     19.13(15)\(^{\mathsection}\) \\
    \bottomrule
  \end{tabular}
  \caption{Results for $t_0/a^2$ and $r_0/a$ for different values of $\beta$. 
  	       For the case of $t_0/a^2$ the relevant references are 
  	       $^\dagger$~\cite{Luscher:2010iy}, $^\ddagger$~\cite{Giusti:2018cmp}, 
  	       $^{\star}$~\cite{Knechtli:2017xgy}. 
  	       For $r_0/a$, instead, the results are from $^{\dagger\dagger}$~\cite{Guagnelli:1998ud} and $^{\star}$~\cite{Knechtli:2017xgy}. The data labeled $^{\mathsection}$ is obtained from the values of $r_c/a$ of~\cite{Necco:2001xg} together with $r_c/r_0=0.5133(24)$ which gives the quoted values of $r_0/a$.} 
  \label{tab:t0data}
\end{table}

\subsection{Determination of $\sqrt{8t_0}\mu_{\rm ref}$: fixed
  renormalization scale}
\label{sec:determ-sqrt8t_0m-ref}

To obtain the sought-after relation $\sqrt{8t_0}\mu_{\rm ref}$ we now
need to compute $\sqrt{8t_0}/a$  
and $a\mu_{\rm had}$ for several values of the lattice spacing 
and extrapolate their product to the continuum limit, i.e.,
\begin{equation}
  \label{eq:dless_ratios}
  \lim_{a\to 0} \bigg(\frac{\sqrt{8t_0}}{a}\bigg)\times a\mu_{\rm had}=
  \sqrt{8t_0}\mu_{\rm had}\,.
\end{equation}  
This is done by determining the finite volume and hadronic scale in
units of the lattice spacing at matching values of the bare coupling, 
$\beta=6/g_0^2$, which guarantees that the value of $a$ is the
same up to scaling violations.
The quantity $t_0/a^2$ is known from the literature
over a wide range of $\beta$. In table \ref{tab:t0data} we collected
the results used in our study and the corresponding
references. Results for $t_0/a^2$ are known for values of
the lattice spacing as fine as $0.025\, {\rm fm}$. Simulations at very
fine lattice spacings deal with the problem of topology
freezing~\cite{DelDebbio:2004xh} either
by simulating very large physical volumes~\cite{Giusti:2018cmp}, or by
using open boundary conditions in the Euclidean time
direction~\cite{Knechtli:2017xgy}.  

In order to determine
$a\mu_{\rm had}$, instead, for a given set of lattice sizes $L/a$, 
we must find the values of the bare coupling, $\beta=\beta_{\rm had}(L/a)$,
for which the conditions (\ref{eq:muhad}) are satisfied. At these 
$\beta$'s we then have: $a\mu_{\rm had}=a/(cL)$ 
(cf.~eqs.~(\ref{eq:GFCouplings}),(\ref{eq:GFCouplings2})). 
Given our large set of data we can consider lattices with sizes
$L/a\in[12,32]$.%
\footnote{In principle data with $L/a=48$ is also available. However, fixing the value
of the coupling to $\bar g^2_{\rm GF}=10.95$ requires in this case an 
\emph{interpolation} of data points which are significantly distant
from the target value  (cf.~table~\ref{tab:la48}). We thus prefer not to 
include this data in our analysis. Nonetheless, we have checked the effect of
including it and found complete agreement for our final result, but
with a slightly smaller error.}
The values of $\beta_{\rm had}(L/a)$ are then easily found by performing 
interpolations of the data in $\beta$ at fixed $L/a$. The 
values of $\beta_{\rm had}(L/a)$ determined this way are collected in 
table~\ref{tab:bt_interpol}, where the results for different lattice 
discretizations of the GF couplings are given.  

From tables \ref{tab:t0data} and \ref{tab:bt_interpol}, it is clear
that the values of $\beta$ for which the large volume quantities
$t_0/a^2$ are available do not match exactly those which 
correspond to $\bar g_{\rm GF}^2(\mu_{\rm had})=10.95$ for our $L/a$'s.  
To obtain the products (\ref{eq:dless_ratios}) at matching $\beta$-values
we can thus proceed in either of the following two ways:

\begin{enumerate}
\item Given the results of table \ref{tab:t0data}, we fit the data for 
 $\log(t_0/a^2)$ as a function of $g_0^2$. In practice, 
 we obtain a good description of the data by using a simple polynomial 
 fit of degree 5. 
 In this way we can infer the values of $\log(t_0/a^2)$ 
 at the bare couplings of table~\ref{tab:bt_interpol}, corresponding to 
 the finite volume results for $a\mu_{\rm had}$.
\item We fit the results for $\beta_{\rm had}(L/a)$ as a function of $L/a$,
 or equivalently $a\mu_{\rm had}$. In this case, we obtain good fits using 
 polynomials of degree 3. Given this functional form, we can then find the 
 value of $a\mu_{\rm had}$ at the bare couplings where the large volume 
 quantities $t_0/a^2$ have been computed (cf.~table~\ref{tab:t0data}). 
\end{enumerate}

\begin{figure}
  \centering
  \includegraphics[width=0.9\textwidth]{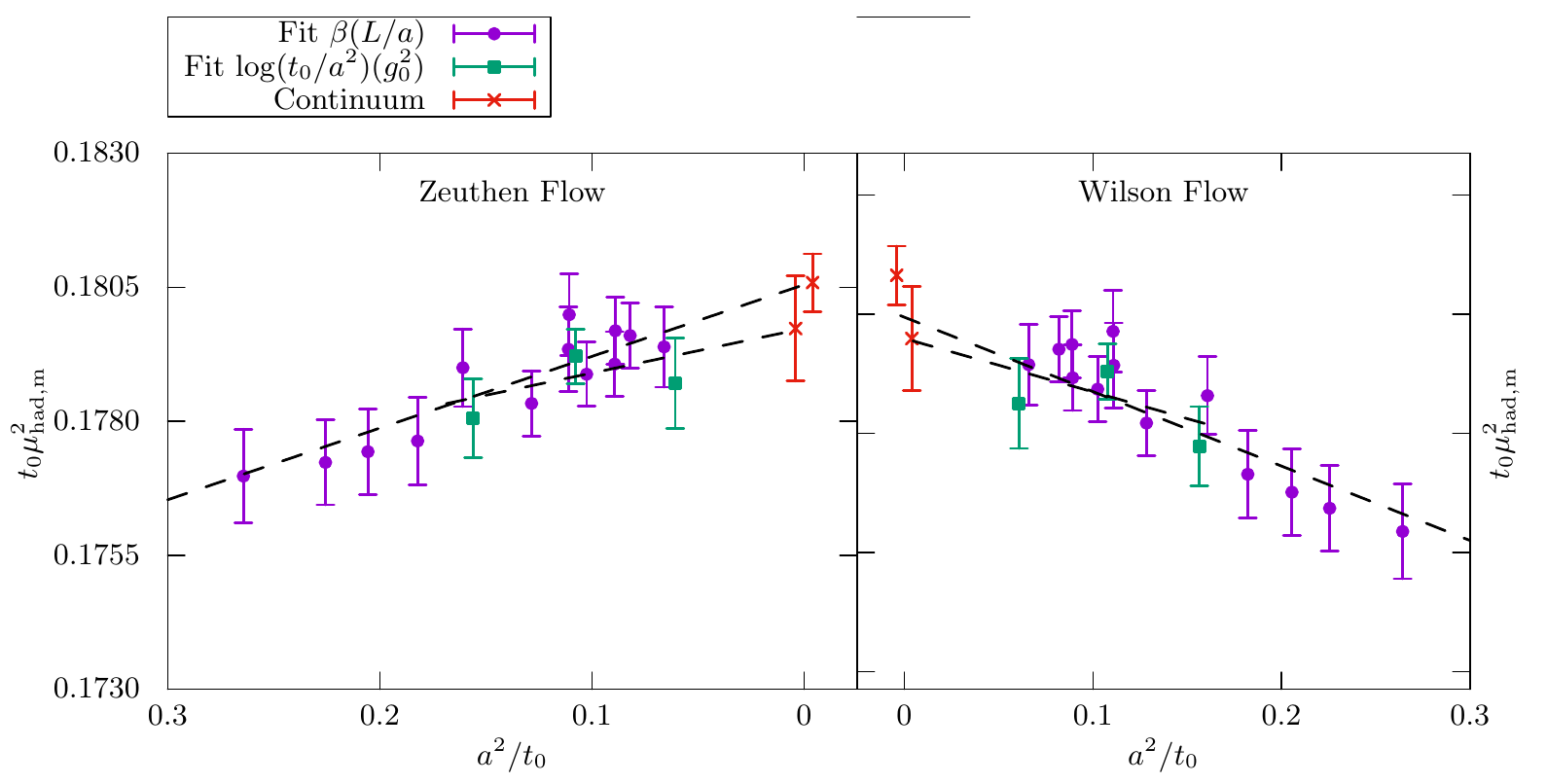}
  \caption{$t_0\mu_{\rm had,m}^2$.}
  \label{fig:hat_t0}
\caption{Continuum extrapolations for $t_0\mu_{\rm had,m}^2$ using
  different strategies and discretizations for the GF couplings  (see
  text for more details).} 
\label{fig:had}
\end{figure}

Figure~\ref{fig:had} shows the continuum extrapolations of $t_0\mu_{\rm ref, m}^2$
determined according to the above strategies, and
for different lattice discretizations for the GF couplings. For the
case of the strategy 1 above we only use the data with $L/a>16$. The
agreement among the different analysis techniques and the different discretizations is evident. 
We choose as final result the determination obtained following
strategy 1 and based on the Zeuthen-flow data. This has the largest error and gives:
\begin{alignat}{2}
  \label{eq:t0muhad}
  \sqrt{8t_0}\mu_{\rm had, m} &=      1.1961(35)\qquad     [0.29\%]\,,
  \qquad
  \sqrt{8t_0}\mu_{\rm had, e} &=       1.1991(33)\qquad     [0.27\%]\,.
\end{alignat}
Note the remarkable precision we obtain, i.e., around a 3 per-mille.
Combining these results with those of eq.~(\ref{eq:murat}) we 
can quote
\begin{equation}
  \label{eq:t0muref}  
  \sqrt{8t_0}\mu_{\rm ref} =        7.808(46) \qquad    [0.59\%]\,,
\end{equation}
which uses the results for the magnetic scheme. The value obtained using
$\mu_{\rm had, e}$ as intermediate scale instead is
$\sqrt{8t_0}\mu_{\rm ref} = 7.814(50)$, which is in good agreement. 

\begin{table}
  \centering
\begin{tabular}{lllllll}
  \toprule
  &&& \multicolumn{4}{c}{$\beta_{\rm had,\,e/m}(L/a)$}\\
  \cmidrule(lr){4-7}
  & &&\multicolumn{2}{c}{Wilson Flow} &\multicolumn{2}{c}{Zeuthen Flow}
  \\
  \cmidrule(lr){4-5} \cmidrule(lr){6-7}
  $L/a$ &$a\mu_{\rm had, e/m}$&$\bar g^2_{\rm GF,\,e/m}(\mu_{\rm had, e/m})$ 
  & Magnetic & Electric & Magnetic & Electric\\
  \cmidrule(lr){1-1}\cmidrule(lr){2-2} \cmidrule(lr){3-3}
  \cmidrule(lr){4-4} \cmidrule(lr){5-5} \cmidrule(lr){6-6}
  \cmidrule(lr){7-7}
32 & 0.1042& 10.95 & 6.5609(17) & 6.5588(18) & 6.5617(17) & 6.5598(19)\\
24 & 0.1389& 10.95 & 6.3512(10) & 6.3497(11) & 6.35170(99) & 6.3504(11)\\
20 & 0.1667& 10.95 & 6.2209(14) & 6.2191(16) & 6.2219(14) & 6.2204(15)\\
16 & 0.2083& 10.95 & 6.0722(14) & 6.0712(15) & 6.0739(14) & 6.0730(14)\\
12 & 0.2778& 10.95 & 5.9004(16) & 5.8992(17) & 5.9033(16) & 5.9023(16)\\
  \bottomrule
\end{tabular}
  \caption{Values of the bare coupling $\beta_{\rm had}(L/a)$ for different
           lattice sizes and discretizations which correspond to a fixed coupling 
           $\bar g^2_{\rm GF}(\mu_{\rm had}) = 10.95$.}
  \label{tab:bt_interpol}
\end{table}

\subsection{Determination of $\sqrt{8t_0}\mu_{\rm ref}$: global analysis}
\label{sec:determ-sqrt8t_0m-ref-1}

As can be seen from table \ref{tab:t0data}, the flow scale  in lattice
units $t_0/a^2$ 
has been determined precisely even at very fine lattice spacings. This
data, together with our results for the $\beta$-function at low energy
in Sect.~\ref{sec:beta-function-le}, calls for an alternative strategy
to determine directly the product $\sqrt{8t_0}\mu_{\rm ref}$. 

We start by performing once again a fit to the data for $\log (t_0/a^2)$ 
of table \ref{tab:t0data} by a polynomial in $\beta = 6/g_0^2$ of
degree 5. This gives us a parametrization:
\begin{equation}
  \label{eq:t0_param}
  \log (t_0/a^2) = P(g_0^2) = \sum_{k=0}^{5} c_k g_0^{2k}\,,
\end{equation}
in the range $\beta \in [5.96, 7.01]$. We then consider our GF
coupling data in the very same range of $\beta$, for lattice sizes $L/a=16,20,24,32,48$. 
For the ease of presentation we focus on the results for the magnetic components of the 
coupling; the analysis using the electric scheme gives compatible results.
We will denote these generically as $u_{\rm FV}=\bar{g}^2_{\rm
GF,m}(\mu_{\rm FV};a/L,g_0)$. Using 
the parametrization of the $\beta$-function, eq.~(\ref{eq:beta_low}), valid in this range
of couplings, we can compute the scale factor between the reference scale $\mu_{\rm ref}$ 
and the scale $\mu_{\rm FV}$ corresponding to a given coupling $u_{\rm FV}=
\bar g_{\rm GF}^2(\mu_{\rm FV})$. This is simply given by (cf.~eq.~(\ref{eq:F2})):
\begin{equation}
  \label{eq:RFV}
  \log\big(R(u_{\rm FV}, a/L, g_0^2)\big) =  F(u_{\rm ref},u_{\rm FV})=
  \log\bigg(\frac{\mu_{\rm ref}}{\mu_{\rm FV}}\bigg)  + \mathcal O(a^2)\,,
\end{equation}
where $u_{\rm ref}\equiv\bar{g}^2_{\rm GF}(\mu_{\rm ref})$, of eq.~(\ref{eq:gmuref_m}).
Note that the scale factor associated to a given $u_{\rm FV}$ is 
only given up to discretization errors, the reason being that we are considering the value 
of the coupling at a certain $L/a$ and $g_0$. Given the results for $R$, we recall that 
$a\mu_{\rm FV} = a/(cL)$. The parametrization (\ref{eq:t0_param}) then allows us to 
determine the second matching factor:
\begin{equation}
  \label{eq:QFV}
  Q(u_{\rm FV}, a/L, g_0^2) =
  \sqrt{8}\exp\left(\frac{P(g_0^2)}{2} \right)\times \bigg(\frac{a}{cL}\bigg) =
  \sqrt{8t_0}\mu_{\rm FV} + \mathcal O(a^2)\,,
\end{equation}
which relates the given finite-volume scale $a\mu_{\rm FV}$ with our hadronic quantity in 
lattice units. 

\begin{figure}
  \centering
  \includegraphics[width=\textwidth]{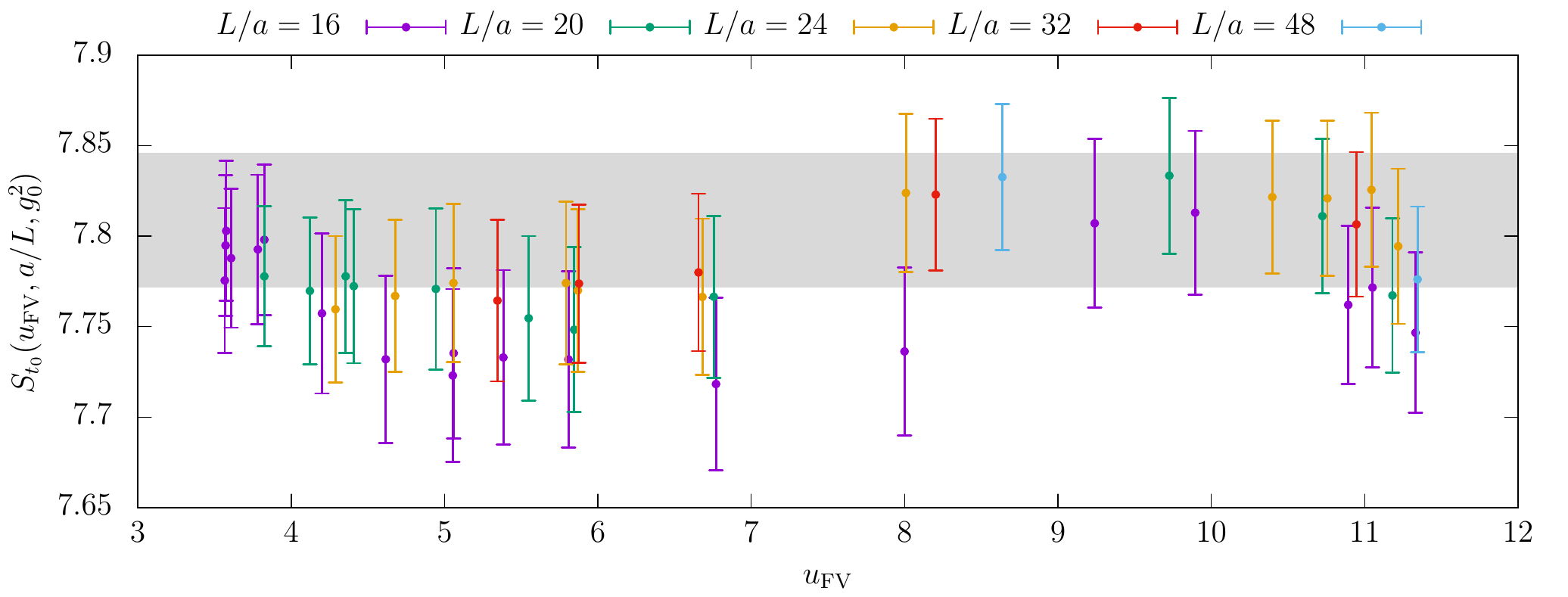} 
  \caption{The product $S_{t_0}$ of the running factor $R$ and the matching factor $Q$ 
  	(see eq.~(\ref{eq:RQ})). This has to be constant and equal to $\sqrt{8t_0} \mu_{\rm ref}$ 
  	up to discretization errors. The plot shows that these are very small. 
    The error band shows the result of the fit, eq.~(\ref{eq:res_gl}). }
  \label{fig:global}
\end{figure}

Combining finally the factor $R$ of eq.~(\ref{eq:RFV}) with $Q$ of (\ref{eq:QFV}) we obtain:
\begin{equation}
  \label{eq:RQ}
  S_{t_0}(u_{\rm FV}, a/L, g_0^2) = R(u_{\rm FV}, a/L, g_0^2) \times Q(u_{\rm FV}, a/L, g_0^2) =
  \sqrt{8t_0} \mu_{\rm ref} + \mathcal O(a^2)\,,
\end{equation}
which, up to discretization errors, is the product $\sqrt{8t_0}
\mu_{\rm ref}$ we were after. Note that by construction the dependence
on $u_{\rm FV}$ cancels in the leading term of the product,
eq.~(\ref{eq:RQ}): only discretization errors thus depend on $u_{\rm
FV}$. The latter turn out to be relatively small.  This can be
appreciated in figure \ref{fig:global}, which shows the different
results for the factor $S_{t_0}$. As we can see from the figure, the product
$S_{t_0}$ is approximately constant as a function of $u_{\rm FV}$ with
variations of at most 2\% when changing $u_{\rm FV}\sim[3.5
,11.5]$. This observation suggests to fit our results for
$S_{t_0}(u_{\rm FV},a/L, g_0^2)$ as: 
\begin{equation}
  \label{eq:Sfit}
  S_{t_0}(u_{\rm FV},a/L, g_0^2) =  a_0 + \sum_{k=1}^{n_c} \rho_k u_{\rm FV}^k \Big(\frac{a}{L}\Big)^2\,, 
\end{equation}
where $a_0$ and $\rho_k$ are fitting parameters. The continuum result
$\sqrt{8t_0}\mu_{\rm ref}$,  
corresponding to the coefficient $a_0$ of the fit, shows very little
dependence on the degree 
of the polynomial describing the cutoff effects, as long as $n_c>1$. As
our final estimate we quote:
\begin{equation}
  \label{eq:res_gl}
  \sqrt{8t_0}\mu_{\rm ref} = 7.809(37)
  \qquad [0.5\%] \,,
\end{equation}
which is based on the Zeuthen-flow data in the magnetic scheme for
$L/a\geq16$ and uses $n_c=2$.
This result is in agreement with our previous result, eq.~(\ref{eq:t0muref}).
We note that including also lattices with $L/a<16$ gives completely compatible results 
with eq.~(\ref{eq:res_gl}), but with smaller uncertainties.

\subsection{Determination of $r_0\mu_{\rm had}$}
\label{sec:determ-r_0muhad}

The strategies we used to compute $\sqrt{8t_0}\mu_{\rm had}$ can also  
be applied for the determination of $r_0\mu_{\rm had}$. In this 
case, however, the situation is a little complicated by a few technical issues. 
First of all, the different results for $r_0/a$ available in the literature which 
we could consider~\cite{Knechtli:2017xgy,Necco:2001xg, Guagnelli:1998ud} use different 
discretizations for the relevant observables. Secondly, two of these determinations, 
refs.~\cite{Necco:2001xg, Guagnelli:1998ud}, are more than 20 years old. A note of 
concern in this case is thus the issue of topology freezing~\cite{DelDebbio:2004xh}, 
which was not known at the time of these computations. This issue is potentially more 
significant at the very fine lattice spacings simulated in ref.~\cite{Necco:2001xg}, 
while the results of ref.~\cite{Guagnelli:1998ud} are confined to substantially larger
spacings. The more recent determination of $r_0/a$ in~\cite{Knechtli:2017xgy}, on the other
hand, employs open boundary conditions. In addition, their values for $r_0/a$ cover a factor
two in lattice spacings and reach down to small spacings comparable to those of~\cite{Necco:2001xg} 
(cf.~table~\ref{tab:t0data}). For these reasons we consider safer to restrict  
our attention in the following only to the results of~\cite{Knechtli:2017xgy,Guagnelli:1998ud}.

\begin{figure}[t]
  \includegraphics[width=0.9\textwidth]{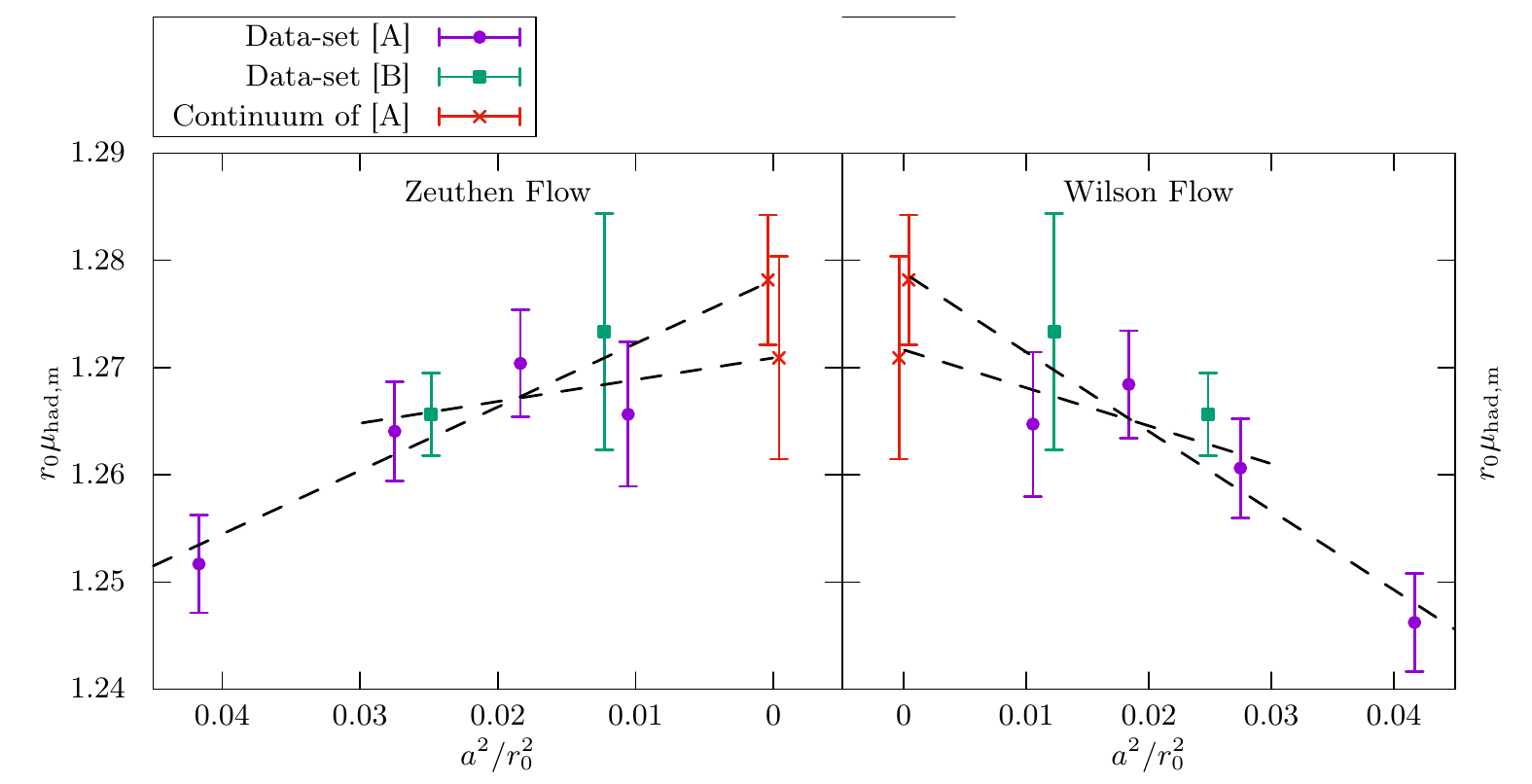}
  \caption{Continuum extrapolations of $r_0\mu_{\rm had,m}$ using the
    data set [A]~\cite{Guagnelli:1998ud} and the results of
    table~\ref{tab:bt_interpol}. For illustration, the two points of 
    data set [B]~\cite{Knechtli:2017xgy} in the range of $\beta$ covered 
    by the finite volume simulations (table~\ref{tab:bt_interpol}) are 
    also plotted. These however are not used for any continuum extrapolation 
    (see main text for the details).}
  \label{fig:had_r0}
\end{figure}

Most of the results of~\cite{Knechtli:2017xgy} have been obtained at values of the 
bare coupling $g_0^2$ which lie outside the range of bare couplings of table~\ref{tab:bt_interpol} 
(cf.~table~\ref{tab:t0data} where the results of~\cite{Knechtli:2017xgy} are labelled with a $\star$). 
This means that  the fixed scale strategy considered in Sect.~\ref{sec:determ-sqrt8t_0m-ref} cannot 
really be applied to this data set. A fixed scale determination can only be considered for
the results of~\cite{Guagnelli:1998ud} (which are labelled by $^{\dagger\dagger}$ in table~\ref{tab:t0data}). 
For these we opt for \emph{strategy 2} of Section~\ref{sec:determ-sqrt8t_0m-ref}. We thus use 
the same fit for $\beta_{\rm had}(L/a)$ as a function of $L/a$ considered there, which is based on a
 3rd degree polynomial. Given this functional form, we then find the value of $a\mu_{\rm had} = a/(cL)$ 
at the values of $g_0^2$ where $r_0/a$ is known, and compute $r_0\mu_{\rm had}$. The continuum extrapolation
\begin{equation}
    \qquad
  \lim_{a\to 0} \bigg(\frac{{r_0}}{a}\bigg)\times a\mu_{\rm had}=
  {r_0}\mu_{\rm had}\,,
\end{equation}
corresponding to the $r_0/a$ data of ref.~\cite{Guagnelli:1998ud} are
shown in 
figure~\ref{fig:had_r0}. Using the results of these extrapolations and
those of eq.~(\ref{eq:murat}) we find:
\begin{eqnarray} \text{4-point extrapolation:} & r_0 \mu_{\rm ref} =
8.327(58)\qquad [0.70\%]\,,\\
  \label{eq:r03pt} \text{3-point extrapolation:} & r_0 \mu_{\rm ref} =
8.277(75)\qquad [0.90\%]\,.
\end{eqnarray}%

\begin{figure}[t!]  \centering
\includegraphics[width=\textwidth]{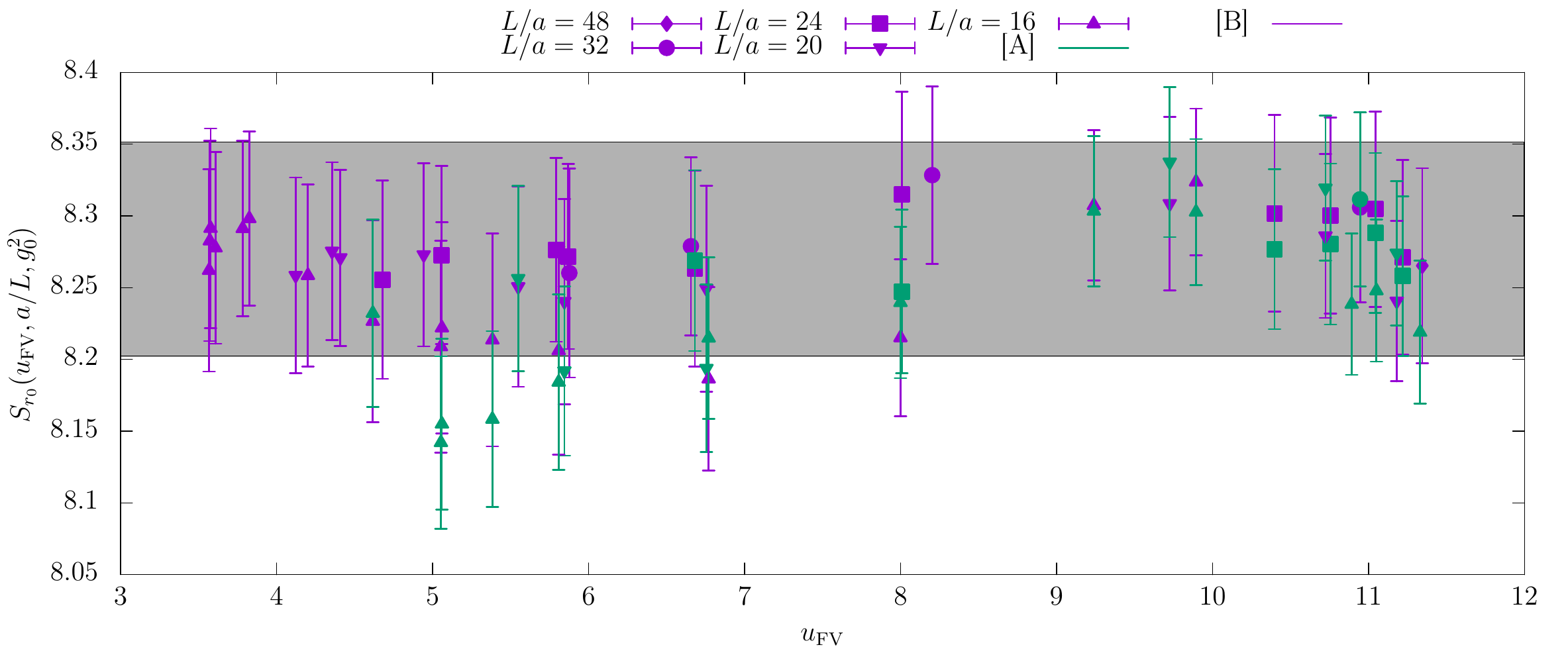}
  \caption{The product $S_{r_0}(u_{\rm FV},a/L, g_0^2)$ of the running
factor $R$ and the matching factor $Q_{r_0}$ (analogous to
eq.~(\ref{eq:QFV}) for $r_0$). This has to be constant and equal to
$r_0 \mu_{\rm ref}$ up to discretization errors; the plot clearly shows that
these are very small for both data sets. In different colours we have the two data
sets: [A] is ref~\cite{Guagnelli:1998ud}, [B] is ref~\cite{Knechtli:2017xgy}
(cf.~table \ref{tab:t0data}), while different symbols correspond to
different lattice sizes. The grey error band is the result of
eq.~(\ref{eq:r03pt}) (see main text for more details). }
  \label{fig:r0_comp}
\end{figure}

The global approach described in Section~\ref{sec:determ-sqrt8t_0m-ref-1} can on the other 
hand also be used for the $r_0/a$ data of ref.~\cite{Knechtli:2017xgy}. We thus  begin 
by fitting the data for $r_0/a$ as a function of the bare coupling $g_0^2$. We do this separately 
for the data set of~\cite{Knechtli:2017xgy} (labelled [B] in figure~\ref{fig:r0_comp})
and for the data set of~\cite{Guagnelli:1998ud} previously used (labelled [A] in figure~\ref{fig:r0_comp});
this because the two computations use different observable discretizations. 
With these functional forms at hand, we then determine the product of the
running factor $\mu_{\rm ref}/\mu_{\rm FV}$ (cf.~eq.~(\ref{eq:RFV}))
and the matching factor $r_0\mu_{\rm FV}$
(cf.~eq.~(\ref{eq:QFV})). The result:
\begin{equation}
  S_{r_0}(u_{\rm FV},a/L, g_0^2) = r_0\mu_{\rm ref} +
\mathcal O(a^2)\,,
\end{equation}
is expected to be constant, up to scaling
violations. Figure~\ref{fig:r0_comp} shows the quantity $S_{r_0}$ for
the two data sets. It is clear that scaling violations within
each data set are very small. Looking at the continuum extrapolated
values, obtained after fitting separately the two data sets to a constant 
with cutoff effects parametrized by a 1st degree polynomial
(cf.~eq.~(\ref{eq:Sfit})), we find:
\begin{eqnarray} \text{Data set [A]:} & r_0 \mu_{\rm ref} =
                                        8.294(62)\,,\\
    \label{eq:r0_gl_S} \text{Data set [B]:} & r_0 \mu_{\rm ref} =
8.297(64)\,.
\end{eqnarray}
which are in very good agreement.

To conclude with the determination of $r_0 \mu_{\rm ref}$, a conservative option 
for us is to quote the result of the 3-point extrapolation, eq.~(\ref{eq:r03pt}).
Among the consistent analysis that we showed, this gives the result with the 
largest uncertainty. 

%%% Local Variables:
%%% mode: latex
%%% TeX-master: "paper"
%%% End:

\section{The $\Lambda$-parameter}
\label{sec:lambda-parameter}

We are now ready to express $\Lambda_{\overline{\rm MS}}$ in terms of
the hadronic scales $t_0$ and $r_0$. Our first main result is:
\begin{equation}
  \label{eq:final_result_t0} \sqrt{8t_0}\, \Lambda_{\overline{\rm MS}}
= \frac{\Lambda_{\overline{\rm MS}}}{\mu_{\rm ref}}\times
\frac{\mu_{\rm ref}}{\mu_{\rm had, m}}\times \sqrt{8t_0}\mu_{\rm had,
m} = 0.6227(98) \qquad [1.57\%]\,.
\end{equation} 
The first factor on the r.h.s.~of this equation comes
from the non-perturbative high-energy determination of the GF
$\beta$-function in the electric scheme, combined with a
non-perturbative matching of the GF and SF schemes. Thanks to this
strategy, perturbation theory in the SF scheme could be safely used to
run to infinite energy and obtain the result of
eq.~(\ref{eq:res_he_sf}). The second factor derives instead from our
determination of the $\beta$-function at low-energies in the magnetic
GF scheme (cf.~eq.~(\ref{eq:murat})). The third and last factor is
the non-perturbative matching of the magnetic GF coupling with
the gradient flow scale $t_0$, eq.~(\ref{eq:t0muhad}). The result in
eq.~(\ref{eq:final_result_t0}) would not practically change, if we
were to choose the scale $\mu_{\rm had, e}$ defined in the electric
scheme in the second and third factor. We would also obtain a
perfectly compatible result ($\sqrt{8t_0}\, \Lambda_{\overline{\rm
MS}} = 0.6227(94)$), if we replaced the last two factors with
$\sqrt{8t_0}\mu_{\rm ref}$ determined via the global approach,
eq.~(\ref{eq:res_gl}).

Our final error estimate ($\sim 1.6\%$) is very conservative. All
three factors are determined by dropping the two coarsest lattice
spacings at our disposal, and we typically chose as final result the
analysis with the largest uncertainty.  The only exception was when we
favoured the result of eq.~(\ref{eq:res_he_sf}) over
eq.~(\ref{eq:res_he}). We found however compelling to match with the
asymptotic perturbative regime in the SF rather than the GF
scheme. Using the GF scheme would have indeed increased our final
error by more than 50\%, due to the bad convergence of its
perturbation theory; \emph{this despite of the fact that we had access
to the non-perturbative running of the coupling up to very large
energy scales, where $\bar g_{\rm GF}^2\sim 1$}.

\begin{figure}[t!]  \centering
\includegraphics[width=\textwidth]{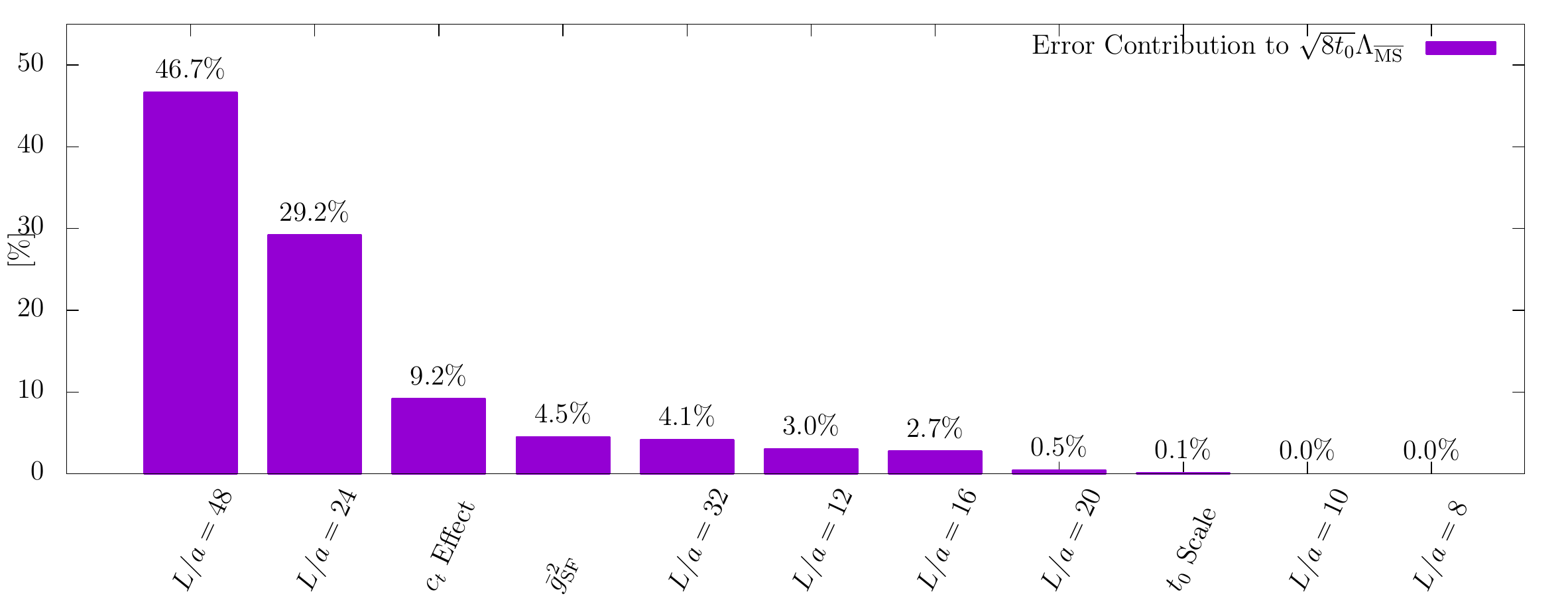}
  \caption{Separate contributions to the square error of
$\sqrt{8t_0}\, \Lambda_{\overline{\rm MS}}$,
eq.~(\ref{eq:final_result_t0}).  Most of the error comes from the
simulations on the largest lattices.  The effect of the systematic
uncertainty associated with $\mathcal O(a)$ terms (labelled $c_t$
effect) is only a small fraction of the error.  Also the simulations
that determined the SF coupling (labelled $\bar g^2_{\rm SF}$), give a
small contribution to the error.}
  \label{fig:error_src}
\end{figure}

Looking at the different contributions to our final uncertainty, it is
clear from eq.~(\ref{eq:res_he_sf}) that most of the uncertainty comes
from the determination of the non-perturbative running from
$\alpha=0.2$ to $\alpha\sim 0.08$.  Given the high-precision we aimed
for, however, we found compulsory to reach, non-perturbatively, these
high-energy scales, and accurately test the applicability of
perturbation theory. Figure~\ref{fig:error_src} then illustrates the
separate contribution to the total error squared from the different
simulations and some other sources. Most of our error comes from the
most expensive simulations, while for instance the contribution from
the uncertainty on $t_0$ is completely negligible. Also the systematic
uncertainty deriving from our ignorance of the boundary $\mathcal
O(a)$ counterterm (labelled as "$c_t$ effect" in the figure)
contributes only little to the final uncertainty. Moreover, in our
strategy the SF coupling is only used at very high energies, where it
is measured very precisely and with negligible cutoff
effects. Consequently, its contribution to the final error is very
small.  In summary, our final uncertainty is completely dominated by
the statistical uncertainty in the measurements of the GF coupling on
our finest/largest lattices. The results could thus be improved
significantly by just investing more computer time on these
simulations.

A completely analogous analysis leads to:
\begin{equation}
  \label{eq:final_result_r0} r_0\, \Lambda_{\overline{\rm MS}} =
0.660(11) \qquad [1.7\%]\,,
\end{equation} where we have used the result for $r_0\mu_{\rm ref}$
based on the 3-point extrapolation of data set [A],
eq.~(\ref{eq:r03pt}).
\begin{figure}[t!]  \centering
\includegraphics[width=\textwidth]{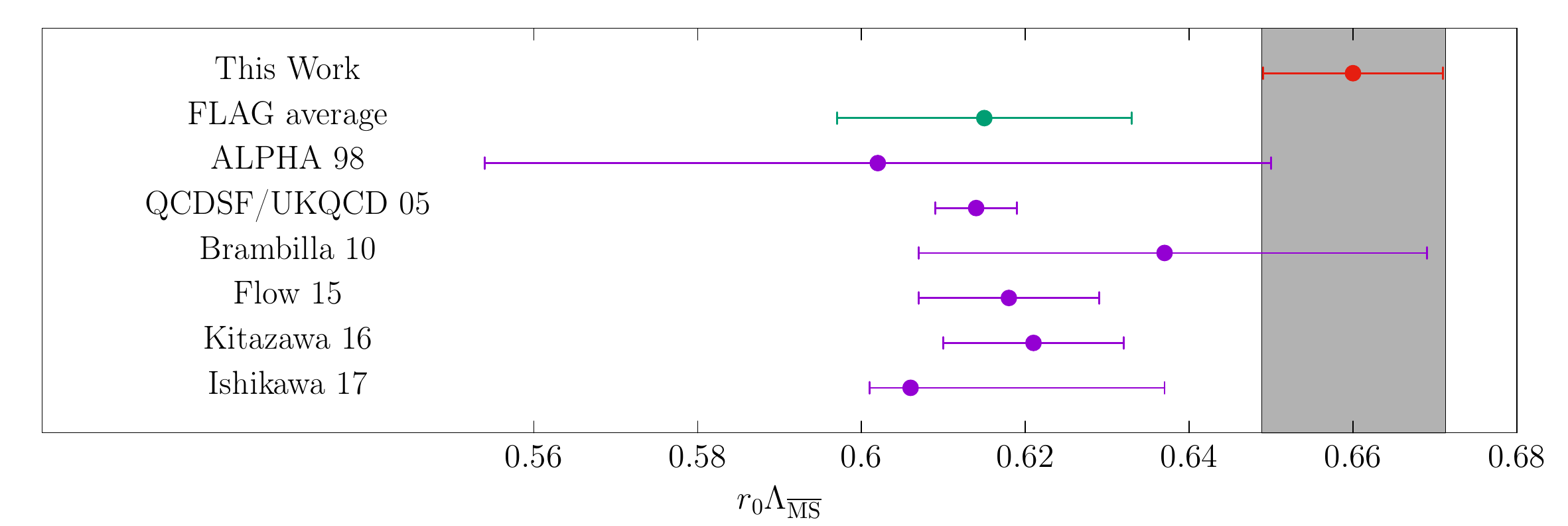}
  \caption{Comparison of our results for $r_0\Lambda_{\overline{\rm
MS}}$ with results from the literature. For comparison we include all
results that pass the FLAG criteria and thus enter their final
average~\cite{Aoki:2019cca}. These are: ALPHA
98~\cite{Capitani:1998mq}, QCDSF/UKQCD 05~\cite{Gockeler:2005rv},
Brambilla~\cite{Brambilla:2010pp}, Flow 15~\cite{Asakawa:2015vta},
Kitazawa 16~\cite{Kitazawa:2016dsl}, Ishikawa
17~\cite{Ishikawa:2017xam}.}
  \label{fig:comp}
\end{figure} Figure~\ref{fig:comp} shows a comparison of our
computations with some other determinations available in the
literature. More precisely, we have included the results entering the
last FLAG average~\cite{Aoki:2019cca}. We find a significant
discrepancy with some of these determinations, in particular with
those of QCDSF/UKQCD 05~\cite{Gockeler:2005rv}, Flow
15~\cite{Asakawa:2015vta} and Kitazawa
16~\cite{Kitazawa:2016dsl}. These determinations extract the
$\Lambda$-parameter from measurements of Wilson loops, and rely on
2-loop bare lattice perturbation theory at a scale of a few GeVs.
Here we recall that our final results
use continuum extrapolations performed with data that cover a factor
two in lattice spacings, with two extra lattice spacings (reaching a
factor 3 change in $a$) used to check the consistency of our
results. The non-perturbative running is performed up to very high
energy scales where $\alpha<0.1$, and our matching with perturbation
theory has been performed in several renormalization schemes. Our
determination satisfies the most stringent of the FLAG
criteria. Nevertheless there is a significant discrepancy with the
FLAG average. Although we think that the FLAG criteria are
conservative, this work shows that even computations that meet these
criteria can differ by more than the quoted uncertainties. This
discrepancies need further investigation, and once clarified the
criteria used to rate different lattice determinations of the strong
coupling might need to be readjusted. Also the authors believe that
given the level of precision reached by current computations, one
should probably consider $\sqrt{8t_0}\Lambda_{\overline{\rm MS} }$
rather than $r_0\Lambda_{\overline{\rm MS} }$ as the standard quantity
of comparison.

%%% Local Variables:
%%% mode: latex
%%% TeX-master: "paper"
%%% End:

\section{Conclusions}
\label{sec:conclusions}

Renormalization schemes based on the gradient flow have many
attractive properties for applications in lattice QCD. Renormalized
couplings are defined via observables with very small variance, which
allows to attain great statistical precision. In addition, the
coupling is given directly by an expectation value. Hence, there are
no systematic errors associated with extracting properties from the
large Euclidean time behaviour of a correlator. By using a
renormalization scheme based on the gradient flow we have obtained a
rather precise determination of the $\Lambda$-parameter in the $SU(3)$
gauge theory. Along the way we have learned several important lessons
that will come useful when applying this technology to the more
relevant case of QCD.

Using finite size scaling techniques we have determined
non-perturbatively the $\beta$-function in a range of couplings
$\alpha\sim 0.1-1$. Our results indicate that at energy scales where
$\alpha\sim 0.1$ contact with 3-loop perturbation theory is not safe
for GF schemes if one aims at a precision $\sim 0.5\%$ in $\alpha_s$.

One might argue that our particular setup (finite volume
renormalization schemes with Schr\"odinger functional boundary
conditions in the pure gauge theory), can cast some doubts on the
general validity of our conclusions. In this respect one should first
note that we have checked two different renormalization schemes (based
on the electric and magnetic components of the energy density at
positive flow time, respectively). The scheme based on the electric
components, in particular, has a very similar perturbative behaviour
to the corresponding infinite volume scheme (i.e.~their non-universal
3-loop coefficients of the $\beta$-function are pretty close; see 
e.g.~figure \ref{fig:beta_ele}).
Secondly, if one considers the parametric uncertainty
originating from the missing perturbative orders in extracting
$\alpha_s$ from the infinite-volume GF coupling in QCD one reaches
similar conclusions to our non-perturbative study: the extraction of
$\alpha_s$ at the electroweak scale in the GF scheme carries a 0.5\%
theoretical uncertainty. If the extraction is performed at a few GeV
(the energy scale typically accessible to large volume simulations),
the theoretical uncertainty increases to $\sim 2-3\%$. 
Quark effects are absent in our study, but
perturbatively their effect at high energy is small compared to the
effect of the gluons~\cite{Harlander:2016vzb}. The presence of quarks
is thus expected not to change the picture very much. We conclude that
the qualitative conclusions of our study are indeed general.

Our work allows us to precisely determine the pure gauge
$\Lambda$-parameter in units of a typical hadronic scale (we
considered both $t_0$ and $r_0$). Our result for
$r_0\Lambda_{\overline{\rm MS}}$, with a precision $\sim 1.7\%$, shows
some tension with other recent lattice computations, in particular
with those where the $\overline{\rm MS}$-coupling is extracted from
Wilson loops at an energy scale set by the lattice cutoff.  One
drawback of the GF couplings are the relatively large cutoff effects,
which have been observed in many different applications
(see~\cite{Ramos:2015dla} for an overview). Despite the fact that we
have a solid theoretical understanding of the nature of these cutoff
effects~\cite{Ramos:2015baa}, they are still the main source of
concern when considering GF-based observables. In order to have
discretization effects under control we used 5 different lattice
resolutions which cover a factor 3 in the spacing, and two different
discretizations to integrate the flow equations and compute our
observables. We see no deviation from pure $\mathcal O(a^2)$ scaling violations. 
Despite these observations, we quote results where the two coarsest lattice
spacings are discarded, and we add a generous estimate for the
$\mathcal{O}(a)$ boundary effects.  We recall that we have performed
the running non-perturbatively up to very large energy scales, we have
matched with the perturbative behaviour in four different
renormalization schemes (with their respective $\Lambda$-parameters
varying by factors of two), and used at least two different methods to
match with a large volume hadronic scale. All in all, the significant
discrepancy with 
the results in the literature shows the difficulty in extracting the
fundamental parameters of the Standard Model with high
precision.

We conclude by pointing out that the results of this work represent a
serious warning for any attempt of reducing the current uncertainty of
the world average value of $\alpha_s$ using lattice QCD and the GF
schemes, especially if one aims at an infinite volume
determination. Here the range of scales that can be explored is
limited to $\alpha \gtrsim 0.25$, completely insufficient to quote
sub-percent precision in $\alpha_s$. On the positive side we have
shown a viable strategy to reach a precision of $0.3\%$ in
$\alpha_s$. It combines the use of the GF schemes to determine the
running non-perturbatively, and a non-perturbative matching at
high-energy with the traditional SF schemes, that show small effects
in the truncation of the perturbative series. Such a project would
also require a precise and accurate determination of the hadronic
scale in a theory with three or more active
quarks~\cite{Athenodorou:2018wpk}.

%%% Local Variables:
%%% mode: latex
%%% TeX-master: "paper"
%%% End:

\section*{Acknowledgments}
\addcontentsline{toc}{section}{Acknowledgments}
We thank Rainer Sommer and Stefan Sint for many discussions and
constant encouragement at all stages of this work, including a
critical reading of the manuscript. We warmly thank
A. Rubeo and S. Sint for sharing their codes to determine the
tree-level coupling norm in lattice perturbation theory with our
numerical setup. AR has profited from discussions with C. Pena and
M. Garc\'ia-Perez.

This work has been possible thanks to the computer time provided by
these HPC systems:
The HPC cluster at CERN, PAX at DESY-Zeuthen,
Altamira, provided by IFCA at the University
of Cantabria, and the FinisTerrae II machine provided by CESGA
(Galicia Supercomputing Centre) and funded by the Xunta de Galicia and
the Spanish MINECO under the 2007-2013 Spanish ERDF Programme.

\cleardoublepage

\appendix

\section{Continuum limit of GF couplings}
\label{sec:UbyU}

Several studies have shown that renormalized couplings defined 
through the gradient flow are affected by significant cutoff 
effects (see e.g. ref.~\cite{Ramos:2015dla} and references therein). 
These cutoff effects have been first carefully studied at tree-level 
of perturbation theory~\cite{Fodor:2014cpa}, and later more systematically 
in the context of Symanzik' effective theory~\cite{Ramos:2015baa}. 
Despite these efforts, however, they remain the main source of concern 
in applications of running couplings derived from the gradient flow. 

\begin{figure}
	\centering
	\begin{subfigure}{.45\textwidth}
		\centering
		\includegraphics[scale=0.6]{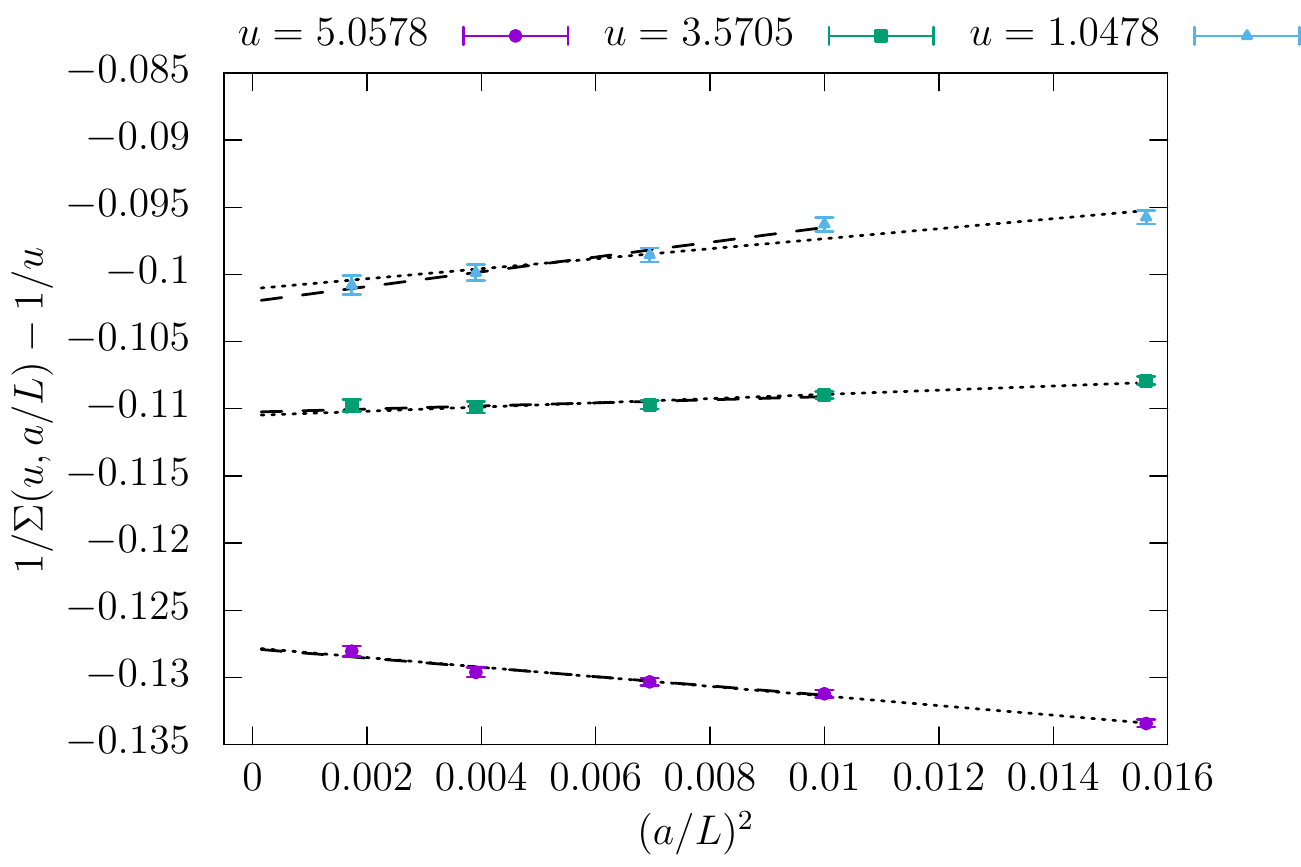}
		\caption{Continuum extrapolations of the lattice SSF 
			     of the magnetic GF coupling at 3 target values
			     of the coupling, eq.~(\ref{eq:utg}). (Data from table~\ref{tab:ssf}.) 
			     We consider the lattice resolutions: $L/a=8,10,12,16,24$,
			     and perform linear fits of the data in $(a/L)^2$, 
			     with and without the coarsest lattice, $L/a=8$.}
		\label{fig:ssf_1}
	\end{subfigure}
	\begin{subfigure}{.45\textwidth}
		\centering
		\vspace*{-7mm}
		\includegraphics[scale=0.575]{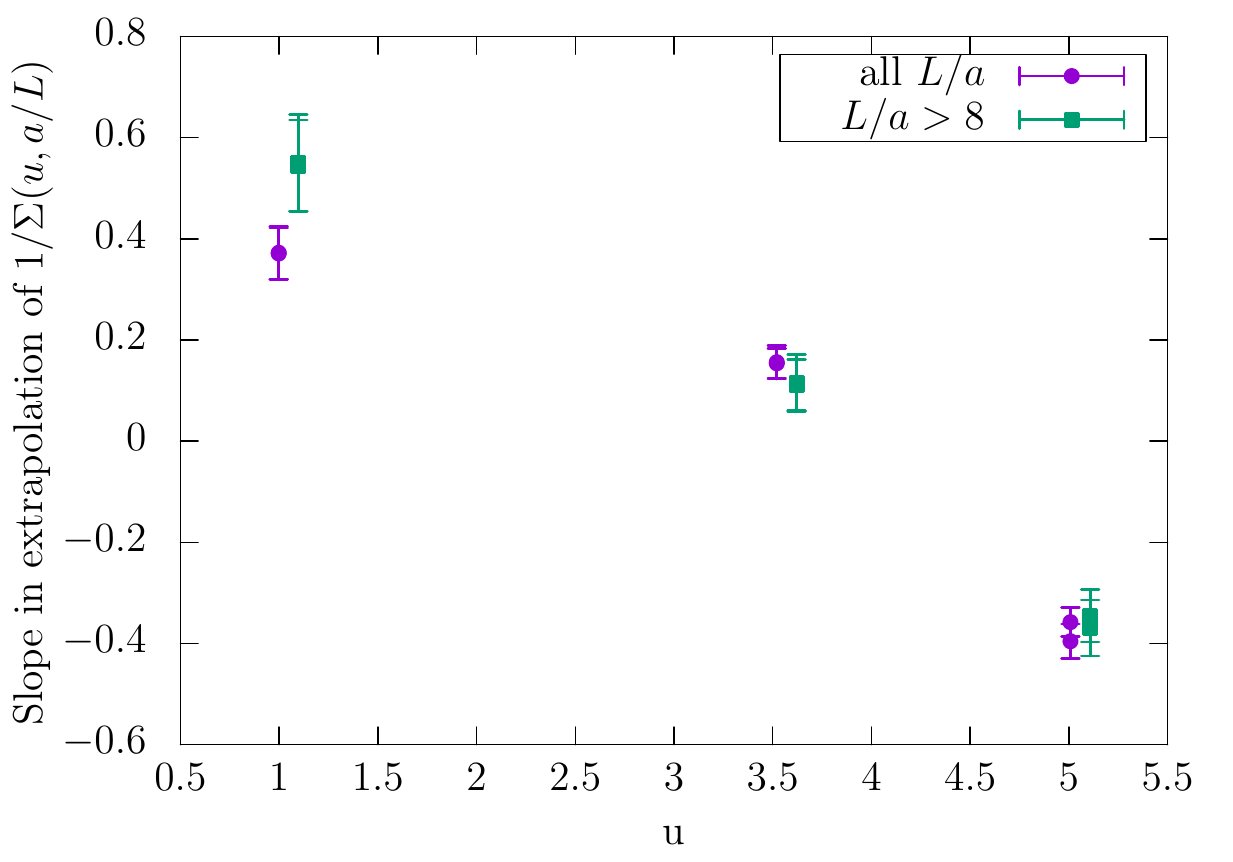}
		\caption{Slopes $\tilde{\rho}$ of the continuum extrapolations 
			     of $1/\Sigma_2$ (cf.~eq.~(\ref{eq:UbyuFits})). (Data from 
			     table~\ref{tab:ssf}.) As we can see, 
			     the slope is positive at weak couplings, while turns 
			     negative at strong couplings.}
		\label{fig:ssf_2}
	\end{subfigure}
\end{figure}

In order to establish that the lattice resolutions employed in our study are 
fine enough to obtain accurate continuum extrapolations, we here analyse 
in detail the continuum extrapolation of the lattice step scaling function,
$\Sigma_{2}(u,a/L)$, for the magnetic GF scheme, at 3 representative 
values of the coupling; these are:
\begin{equation}
\label{eq:utg}
  u_1 = 1.04784, \qquad u_2 = 3.5705,\qquad u_3 = 5.0578\,.
\end{equation}
Note that we shall focus on our preferred choice of discretization for the observable, 
i.e., Zeuthen flow/improved definition (cf.~Sect.~\ref{sec:LatticeSetup}). To perform 
the continuum limit of the lattice SSF at these coupling values, we first need to find 
the values of the bare coupling $\beta$ for the chosen lattice sizes, for which the 
renormalized coupling, $u=\bar g^2_{\rm GF,\,m}(\mu)$,  is equal to the target values 
(\ref{eq:utg}). We did this for lattice sizes: $L/a=8,10,12,16,24$. The results 
of this tuning are given in the third and fourth columns of table~\ref{tab:ssf}.
Once the values of the bare coupling are determined, at these $\beta$-values
we compute  the GF coupling on lattices of double size, i.e. with $L/a=16,20,24,32,48$,  
respectively. The corresponding results are given in the last column of table~\ref{tab:ssf}. 
Finally, the lattice SSFs so obtained can be extrapolated to the continuum limit
(see figure~\ref{fig:ssf_1}).

\begin{table}[h]
  \centering
  \begin{tabular}{lllll}
    \toprule
    $u$&$L/a$&$\beta$& $\bar g^2_{\rm GF\,m}(\mu)$ & $\bar g^2_{\rm GF\,m}(\mu/2)$ \\
    \midrule
    \multirow{9}{*}{$u_1$} 
       &  8 &10.1106 & 1.04784(18) & \phantom{1}1.16469(62) \\
       & 10 &10.2830 & 1.04784(29) & \phantom{1}1.16540(63) \\
       & 12 &10.4258 & 1.04784(26) & \phantom{1}1.16851(64) \\
       & 16 &10.6586 & 1.04784(24) & \phantom{1}1.17030(76) \\
       & 24 &11.0000 & 1.04784(54) & \phantom{1}1.17157(68) \\
       &&&&\\
       & $\infty$ (Fit $1/\Sigma_2$; All $L/a$) &    & 1.04784     & \phantom{1}1.17195(70) \\
       & $\infty$ (Fit $\phantom{1/}\Sigma_2$; All $L/a$) &    & 1.04784     & \phantom{1}1.17195(69) \\
       & $\infty$ (Fit $1/\Sigma_2$; $L/a>8$) &    & 1.04784     & \phantom{1}1.17325(91) \\
       & $\infty$ (Fit $\phantom{1/}\Sigma_2$; $L/a>8$) &    & 1.04784     & \phantom{1}1.17322(86) \\
    \midrule
    \multirow{9}{*}{$u_2$} 
       &  8 &\phantom{1}6.3598 & 3.5705(24) & \phantom{1}5.808(8) \\
       & 10 &\phantom{1}6.5197 & 3.5705(13) & \phantom{1}5.844(8) \\
       & 12 &\phantom{1}6.6559 & 3.5705(18) & \phantom{1}5.869(9) \\
       & 16 &\phantom{1}6.8786 & 3.5705(17) & \phantom{1}5.876(13) \\
       & 24 &\phantom{1}7.2000 & 3.5705(31) & \phantom{1}5.871(13) \\
       &&&&\\
       & $\infty$ (Fit $1/\Sigma_2$; All $L/a$) &    & 3.5705     & \phantom{1}5.897(12) \\
       & $\infty$ (Fit $\phantom{1/}\Sigma_2$; All $L/a$) &    & 3.5705         & \phantom{1}5.896(11) \\
       & $\infty$ (Fit $1/\Sigma_2$; $L/a>8$) &    & 3.5705     & \phantom{1}5.888(15) \\
       & $\infty$ (Fit $\phantom{1/}\Sigma_2$; $L/a>8$) &    & 3.5705     & \phantom{1}5.888(13) \\
    \midrule
    \multirow{9}{*}{$u_3$} 
       & 8  &\phantom{1}5.9900 & 5.0578(21) & 15.555(61) \\
       & 10 &\phantom{1}6.1365 & 5.0578(24) & 15.037(59) \\
       & 12 &\phantom{1}6.2654 & 5.0578(26) & 14.839(59) \\
       & 16 &\phantom{1}6.4742 & 5.0578(26) & 14.686(70) \\
       & 24 &\phantom{1}6.7859 & 5.0578(61) & 14.353(77) \\
       &&&&\\
       & $\infty$ (Fit $1/\Sigma_2$; All $L/a$) &    & 5.0578     & \phantom{1}14.303(58) \\
       & $\infty$ (Fit $\phantom{1/}\Sigma_2$; All $L/a$) &    & 5.0578     & \phantom{1}14.278(60) \\
       & $\infty$ (Fit $1/\Sigma_2$; $L/a>8$) &    & 5.0578     & \phantom{1}14.317(75) \\
       & $\infty$ (Fit $\phantom{1/}\Sigma_2$; $L/a>8$) &    & 5.0578     & \phantom{1}14.304(74) \\
    \bottomrule
  \end{tabular}
  \caption{Data for the lattice step scaling function, $\Sigma_2(u,a/L)$, of the magnetic 
		   GF coupling, at the target couplings of eq.~(\ref{eq:utg}). The results of different
		   continuum extrapolations are given.}
  \label{tab:ssf}
\end{table}

%%% Local Variables:
%%% mode: latex
%%% TeX-master: "../paper"
%%% End:

We consider for $\Sigma_2$  continuum extrapolations linear in $(a/L)^2$.%
\footnote{Note that we neglect any systematic uncertainty associated 
	      with possible $\mathcal O(a)$ cutoff effects contaminating our 
	      data (cf.~Section \ref{sec:oa}). The only uncertainties entering the
	      fits are hence the statistical ones.} 
Moreover, in order to gain some insight on higher-order discretization errors 
we consider linear extrapolations in  $(a/L)^2$ of $1/\Sigma_2$. We 
expect these two strategies to give  compatible results up to 
$\mathcal{O}((a/L)^4)$ errors. We thus have:
\begin{equation}
  \label{eq:UbyuFits}
  \Sigma_2(u,a/L) = \sigma_2(u) + \rho \left( \frac{a}{L} \right)^2\,,  
  \qquad
  \frac{1}{\Sigma_2(u,a/L)} = \frac{1}{\sigma_2(u)} + \tilde \rho \left( \frac{a}{L} \right)^2 \,.
\end{equation}
As a further test on the scaling properties of our data, we also
study the effect of discarding our coarser lattice with $L/a=8$ from the
extrapolations. 

Having this noticed, all our fits have good quality, and the agreement among 
different determinations of the continuum SSF, $\sigma_2(u)$, is 
in fact quite good (compare the rows marked by $L/a=\infty$ in table \ref{tab:ssf}).%
\footnote{The data at the smallest coupling $u_1$ shows a $1\sigma$
          deviation between the fits where the coarsest lattice, $L/a=8$,
          is included or not. This difference is however not statistically
          significant.}
We thus conclude that for our choice of discretization, our data show no 
significant deviation from $\mathcal O(a^2)$ scaling. In addition,
figure \ref{fig:ssf_1} shows that the slope of the continuum extrapolations
is positive at weak couplings, while changes to negative at strong couplings. 
Somewhere around $u\sim 4$, the data have no significant cutoff 
effects.

In summary, the detailed study presented in this appendix shows that 
once lattice sizes in the range $L/a=8-24$ are considered, within the whole
range of couplings $u\in[1,5]$ the continuum extrapolations of the 
lattice step scaling function of the GF coupling present no significant 
deviation from $\mathcal O(a^2)$ scaling within our precision.

%%% Local Variables:
%%% mode: latex
%%% TeX-master: "paper"
%%% End:

\section{Boundary $\mathcal{O}(a)$ effects}
\label{sec:oa}

It is well-known that due to the breaking of translational invariance
in the time direction, even the pure Yang-Mills theory with Schr\"odinger 
functional boundary conditions suffers from $\mathcal O(a)$ discretizations
effects~\cite{Luscher:1992an}. These can in principle be entirely 
removed  by a proper tuning of a single boundary counterterm coefficient,
$c_t(g_0)$ (cf.~eq.~(\ref{eq:Waction})). Unfortunately, however, in practice
there is no compelling method to determine $c_t$ non-perturbatively. As a 
result, at present, given our choice of Wilson gauge action, $c_t$ is only 
known in perturbation theory to two-loop order~\cite{Bode:1998hd,Bode:1999sm}. 
The leading discretization errors in our data are thus parametrically of 
$\mathcal O(g_0^6a/L)$. On the other hand, the results of the investigation 
in Appendix \ref{sec:UbyU}, where we ignored any $\mathcal O(a)$ effects in 
the data, support the conclusion that these effects are in practice very small, 
and below our statistical precision. Nonetheless, in order to guarantee the 
high-precision of our results, we here want to address this potential source of 
systematic effects in detail. To this end, we measured the GF couplings for 
several values of $c_t$ which are shifted from the 2-loop result,  $c_t^\star$,
used in the simulations:
\begin{equation}
  \label{eq:TwoLoopCt}
  c_t^\star(g_0) = 1 - 0.08900 \times g_0^2 - 0.0294 \times g_0^4
  \qquad 
  (g^2_0=6/\beta)\,. 
\end{equation}
We did this for $L/a=8,10,12$, and at the $\beta$-values corresponding to 
the 3 couplings of eq.~\eqref{eq:utg}. The results we obtained are 
collected in   table \ref{tab:rawoa}. The deviation, $\Delta \bar{g}^2_{\rm GF,\,e/m}$,
of these results from the couplings of eq.~\eqref{eq:utg}, is a clear measure of 
the sensitivity of the coupling on the boundary improvement coefficient $c_t$. 
The data of table  \ref{tab:rawoa} is well represented by a fit:
\begin{equation}
  \bar g^2_{{\rm GF},\,c_t^\star + \Delta c_t} = \bar g^2_{{\rm GF},\,c_t^\star}
  + \frac{a}{L}\left(a_0 \bar g^2_{{\rm GF},\,c_t^\star} + a_1\bar g^4_{{\rm GF},\,c_t^\star}\right) 
  \Delta c_t\,,
\end{equation}
where $\bar g^2_{{\rm GF},\,c_t}$ is the GF coupling (either electric or magnetic) measured
for a given $c_t$, and $\Delta c_t = c_t - c_t^\star$. For the fit coefficients $a_0, a_1$ 
we obtain the results:
\begin{subequations}
\label{eq:ctcoef}
\begin{eqnarray}
   \bar g^2_{\rm GF,\,m}:\qquad &a_0 = -0.14(5)\,,& \quad a_1 =
  -0.26(3)\,. \\
  \bar g^2_{\rm GF,\,e}:\qquad &a_0 = -0.48(5)\,,& \quad a_1 =
  -0.25(3)\,. 
\end{eqnarray}
\end{subequations}
As expected from general considerations, the electric components are the 
most affected by boundary $\mathcal{O}(a)$ effects~\cite{Ramos:2015dla}. 
The values of these fit coefficients, on the other hand, are basically the 
same for our two choices of discretization of the observable, i.e., 
Wilson flow/clover and Zeuthen flow/improved observable.

Having established the sensitivity of the GF couplings around the two-loop value
of $c_t$, eq.~(\ref{eq:TwoLoopCt}), in order to estimate the uncertainty to attribute 
to our data for the incomplete tuning of $c_t$, we now need an estimate for the 
difference between $c_t^\star$ and the non-perturbative value of $c_t$. Having no
information about the latter, a reasonable guess is to take for this deviation the 
full two-loop term of the series eq.~(\ref{eq:TwoLoopCt}). Given the fact that the 
coefficients of the series (\ref{eq:TwoLoopCt}) appear to decrease with the order, 
our estimate can be considered a conservative one. 

In conclusions, we add in quadrature to the statistical uncertainty of the 
GF couplings computed at $c_t=c_t^\star$ and for a given $g_0$ and $L/a$,
the systematic uncertainty: 
\begin{equation}
\label{eq:dltct}
  \delta_{c_t} \bar g^2_{\rm GF} = \frac{a}{L}\left(a_0 \bar g^2_{{\rm GF},\,{c_t^\star}} +
    a_1\bar g_{{\rm GF},\,{c_t^\star}}^4\right) \times 0.0294\,g_0^4 \,,
\end{equation}
with $a_0,a_1$ given by eqs.~\eqref{eq:ctcoef}. We stress once again that this 
is done in order to take into account possible $\mathcal O(a)$ effects in our 
data that might arise from the mistuning of the boundary conterterm coefficient
$c_t$. With the exception of our coarsest lattices, this effect turns out to be 
sub-dominant to the statistical errors.

\begin{table}
  \centering
\begin{tabular}{rrrrrr}
  \toprule
\(L/a\) & \(\Delta c_t\) & \(\bar g^2_{\rm GF,\,m}\) & \(\Delta\bar g^2_{\rm GF,\,m}\) &
\(\bar g^2_{\rm GF,\,e}\) & \(\Delta\bar g^2_{\rm GF,\,e}\)\\
  \midrule
8 & -0.02949825 & 5.0608(51) & \phantom{-}0.0249(73) & 4.9533(50) & \phantom{-}0.0307(72)\\
8 & \phantom{-}0.02949825 & 5.0608(51) & -0.0327(71) & 4.9533(50) & -0.0438(70)\\
8 & -0.02616754 & 3.5659(31) & \phantom{-}0.0191(44) & 3.4962(31) & \phantom{-}0.0168(43)\\
8 & \phantom{-}0.02616754 & 3.5659(31) & -0.0064(44) & 3.4962(31) & -0.0115(44)\\
8 & -0.04141484 & 1.04846(48) & \phantom{-}0.00265(68) & 1.03689(48) & \phantom{-}0.00395(68)\\
8 & -0.03106113 & 1.04846(48) & \phantom{-}0.00141(68) & 1.03689(48) & \phantom{-}0.00271(68)\\
8 & -0.02070742 & 1.04846(48) & \phantom{-}0.00010(68) & 1.03689(48) & \phantom{-}0.00212(69)\\
8 & -0.01035371 & 1.04846(48) & \phantom{-}0.00064(68) & 1.03689(48) & \phantom{-}0.00065(68)\\
8 & \phantom{-}0.01035371 & 1.04846(48) & -0.00053(68) & 1.03689(48) & -0.00100(68)\\
8 & \phantom{-}0.02070742 & 1.04846(48) & -0.00061(68) & 1.03689(48) & -0.00215(68)\\
8 & \phantom{-}0.03106113 & 1.04846(48) & -0.00174(67) & 1.03689(48) & -0.00232(68)\\
8 & \phantom{-}0.04141484 & 1.04846(48) & -0.00225(67) & 1.03689(48) & -0.00446(67)\\
\midrule
10 & -0.02810660 & 5.0648(49) & \phantom{-}0.0179(69) & 4.9533(51) & \phantom{-}0.02200(72)\\
10 & \phantom{-}0.02810660 & 5.0648(49) & -0.0200(69) & 4.9533(51) & -0.02535(72)\\
10 & -0.04979946 & 3.5699(31) & \phantom{-}0.0224(44) & 3.4933(32) & \phantom{-}0.02485(45)\\
10 & -0.02489973 & 3.5699(31) & \phantom{-}0.0112(44) & 3.4933(32) & \phantom{-}0.01209(45)\\
10 & \phantom{-}0.02489973 & 3.5699(31) & -0.0044(45) & 3.4933(32) & -0.00749(45)\\
10 & \phantom{-}0.04979946 & 3.5699(31) & -0.0171(44) & 3.4933(32) & -0.02294(45)\\
10 & -0.04003779 & 1.04737(55) & \phantom{-}0.00140(78) & 1.03449(57) & \phantom{-}0.00380(81)\\
10 & -0.03002835 & 1.04737(55) & \phantom{-}0.00105(78) & 1.03449(57) & \phantom{-}0.00248(80)\\
10 & -0.02001890 & 1.04737(55) & \phantom{-}0.00175(79) & 1.03449(57) & \phantom{-}0.00352(81)\\
10 & -0.01000945 & 1.04737(55) & \phantom{-}0.00071(78) & 1.03449(57) & \phantom{-}0.00224(81)\\
10 & \phantom{-}0.01000945 & 1.04737(55) & -0.00038(77) & 1.03449(57) & -0.00033(81)\\
10 & \phantom{-}0.02001890 & 1.04737(55) & -0.00119(78) & 1.03449(57) & -0.00110(80)\\
10 & \phantom{-}0.03002835 & 1.04737(55) & -0.00182(78) & 1.03449(57) & -0.00235(80)\\
10 & \phantom{-}0.04003779 & 1.04737(55) & -0.00160(77) & 1.03449(57) & -0.00215(80)\\
\midrule
12 & -0.02696201 & 5.0533(47) & \phantom{-}0.0127(68) & 4.9508(48) & \phantom{-}0.0211(71)\\
12 & \phantom{-}0.02696201 & 5.0533(47) & -0.0227(68) & 4.9508(48) & -0.0237(70)\\
12 & -0.04778221 & 3.5670(31) & \phantom{-}0.0221(44) & 3.4910(32) & \phantom{-}0.0196(45)\\
12 & -0.02389111 & 3.5670(31) & \phantom{-}0.0105(43) & 3.4910(32) & \phantom{-}0.0124(45)\\
12 & \phantom{-}0.02389111 & 3.5670(31) & -0.0070(43) & 3.4910(32) & -0.0048(44)\\
12 & \phantom{-}0.04778221 & 3.5670(31) & -0.0054(44) & 3.4910(32) & -0.0103(45)\\
12 & -0.04868566 & 1.04824(54) & \phantom{-}0.00119(78) & 1.03495(59) & \phantom{-}0.00332(83)\\
12 & -0.03894853 & 1.04824(54) & \phantom{-}0.00095(78) & 1.03495(59) & \phantom{-}0.00207(83)\\
12 & -0.02921140 & 1.04824(54) & \phantom{-}0.00085(78) & 1.03495(59) & \phantom{-}0.00237(84)\\
12 & -0.01947426 & 1.04824(54) & -0.00022(77) & 1.03495(59) & \phantom{-}0.00177(83)\\
12 & -0.00973713 & 1.04824(54) & -0.00050(78) & 1.03495(59) & \phantom{-}0.00011(83)\\
12 & \phantom{-}0.00973713 & 1.04824(54) & -0.00068(78) & 1.03495(59) & -0.00016(82)\\
12 & \phantom{-}0.01947426 & 1.04824(54) & -0.00068(78) & 1.03495(59) & -0.00025(83)\\
12 & \phantom{-}0.02921140 & 1.04824(54) & -0.00187(77) & 1.03495(59) & -0.00209(83)\\
12 & \phantom{-}0.03894853 & 1.04824(54) & -0.00202(78) & 1.03495(59) & -0.00341(83)\\
12 & \phantom{-}0.04868566 & 1.04824(54) & -0.00285(78) & 1.03495(59) & -0.00250(84)\\
\bottomrule
\end{tabular}
  \caption{Data to study the $c_t$ dependence of the GF couplings. The columns labelled
		   by $\Delta \bar{g}^2_{\rm GF,\,e/m}$ give the deviation of the GF couplings
		   measured at $c_t=c_t^\star+\Delta c_t$, from the values of eq.~(\ref{eq:utg})
		   measured at $c_t=c_t^\star$. }
  \label{tab:rawoa}
\end{table}

%%% Local Variables:
%%% mode: latex
%%% TeX-master: "paper"
%%% End:

\section{Step-scaling determination of $\Lambda_{\overline{\rm MS}}/\mu_{\rm ref}$}
\label{sec:ss}

As an alternative determination of $\Lambda_{\overline{\rm MS}}/\mu_{\rm ref}$
we also consider a more traditional approach based on the computation
of the continuum step-scaling function of the GF coupling, combined with 
a non-perturbative matching of the GF and SF couplings. For the determination 
of the continuum SSF we employ the very same data for 
$\Sigma_2(u,a/L)$ entering the computation of the corresponding $\beta$-function 
(cf.~Sect.~\ref{sec:beta-function-he}).% 
\footnote{Note that combining the $s=3/2,2$ results is in this case less
	      trivial than for the computation based on the $\beta$-function.
	      Moreover, from the latter we expect that including the $s=3/2$
	      data will improve only little the precision of the final result. 
	      For these reasons, and in order to keep the presentation simple, we 
	      thus focus here on the $s=2$ data only.}
More precisely, we consider data for the Zeuthen flow/improved observable
discretization, we study both the electric and magnetic definition, and restrict
our attention to couplings: $u\leq \bar{g}^2_{\rm GF,\,ref}$, where 
$\bar{g}^2_{\rm GF,\,ref}\equiv\bar{g}^2_{\rm GF,\,ref,e/m}$, 
depending on the chosen scheme.%
\footnote{As usual, for ease of notation we will use in general a unique symbol for 
		  the couplings and SSFs of the electric and magnetic components.}
 The lattice SSF, 
$\Sigma_2(u,a/L)$, is then fitted according to the functional form:
\begin{equation}
	\Sigma_2(u,a/L) = \sigma_2(u)+ \rho^{(2)}(u)\Big({a\over L}\Big)^2,
\end{equation}
where $\sigma_2$ is a parametrization of the continuum SSF:
{\openup 1.5\jot 
\begin{gather}
	\sigma_2(u) = u+ s_0 u^2 + s_1 u^3 + s_2 u^4 + \sum_{k=5}^{n_\sigma} p_k u^k, \\
	s_0=2b_0\ln 2,
	\quad
	s_1=s_0^2+2b_1\ln 2,
	\quad
	s_2=s_0^3+10b_0b_1(\ln 2)^2+2b_2\ln 2,
\end{gather}}%
with coefficients $s_0,s_1,s_2$ fixed to their perturbative values 
(cf.~eqs.~(\ref{eq:beta_univ}),(\ref{eq:b2GF})), while the function $\rho^{(2)}$ models 
the coupling dependence of the leading discretization errors in the data:
\begin{equation}
	\rho^{(2)}(u) = \sum_{i=2}^{n_c} \rho_i^{(2)} u^{i}.
\end{equation}
(We recall that we assume the leading discretization errors to be $\mathcal{O}(a^2)$ 
as the $\mathcal{O}(a)$ boundary effects are taken into account as systematic uncertainties
(cf.~Sect.~\ref{sec:oa}).)
The data with $L/a>8$ is well described by the above functional form for any 
combination of $n_\sigma=6,7$ and $n_c=2,3$, giving a $\chi^2/{\rm dof}\sim 0.5-1$,
depending on the exact fit. In particular fits to the electric GF coupling data always
have smaller $\chi^2$ than those involving the magnetic ones. Similarly to what we did 
for the $\beta$-function, in order to be conservative, we take as our preferred 
fits, fits to the data with $L/a>10$. We then choose $n_\sigma=6$ and $n_c=3$. These fits
yet have excellent $\chi^2$'s, but have larger errors. 

The second step is the determination of the non-perturbative matching between the 
GF and SF couplings. More precisely, we here consider three different SF
coupling definitions, corresponding to $\nu=-0.3,0,0.3$. Similarly to what 
discussed in Sect.~\ref{subsec:GF2SF} for the $\nu=0$ case, fits of the form (\ref{eq:gf2sf}) give a
very good description of the data; moreover also in this case the results
for $\Lambda_{\overline{\rm MS}}/\mu_{\rm ref}$ depend very little on the exact
choice we make. We thus settle on fits where $f_0$ and $f_1$ are fixed to their
perturbative values,  $n_f=3$, and $n_{\tilde{\rho}}=2$. 
Note that, in being once again conservative, we neglect the $L/a=6$ SF and $L/a=12$ GF
coupling results in determining the matching, even though these are well described
by the fit function too. We then opt for not considering the PT improvement of the SF 
data as this has anyway no significant effect once $L/a\geq 8$. 

Having all the basic ingredients, we can now proceed with the determination of 
$\Lambda_{\overline{\rm MS}}/\mu_{\rm ref}$. Starting from the value of the GF
coupling, $u_0=\bar{g}^2_{\rm GF,\,ref}$, the knowledge of the continuum SSF allows
us to infer the value of the coupling at the energies scales 
$\mu_n=2^n\mu_{\rm ref}$, $n=1,2,\ldots$, by solving the recursion relation
($\bar{g}^2_{\rm GF}\equiv \bar{g}^2_{\rm GF,\,e/m}$, $\sigma_2\equiv \sigma_{2}^{\rm GF,\,e/m}$):
\begin{equation}
	\label{eq:StepScaling}
	\bar{g}^2_{\rm GF,\,ref} = u_0 ,
	\qquad
	u_k = \sigma_2(u_{k+1}) = \bar{g}^2_{\rm GF}(2^{k}\mu_{\rm ref}),
	\qquad
	k = 0, 1, \ldots,n .
\end{equation}
Considering both the range of GF couplings we covered and the range where the 
non-perturbative matching with the SF couplings is available, our data allow us 
to perform $n=6$ steps, i.e., starting from $\mu_{\rm ref}$, we are able to increase
the energy scale by a factor 64. For a given value of $u_n=\bar{g}^2_{\rm GF,\,e/m}(\mu_n)$ 
determined this way, we can now compute $\Lambda_{\overline{\rm MS}}/\mu_{\rm ref}$ through:
\begin{equation}
  \frac{\Lambda_{\overline{\rm MS}}}{\mu_{\rm ref}} = 
  s_n\frac{\Lambda_{\overline{\rm MS}}}{\Lambda_{\rm X}}
  ( b_0 \bar g_{{\rm X},n}^2)^{-\frac{b_1}{2b_0^2}}\,{\rm e}^{-\frac{1}{2b_0 \bar{g}_{{\rm X},n}^2}}  
  \times
  \exp\{-I^{{\rm X},\,3}_g(\bar{g}_{{\rm X},n},0)\}\,,
\end{equation}
where either ${\rm X=GF}$, $\bar{g}_{{\rm X},n}=\bar{g}_{\rm GF,\,e/m}(\mu_n)$ and $s_n=2^n$,
or ${\rm X=SF}$, $\bar{g}_{{\rm X},n}=\bar{g}_{\rm SF}(2c\mu_n)$ and $s_n=2^{n+1}c$. 
The value of $\bar{g}_{\rm SF}(2c\mu_n)$ is of course inferred from that of 
$\bar{g}_{\rm GF,\,e/m}(\mu_n)$, using the non-perturbative matching relation between 
the SF and GF schemes previously established. We then expect that the results for different values of $n$ 
and/or schemes should all agree, up to O($\bar{g}^4_{{\rm X},n}$) corrections, 
as $\bar{g}_{{\rm X},n}\to0$.

\begin{figure}[hptb]
	\centering
	\includegraphics[scale=1]{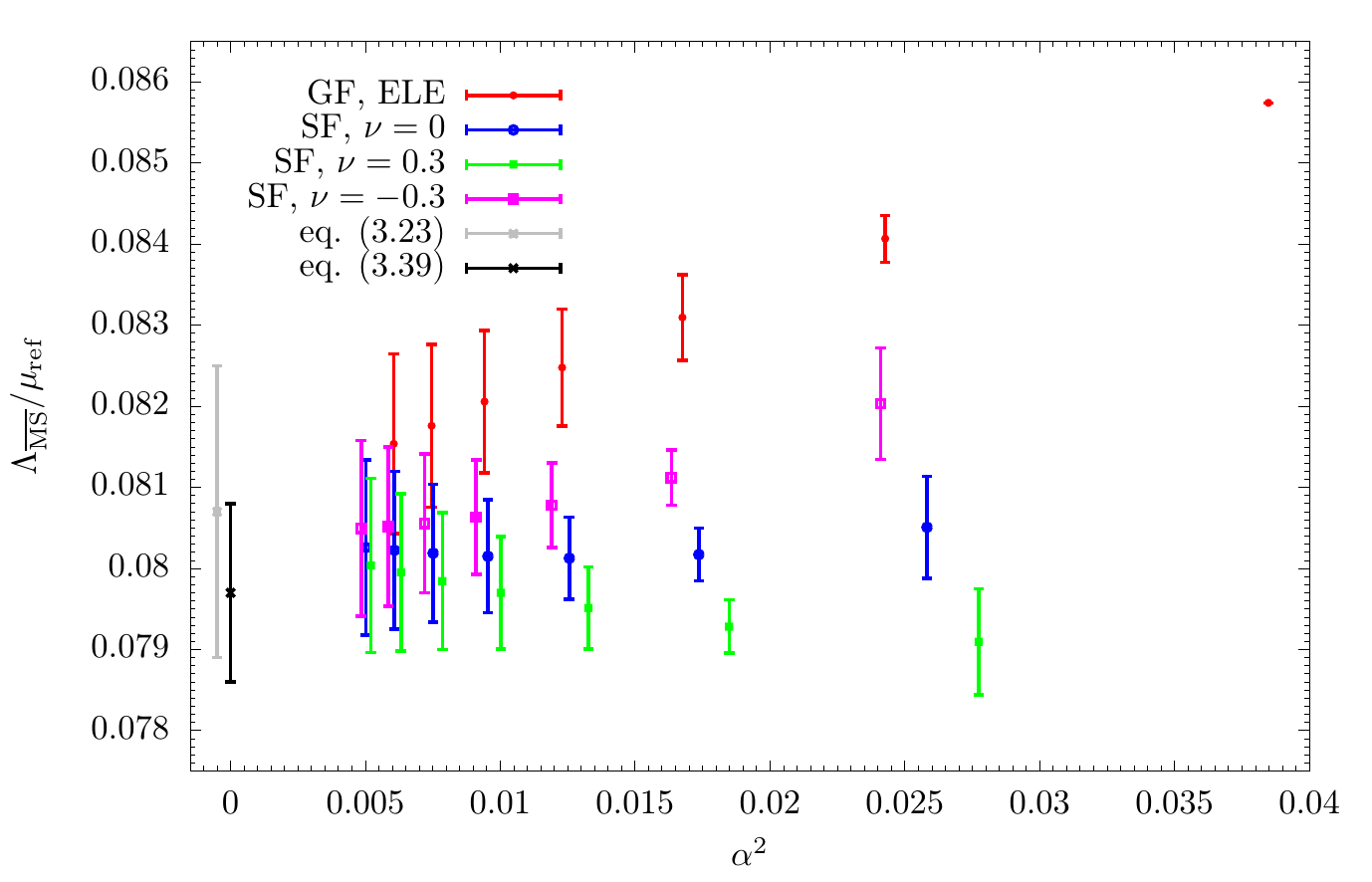}
	\caption{Results for the extraction of $\Lambda_{\overline{\rm MS}}/\mu_{\rm ref}$ 
	based on several schemes, and for different values of the corresponding couplings. 
	The results from the GF coupling refer to its electric scheme and are 
	computed following a step-scaling analysis. At each value of the 
	GF coupling coming from step-scaling, the GF scheme is matched non-perturbatively 
	to three different SF schemes corresponding to $\nu=-0.3,0,0.3$, which are then 
	used to also extract  $\Lambda_{\overline{\rm MS}}/\mu_{\rm ref}$.
	A comparison with our final estimates eqs.~(\ref{eq:res_he}) and 
	(\ref{eq:res_he_sf}) is given.}
	\label{fig:Lmb-GFele-SFnu}
\end{figure}

In Fig.~\ref{fig:Lmb-GFele-SFnu} we present the results based on the SSF
of the electric GF scheme. As one can see from the figure, the determination
of $\Lambda_{\overline{\rm MS}}/\mu_{\rm ref}$ obtained directly from 
the GF coupling suffers from large $\mathcal{O}(\alpha^2)$ corrections. The situation
is of course completely analogous to what we already discussed in Sect.~\ref{sec:beta-function-he}
in terms of the $\beta$-function (cf.~Fig.~\ref{fig:lovm}). In particular, if we 
read off the value of $\Lambda_{\overline{\rm MS}}/\mu_{\rm ref}$ for say,
$n=6$, the result is:
\begin{equation}
	{\Lambda_{\overline{\rm MS}}\over\mu_{\rm ref}}= 0.0815(11)\,.
\end{equation}
Although this is compatible with our final estimate, eq.~(\ref{eq:res_he}), 
it is clear that given the trend of the results for different values of $n$,
a reliable determination of both mean value \emph{and} error,
necessarily requires an extrapolation to $\alpha=0$. 

Quite different is the situation for the determinations obtained after switching 
non-perturbatively to the SF couplings. These show indeed much milder $\mathcal{O}(\alpha^2)$
corrections, particularly so for the case $\nu=0$ (cf.~Fig.~\ref{fig:lovmSF}). 
At values of the coupling $\alpha\sim 0.1$, any discrepancy between different
$\nu$-values is negligible within our statistical errors, and we can thus 
safely quote:
\begin{equation}
	\label{eq:res_he_sf_ssf}
	{\Lambda_{\overline{\rm MS}}\over \mu_{\rm ref}}= 0.0802(10)\,,
\end{equation}
which corresponds to the value obtained for $\nu=0$ at $n=5$.
This is in perfect agreement with our final estimate eq.~(\ref{eq:res_he_sf}).

\begin{figure}[hptb]
	\centering
	\includegraphics[scale=1]{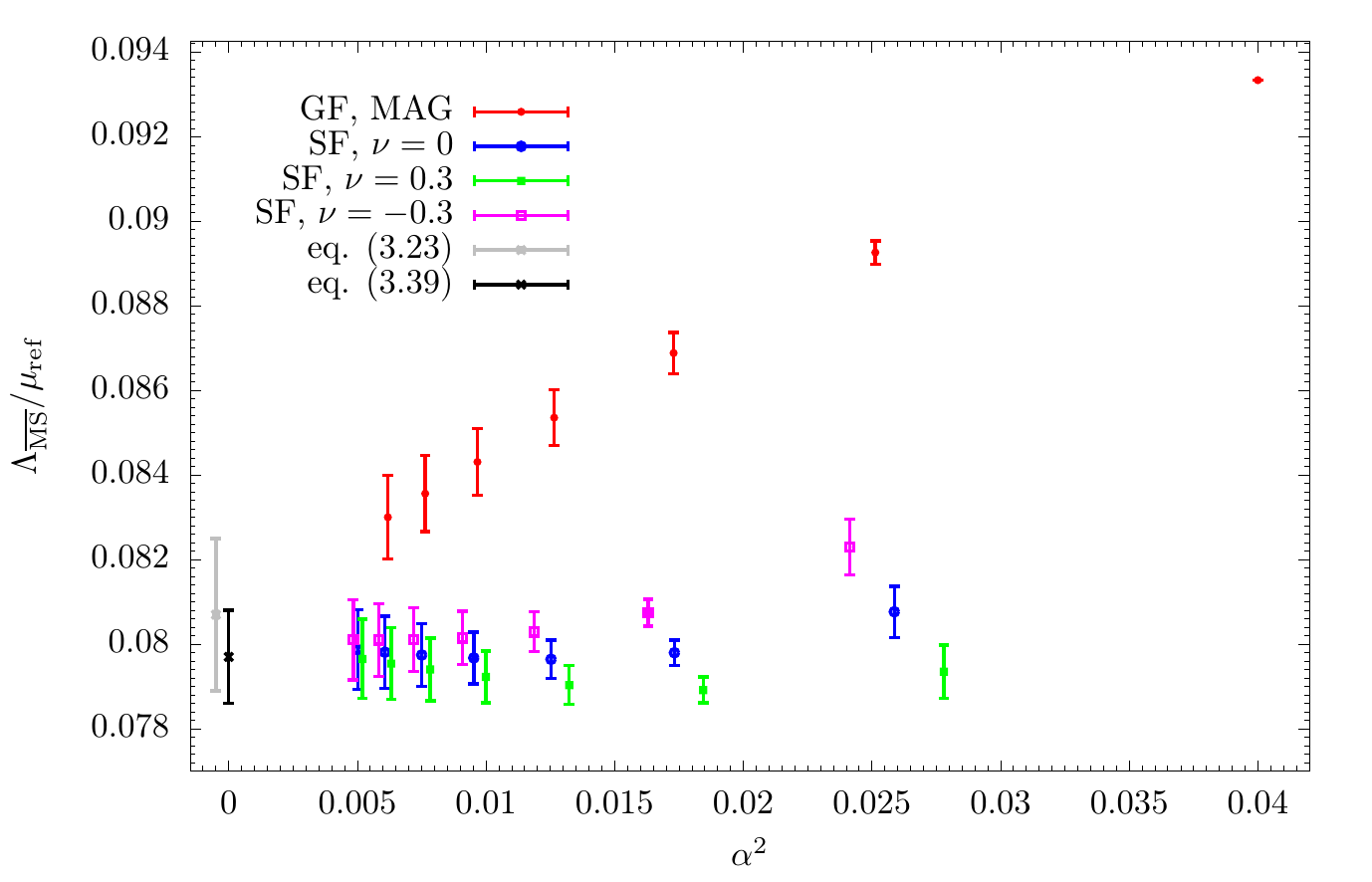}
	\caption{Results for the extraction of $\Lambda_{\overline{\rm MS}}/\mu_{\rm ref}$ 
	based on several schemes, and for different values of the corresponding couplings. 
	The results from the GF coupling refer to its magnetic scheme and are 
	computed following a step-scaling analysis. At each value of the 
	GF coupling coming from step-scaling, the GF scheme is matched non-perturbatively 
	to three different SF schemes corresponding to $\nu=-0.3,0,0.3$, which are then 
	used to also extract  $\Lambda_{\overline{\rm MS}}/\mu_{\rm ref}$.
	A comparison with our final estimates eqs.~(\ref{eq:res_he}) and 
	(\ref{eq:res_he_sf}) is given.}
	\label{fig:Lmb-GFmag-SFnu}
\end{figure}

For completeness we also present in Fig.~\ref{fig:Lmb-GFmag-SFnu} the corresponding
results based on the magnetic GF scheme. It is no surprise that the results
for $\Lambda_{\overline{\rm MS}}/\mu_{\rm ref}$ coming from the magnetic coupling
show larger $\mathcal{O}(\alpha^2)$ corrections than what we have seen for the electric scheme
(cf.~Fig.~\ref{fig:lovm}). Considering for instance the results for $n=6$, we have: 
$\Lambda_{\overline{\rm MS}}/\mu_{\rm ref}=0.0830(10)$, which is clearly biased in
both central value and error if compared to eq.~(\ref{eq:res_he}) (note that
$\alpha_{\rm GF,\,m}(\mu_{n=6})\sim 0.08$!). On the other hand, once again switching non-perturbatively
to the SF schemes solves the issue. Taking for instance the results for $\nu=0$ 
and $n=5$ we obtain: $\Lambda_{\overline{\rm MS}}/\mu_{\rm ref}=0.0798(9)$, well in 
agreement with the result of eqs.~(\ref{eq:res_he_sf}) and (\ref{eq:res_he_sf_ssf}).
In conclusion, the determination based on the SSF of the GF couplings rather than
the $\beta$-function gives perfectly compatible results, reinforcing even further 
the robustness of our analysis and conclusions. 

%%% Local Variables:
%%% mode: latex
%%% TeX-master: "paper"
%%% End:

\section{Perturbative improvement of the SF coupling}
\label{sec:SFPT}

In this appendix we give some details on the perturbative improvement 
of the SF coupling in the $\nu=0$ scheme, employed in Sect.~\ref{subsec:GF2SF}.  

In bare lattice perturbation theory, the SF coupling is given by,
\begin{equation}
 \label{eq:PTgSF}
 \bar{g}^2_{\rm SF}(L)= g_0^2 + m_1(L/a) g_0^4 + m_2(L/a) g_0^6 + \mathcal{O}(g_0^8),
\end{equation}
where the coefficients of this series can be written as:
\begin{align}
 m_1 &= m_1^a + c_{ t}^{(1)} m_1^b,\\
 m_2-m_1^2 &= m_2^a + c_{ t}^{(1)} m_2^b + [c_{ t}^{(1)}]^2 m_2^c + c_{t}^{(2)} m_2^d.
\end{align}
Referring to ref.~\cite{Bode:1998hd} one can easily show that
\begin{equation}
 m_1^b = m_2^d = - {2a\over L},
\end{equation}
and 
\begin{equation}
 m_2^c = {2a\over L}-{4a^2\over L^2} + \mathcal{O}(a^5),
\end{equation}
which approximation should be good enough in practice. The coefficients $m_1^a$, $m_2^a$ and 
$m_2^b$ instead can be found in table 1 of ref.~\cite{Bode:1998hd}.

To work out the cutoff effects in the SF coupling to two-loop order 
we need to know the asymptotic behavior of $m_1$ and $m_2$ for
$L/a\to\infty$. This is given by (see table 2 of ref.~\cite{Bode:1999sm}):\footnote{Note that 
a slightly different value for $m_2^\infty$ was first given in ref.~\cite{Bode:1998hd}.}
\begin{align}
 m_1^\infty &=2b_0\ln(L/a) +  0.36828215(13) + \mathcal{O}(a^2),\\
 m_2^\infty-[m_1^{\infty}]^2 &= 2b_1\ln(L/a) + 0.048091(2) + \mathcal{O}(a^2),
\end{align}
where $b_0$ and $b_1$ are the universal one- and two-loop coefficients 
of the $\beta$-function, eq.~(\ref{eq:beta_univ}).

Given these results we can define,
\begin{equation}
 {\bar{g}^2_{\rm SF}(L) - \bar{g}^{2}_{{\rm SF},\infty}(L)\over \bar{g}^2_{{\rm SF},\infty}(L)}
 = \delta_1(L/a)\,g_0^2 + \delta_2(L/a)\,g_0^4 + \mathcal{O}(g_0^6),
\end{equation}
where $\bar{g}^{2}_{{\rm SF},\infty}(L)$ as an expansion analogous to (\ref{eq:PTgSF}),
with the replacement $m_i\to m_i^\infty$, $i=1,2$. It is then easy to 
show that,
\begin{equation}
 \delta_1(L/a)=\Delta m_1,
 \qquad
 \delta_2(L/a)=\Delta m_2-\delta_1 m_1^\infty,
\end{equation}
where
\begin{equation}
 \Delta m_i(L/a)=m_i(L/a)-m_i^\infty(L/a), 
 \quad
 i=1,2.
\end{equation}
The corresponding results, taking,
\begin{equation}
 c_t^\star  = 1 -0.08900\,g_0^2 -0.0294\,g_0^4,
\end{equation}
are given in table \ref{tab:Deltam}.
A two-loop improved SF coupling can thus be defined as,
\begin{equation}
 \label{eq:SFImproved}
 \bar{g}^2_{\rm SF,I}(L)= {\bar{g}^2_{\rm SF}(L)\over 1+ \delta_1(L/a) g_0^2 + \delta_2(L/a) g_0^4},
\end{equation}
or more simply,
\begin{equation}
 \bar{g}^2_{\rm SF, I}(L)= \bar{g}^2_{\rm SF}(L) - \Delta m_1(L/a) g_0^4 - \Delta m_2(L/a) g_0^6.
\end{equation}
The two are of course equivalent up to $\mathcal{O}(g_0^8)$ terms. 
In Sect.~\ref{subsec:GF2SF} we considered the form eq.~(\ref{eq:SFImproved}).
Note that by using eq.~(\ref{eq:PTgSF}) and the results of Sect.~\ref{sec:RenormSchemes}, 
we could in principle rexpress the perturbative improvement of the SF coupling  
in terms the GF couplings, rather than the bare one. Although this might appear more natural 
when determining the matching between the GF and SF couplings, we prefer not to do so and 
stick with the bare results. In any case, as already mentioned in the main text, the effect
of the perturbative improvement is very small in practice (cf.~table \ref{tab:Deltam}). Choosing 
a different option hence does not make any real difference.

\begin{table}[hpbt]
\centering
\begin{tabular}{llll}
\toprule
 $L/a$ & $\Delta m_1=\delta_1$ & $\Delta m_2$ & $\delta_2$ \\
\midrule
4 & 0.012845313751142 & 0.015321716769734 & 0.008110152038992 \\
5 & 0.007777219114216 & 0.009579837534989 & 0.004971807117380 \\
6 & 0.004974956107720 & 0.006333958932812 & 0.003259913472149 \\
7 & 0.003426899898274 & 0.004486213192082 & 0.002295122656827 \\
8 & 0.002511485605829 & 0.003360322924560 & 0.001707808260234 \\
9 & 0.001927244066032 & 0.002622904409882 & 0.001323186298664 \\
10 & 0.001529846640399 & 0.002110676347599 & 0.001056504097430 \\
11 & 0.001246002180829 & 0.001738571617701 & 0.000863443440673 \\
12 & 0.001035543531126 & 0.001458806955388 & 0.000718941170315 \\
13 & 0.000874854600749 & 0.001242679157694 & 0.000607865198940 \\
14 & 0.000749224845475 & 0.001071984982022 & 0.000520595416728 \\
15 & 0.000649058490722 & 0.000934670808009 & 0.000450759687080 \\
16 & 0.000567859115075 & 0.000822471509072 & 0.000393993493151 \\
17 & 0.000501091899901 & 0.000729553491342 & 0.000347222437166 \\
18 & 0.000445508755510 & 0.000651697453251 & 0.000308228470592 \\
19 & 0.000398731340410 & 0.000585786821706 & 0.000275377878968 \\
20 & 0.000358984476884 & 0.000529476970044 & 0.000247445314836 \\
21 & 0.000324920687670 & 0.000480975306052 & 0.000223496862669 \\
22 & 0.000295501761385 & 0.000438891494416 & 0.000202810480112 \\
23 & 0.000269917024381 & 0.000402133251700 & 0.000184820726059 \\
24 & 0.000247525826056 & 0.000369832464991 & 0.000169079640890 \\
25 & 0.000227816347952 & 0.000341291913117 & 0.000155228589411 \\
26 & 0.000210375634016 & 0.000315946254512 & 0.000142977686219 \\
27 & 0.000194867470794 & 0.000293333060719 & 0.000132090551354 \\
28 & 0.000181015847174 & 0.000273071038353 & 0.000122372875705 \\
29 & 0.000168592437787 & 0.000254843458832 & 0.000113663738570 \\
30 & 0.000157407026808 & 0.000238385422055 & 0.000105828951105 \\
31 & 0.000147300106887 & 0.000223473964266 & 0.000098755896729 \\
32 & 0.000138137105359 & 0.000209920310316 & 0.000092349500925 \\
\bottomrule
\end{tabular}
\caption{Two-loop cutoff effects for the SF coupling with $\nu=0$.}
\label{tab:Deltam}
\end{table}

%%% Local Variables:
%%% mode: latex
%%% TeX-master: "paper"
%%% End:

\clearpage
\section{Raw the measurement data}
\label{sec:raw-data-meas}

\subsection{GF coupling}
\label{sec:Data}

Tables~\ref{tab:la48} to~\ref{tab:la8} collect the results of the measurements of the
GF couplings for the lattice sizes $L/a=48$, $32$, $24$, $20$, $16$, $12$, $8$, 
and the corresponding $\beta$ values we have simulated. We list the results for all
four coupling definitions and give the total statistics collected in the
column labelled by $N_{\rm msm}$. Note that this column also contains in parenthesis 
the number of measurements for which $|Q|>0.5$, and which thus do not enter the
determination of the coupling. Here $Q$ is the topological charge defined through the
Zeuthen flow (cf.~Sect.~\ref{sec:LatticeSetup}); there are only small differences in
the measurements of $Q$ between the Zeuthen and Wilson flow.

\subsection{SF coupling}
\label{sec:sf-coupl-meas}
Table \ref{tab:SF} contains the results of the measurements of the SF couplings
for all the lattice sizes and $\beta$-values we considered, together
with the total number of measurements collected, $N_{\rm msm}$; note that 
all measurements have in this case $Q=0$. We give results for the 
three values of $\nu=-0.3,0,0.3$, which enter the analysis of 
Appendix.~\ref{sec:ss}. We note that by using any two of these definitions,
it is possible to determine the value of the SF coupling and its proper error 
for any other value of $\nu$ (cf.~eq.~(\ref{eq:SFnu}) and ref.~\cite{DallaBrida:2018rfy}).

\begin{table}[pht!]
  \centering	
    \begin{tabular}{llllll}
    \toprule
    $\beta$&$N_{\rm
             msm}$&Zeuthen/mag&Wilson/mag&Zeuthen/ele&Wilson/ele\\
    \midrule
6.7859 & 2740(2048) & 14.353(77) & 14.336(76) & 14.325(81) & 14.307(81)\\
6.8637 & 3880(2619) & 11.345(63) & 11.336(63) & 11.363(67) & 11.354(67)\\
6.9595 & 2040(811) & 8.637(39) & 8.635(39) & 8.539(41) & 8.536(41)\\
7.1146 & 1560(0) & 6.551(17) & 6.551(17) & 6.438(20) & 6.437(20)\\
7.2000 & 2100(41) & 5.872(13) & 5.872(13) & 5.765(14) & 5.764(14)\\
7.6000 & 2300(0) & 4.0120(74) & 4.0132(74) & 3.9334(80) & 3.9339(80)\\
8.0000 & 3660(0) & 3.0886(47) & 3.0897(47) & 3.0260(51) & 3.0267(52)\\
8.5000 & 4720(0) & 2.4080(29) & 2.4090(29) & 2.3606(32) & 2.3612(32)\\
9.0000 & 7220(0) & 1.9800(19) & 1.9808(19) & 1.9448(21) & 1.9454(21)\\
9.5000 & 9280(0) & 1.6860(14) & 1.6867(14) & 1.6558(16) & 1.6563(16)\\
10.0000 & 11080(0) & 1.4691(11) & 1.4697(11) & 1.4470(12) & 1.4474(12)\\
10.5000 & 14700(0) & 1.30170(83) & 1.30223(83) & 1.28435(91) & 1.28474(92)\\
11.0000 & 18120(0) & 1.17159(68) & 1.17204(68) & 1.15554(74) & 1.15587(75)\\
\bottomrule
  \centering
  \end{tabular}

  \caption{GF coupling data for $L/a=48$.}
  \label{tab:la48}
\end{table}

\begin{table}[pht!]
  \centering	
    \begin{tabular}{llllll}
    \toprule
    $\beta$&$N_{\rm
             msm}$&Zeuthen/mag&Wilson/mag&Zeuthen/ele&Wilson/ele\\
    \midrule
6.4740 & 3300(2354) & 14.686(70) & 14.640(69) & 14.690(74) & 14.644(73)\\
6.5619 & 15350(8273) & 10.946(24) & 10.930(24) & 10.907(25) & 10.888(25)\\
6.6669 & 1650(246) & 8.203(27) & 8.201(27) & 8.092(32) & 8.088(32)\\
6.7859 & 1700(9) & 6.656(18) & 6.657(17) & 6.539(19) & 6.538(19)\\
6.8786 & 1800(10) & 5.877(14) & 5.879(14) & 5.744(16) & 5.744(16)\\
6.9595 & 1450(0) & 5.345(14) & 5.348(14) & 5.216(14) & 5.217(15)\\
7.1146 & 2250(0) & 4.5866(92) & 4.5898(92) & 4.4841(99) & 4.4855(99)\\
7.2000 & 2100(0) & 4.2584(87) & 4.2617(87) & 4.1558(92) & 4.1572(92)\\
7.6000 & 4450(0) & 3.2059(41) & 3.2089(42) & 3.1362(47) & 3.1381(47)\\
8.0000 & 7200(0) & 2.5918(26) & 2.5944(26) & 2.5433(28) & 2.5450(28)\\
8.5000 & 7200(0) & 2.0989(20) & 2.1011(20) & 2.0610(22) & 2.0624(22)\\
9.0000 & 8700(0) & 1.7713(15) & 1.7730(15) & 1.7375(16) & 1.7386(17)\\
9.5000 & 12000(0) & 1.5319(11) & 1.5334(11) & 1.5045(12) & 1.5056(12)\\
10.0000 & 13000(0) & 1.35038(91) & 1.35164(91) & 1.3314(10) & 1.3323(10)\\
10.4258 & 6000(0) & 1.2275(13) & 1.2286(13) & 1.2115(13) & 1.2123(13)\\
10.5000 & 17000(0) & 1.20874(71) & 1.20984(71) & 1.19340(78) & 1.19421(79)\\
10.6586 & 14050(0) & 1.17031(76) & 1.17134(76) & 1.15509(83) & 1.15587(84)\\
11.0000 & 19350(0) & 1.09622(60) & 1.09718(60) & 1.08289(67) & 1.08360(67)\\
\bottomrule
  \centering
  \end{tabular}

  \caption{GF coupling data for $L/a=32$.}
  \label{tab:la32}
\end{table}

\begin{table}[pht!]
  \centering		 
    \begin{tabular}{llllll}
    \toprule
    $\beta$&$N_{\rm
             msm}$&Zeuthen/mag&Wilson/mag&Zeuthen/ele&Wilson/ele\\
    \midrule
6.2556 & 5700(4313) & 15.307(56) & 15.192(54) & 15.317(57) & 15.198(56)\\
6.2654 & 4000(2834) & 14.839(59) & 14.740(57) & 14.851(65) & 14.750(63)\\
6.3451 & 3200(1792) & 11.219(55) & 11.193(54) & 11.168(60) & 11.137(60)\\
6.3509 & 3200(1810) & 11.044(49) & 11.018(49) & 10.973(55) & 10.945(55)\\
6.3560 & 3200(1761) & 10.757(53) & 10.723(52) & 10.687(56) & 10.655(54)\\
6.3642 & 4000(2040) & 10.398(39) & 10.380(38) & 10.326(43) & 10.305(43)\\
6.4630 & 3200(635) & 8.008(21) & 8.007(21) & 7.876(23) & 7.871(23)\\
6.5619 & 6300(316) & 6.6813(93) & 6.6852(93) & 6.555(11) & 6.555(11)\\
6.6559 & 4000(51) & 5.8695(94) & 5.8753(94) & 5.744(10) & 5.746(10)\\
6.6669 & 4200(37) & 5.7919(92) & 5.7974(92) & 5.6858(97) & 5.6876(98)\\
6.7859 & 6400(2) & 5.0578(61) & 5.0641(61) & 4.9486(65) & 4.9516(66)\\
6.8637 & 6000(0) & 4.6776(56) & 4.6842(56) & 4.5687(60) & 4.5718(60)\\
6.9595 & 6000(0) & 4.2884(51) & 4.2950(51) & 4.2048(58) & 4.2086(58)\\
7.1146 & 4000(0) & 3.7942(53) & 3.8005(53) & 3.7197(60) & 3.7235(60)\\
7.2000 & 10000(0) & 3.5705(31) & 3.5767(31) & 3.4942(34) & 3.4980(35)\\
7.6000 & 10000(0) & 2.8088(24) & 2.8141(24) & 2.7560(26) & 2.7595(26)\\
8.0000 & 12900(0) & 2.3320(17) & 2.3363(17) & 2.2895(18) & 2.2926(18)\\
8.5000 & 14500(0) & 1.9284(13) & 1.9320(13) & 1.8942(14) & 1.8967(14)\\
9.0000 & 14500(0) & 1.6458(11) & 1.6488(11) & 1.6169(12) & 1.6190(12)\\
9.5000 & 16000(0) & 1.43887(89) & 1.44135(90) & 1.41746(98) & 1.41927(99)\\
10.0000 & 19100(0) & 1.27695(70) & 1.27905(71) & 1.26060(78) & 1.26212(79)\\
10.4258 & 20000(0) & 1.16851(64) & 1.17034(64) & 1.15338(70) & 1.15474(71)\\
10.5000 & 18800(0) & 1.15194(65) & 1.15374(65) & 1.13672(71) & 1.13804(72)\\
11.0000 & 21800(0) & 1.04784(54) & 1.04941(54) & 1.03480(59) & 1.03598(60)\\
\bottomrule
  \centering
  \end{tabular}

  \caption{GF coupling data for $L/a=24$.}
  \label{tab:la24}
\end{table}

\begin{table}[pht!]
  \centering		
    \begin{tabular}{llllll}
    \toprule
    $\beta$&$N_{\rm
             msm}$&Zeuthen/mag&Wilson/mag&Zeuthen/ele&Wilson/ele\\
    \midrule
6.1365 & 5000(3710) & 15.037(59) & 14.906(57) & 15.016(63) & 14.876(62)\\
6.1700 & 6000(4110) & 13.310(48) & 13.201(47) & 13.295(51) & 13.182(51)\\
6.2160 & 5000(2930) & 11.181(40) & 11.138(38) & 11.118(42) & 11.066(41)\\
6.2280 & 5000(2818) & 10.724(39) & 10.692(38) & 10.668(43) & 10.633(42)\\
6.2556 & 6000(2658) & 9.726(27) & 9.707(26) & 9.647(30) & 9.624(30)\\
6.4200 & 6000(286) & 6.7554(97) & 6.7615(97) & 6.620(11) & 6.620(11)\\
6.5197 & 5000(85) & 5.8445(84) & 5.8535(84) & 5.7207(92) & 5.7245(93)\\
6.5619 & 6000(50) & 5.5487(71) & 5.5584(71) & 5.4171(76) & 5.4210(76)\\
6.6669 & 6000(17) & 4.9431(61) & 4.9532(61) & 4.8332(67) & 4.8384(68)\\
6.7859 & 6000(0) & 4.4082(55) & 4.4182(55) & 4.3157(56) & 4.3216(57)\\
6.8000 & 6000(11) & 4.3553(53) & 4.3656(54) & 4.2637(55) & 4.2698(56)\\
6.8637 & 6000(4) & 4.1219(48) & 4.1319(48) & 4.0268(53) & 4.0329(53)\\
6.9595 & 6000(0) & 3.8245(44) & 3.8340(44) & 3.7398(50) & 3.7459(51)\\
7.1146 & 6000(0) & 3.4254(39) & 3.4344(39) & 3.3509(40) & 3.3566(41)\\
7.2000 & 6000(0) & 3.2426(36) & 3.2513(36) & 3.1803(40) & 3.1860(41)\\
7.6000 & 6000(0) & 2.6124(29) & 2.6197(29) & 2.5547(30) & 2.5595(30)\\
8.0000 & 8000(0) & 2.1908(20) & 2.1969(20) & 2.1509(21) & 2.1552(22)\\
8.5000 & 14000(0) & 1.8304(12) & 1.8353(12) & 1.8015(13) & 1.8050(13)\\
9.0000 & 14000(0) & 1.5766(10) & 1.5806(10) & 1.5521(11) & 1.5550(11)\\
9.5000 & 14000(0) & 1.38439(90) & 1.38775(90) & 1.36570(98) & 1.36819(100)\\
10.0000 & 16000(0) & 1.23742(76) & 1.24027(76) & 1.22099(81) & 1.22309(83)\\
10.2830 & 20000(0) & 1.16540(63) & 1.16803(63) & 1.15006(70) & 1.15196(71)\\
10.5000 & 20000(0) & 1.11778(61) & 1.12026(61) & 1.10308(65) & 1.10490(66)\\
11.0000 & 24000(0) & 1.01933(51) & 1.02149(51) & 1.00702(54) & 1.00864(55)\\
\bottomrule
  \centering
  \end{tabular}

  \caption{GF coupling data for $L/a=20$.}
  \label{tab:la20}
\end{table}

\begin{table}[pht!]
  \centering	
    \begin{tabular}{llllll}
    \toprule
    $\beta$&$N_{\rm
             msm}$&Zeuthen/mag&Wilson/mag&Zeuthen/ele&Wilson/ele\\
    \midrule
5.9900 & 5000(3782) & 15.555(61) & 15.285(58) & 15.548(62) & 15.263(60)\\
6.0662 & 5750(3439) & 11.331(40) & 11.234(38) & 11.323(42) & 11.221(41)\\
6.0722 & 5000(2898) & 11.051(42) & 10.982(40) & 11.023(44) & 10.942(43)\\
6.0740 & 5000(2850) & 10.892(38) & 10.832(37) & 10.829(41) & 10.757(41)\\
6.1000 & 5000(2400) & 9.894(32) & 9.843(31) & 9.806(35) & 9.749(35)\\
6.1200 & 3750(1497) & 9.239(31) & 9.213(30) & 9.124(34) & 9.089(33)\\
6.1700 & 5750(1272) & 7.999(15) & 8.005(15) & 7.865(17) & 7.861(17)\\
6.2556 & 5750(319) & 6.770(10) & 6.783(10) & 6.638(11) & 6.640(11)\\
6.3598 & 5000(61) & 5.8085(82) & 5.8241(82) & 5.6839(88) & 5.6908(89)\\
6.4200 & 5750(30) & 5.3833(70) & 5.3999(70) & 5.2744(76) & 5.2830(77)\\
6.4740 & 5000(13) & 5.0535(67) & 5.0706(67) & 4.9430(71) & 4.9519(72)\\
6.4741 & 10000(28) & 5.0594(49) & 5.0763(49) & 4.9591(51) & 4.9683(52)\\
6.5619 & 5750(1) & 4.6165(56) & 4.6335(56) & 4.5197(60) & 4.5296(62)\\
6.6669 & 5000(0) & 4.2007(57) & 4.2173(57) & 4.1056(58) & 4.1150(59)\\
6.7859 & 6000(0) & 3.8249(46) & 3.8404(46) & 3.7391(48) & 3.7488(49)\\
6.8000 & 5750(2) & 3.7825(44) & 3.7980(44) & 3.6948(46) & 3.7048(47)\\
6.8637 & 10000(0) & 3.6081(32) & 3.6232(32) & 3.5222(34) & 3.5318(34)\\
6.8776 & 7750(0) & 3.5765(36) & 3.5915(36) & 3.5016(38) & 3.5112(39)\\
6.8786 & 7250(0) & 3.5720(37) & 3.5871(37) & 3.4924(40) & 3.5021(41)\\
6.8787 & 5000(0) & 3.5668(45) & 3.5816(46) & 3.4991(48) & 3.5086(49)\\
6.9595 & 5000(0) & 3.3744(43) & 3.3888(44) & 3.3056(48) & 3.3147(49)\\
7.1146 & 10000(0) & 3.0721(26) & 3.0854(27) & 3.0109(28) & 3.0196(29)\\
7.2000 & 5750(0) & 2.9197(32) & 2.9324(32) & 2.8614(36) & 2.8699(37)\\
7.6000 & 6000(0) & 2.4044(26) & 2.4147(26) & 2.3603(27) & 2.3676(28)\\
8.0000 & 11750(0) & 2.0450(15) & 2.0537(16) & 2.0087(16) & 2.0148(16)\\
8.5000 & 12000(0) & 1.7296(12) & 1.7366(12) & 1.7003(13) & 1.7053(14)\\
9.0000 & 12000(0) & 1.4976(10) & 1.5034(10) & 1.4761(11) & 1.4802(12)\\
9.5000 & 22750(0) & 1.32609(68) & 1.33083(70) & 1.30821(72) & 1.31164(74)\\
10.0000 & 21000(0) & 1.18949(63) & 1.19363(63) & 1.17455(68) & 1.17751(70)\\
10.1106 & 20000(0) & 1.16469(63) & 1.16869(63) & 1.14841(67) & 1.15129(69)\\
10.4382 & 20000(0) & 1.09154(59) & 1.09517(59) & 1.07696(64) & 1.07953(66)\\
10.5000 & 21000(0) & 1.07807(56) & 1.08162(57) & 1.06517(60) & 1.06775(62)\\
10.6581 & 3250(0) & 1.0466(14) & 1.0500(14) & 1.0348(15) & 1.0372(15)\\
10.6586 & 19250(0) & 1.04936(58) & 1.05279(59) & 1.03651(62) & 1.03905(64)\\
10.9270 & 20000(0) & 0.99954(53) & 1.00274(53) & 0.98835(57) & 0.99064(59)\\
11.0000 & 23500(0) & 0.98731(48) & 0.99043(48) & 0.97657(52) & 0.97890(54)\\
\bottomrule
  \centering
  \end{tabular}

  \caption{GF coupling data for $L/a=16$.}
  \label{tab:la16}
\end{table}

\begin{table}[pht!]
  \centering		
    \begin{tabular}{llllll}
    \toprule
    $\beta$&$N_{\rm
             msm}$&Zeuthen/mag&Wilson/mag&Zeuthen/ele&Wilson/ele\\
    \midrule
5.8900 & 8000(4895) & 11.681(37) & 11.483(33) & 11.679(39) & 11.470(36)\\
5.9000 & 4800(2823) & 11.168(48) & 11.003(44) & 11.115(49) & 10.936(45)\\
5.9080 & 1200(644) & 10.787(90) & 10.639(82) & 10.654(91) & 10.476(85)\\
5.9160 & 7200(3813) & 10.291(31) & 10.189(29) & 10.217(32) & 10.105(31)\\
5.9200 & 4400(2143) & 10.077(38) & 9.981(35) & 9.983(41) & 9.881(38)\\
5.9900 & 2000(417) & 7.857(26) & 7.850(25) & 7.730(27) & 7.710(26)\\
6.0000 & 2000(371) & 7.659(26) & 7.657(25) & 7.528(27) & 7.514(27)\\
6.0662 & 10000(680) & 6.6751(79) & 6.6981(78) & 6.5575(81) & 6.5629(82)\\
6.1700 & 10000(118) & 5.6802(58) & 5.7094(58) & 5.5705(61) & 5.5841(63)\\
6.2556 & 10000(52) & 5.1104(49) & 5.1414(49) & 5.0010(54) & 5.0173(56)\\
6.2643 & 10000(38) & 5.0690(50) & 5.1000(50) & 4.9572(51) & 4.9728(52)\\
6.2654 & 10000(37) & 5.0489(50) & 5.0796(50) & 4.9387(51) & 4.9552(52)\\
6.3451 & 4000(4) & 4.6316(71) & 4.6621(71) & 4.5352(72) & 4.5518(74)\\
6.3509 & 10000(12) & 4.6164(43) & 4.6468(44) & 4.5206(45) & 4.5381(47)\\
6.3560 & 4000(2) & 4.5905(68) & 4.6208(69) & 4.4883(73) & 4.5050(75)\\
6.3642 & 10000(9) & 4.5554(44) & 4.5853(44) & 4.4496(45) & 4.4651(46)\\
6.3894 & 8800(3) & 4.4339(45) & 4.4638(45) & 4.3433(46) & 4.3596(48)\\
6.4133 & 10000(1) & 4.3543(40) & 4.3837(40) & 4.2606(42) & 4.2771(43)\\
6.4200 & 10000(6) & 4.3136(40) & 4.3434(40) & 4.2254(41) & 4.2428(43)\\
6.4630 & 10000(2) & 4.1590(38) & 4.1877(38) & 4.0659(39) & 4.0830(42)\\
6.5619 & 10000(0) & 3.8287(34) & 3.8563(34) & 3.7536(35) & 3.7707(36)\\
6.6559 & 10000(0) & 3.5720(31) & 3.5983(32) & 3.4993(32) & 3.5160(34)\\
6.6566 & 10000(1) & 3.5665(31) & 3.5928(31) & 3.4926(34) & 3.5088(35)\\
6.6669 & 10000(1) & 3.5441(31) & 3.5704(31) & 3.4707(32) & 3.4864(33)\\
6.7859 & 10000(0) & 3.2730(30) & 3.2973(30) & 3.2041(29) & 3.2190(31)\\
6.8000 & 10000(0) & 3.2384(28) & 3.2627(28) & 3.1699(29) & 3.1851(30)\\
6.8544 & 8800(0) & 3.1382(29) & 3.1620(30) & 3.0737(30) & 3.0884(31)\\
6.8637 & 10000(0) & 3.1139(27) & 3.1371(27) & 3.0542(28) & 3.0693(29)\\
6.9595 & 10000(0) & 2.9506(25) & 2.9728(25) & 2.8908(26) & 2.9048(27)\\
7.1146 & 10000(0) & 2.7085(23) & 2.7287(23) & 2.6564(24) & 2.6696(26)\\
7.2000 & 10000(0) & 2.6000(22) & 2.6195(22) & 2.5516(23) & 2.5643(24)\\
7.6000 & 10000(0) & 2.1766(18) & 2.1925(18) & 2.1431(19) & 2.1538(19)\\
8.0000 & 10000(0) & 1.8809(15) & 1.8940(15) & 1.8504(15) & 1.8591(16)\\
8.5000 & 10000(0) & 1.6118(13) & 1.6225(13) & 1.5867(13) & 1.5938(14)\\
9.0000 & 18250(0) & 1.41310(81) & 1.42191(82) & 1.39153(85) & 1.39754(90)\\
9.5000 & 18250(0) & 1.25790(72) & 1.26532(73) & 1.24192(74) & 1.24710(78)\\
10.0000 & 20000(0) & 1.13541(61) & 1.14181(61) & 1.12061(64) & 1.12483(67)\\
10.4250 & 10000(0) & 1.04869(78) & 1.05425(79) & 1.03752(84) & 1.04154(89)\\
10.4258 & 18800(0) & 1.04795(57) & 1.05363(58) & 1.03529(63) & 1.03909(66)\\
10.4262 & 20000(0) & 1.04782(56) & 1.05344(56) & 1.03512(58) & 1.03899(61)\\
10.5000 & 40000(0) & 1.03364(39) & 1.03913(39) & 1.02089(41) & 1.02467(43)\\
11.0000 & 20000(0) & 0.95006(50) & 0.95488(50) & 0.93969(53) & 0.94299(56)\\
11.1572 & 10000(0) & 0.92748(69) & 0.93213(70) & 0.91626(73) & 0.91945(76)\\
12.0000 & 10000(0) & 0.81767(61) & 0.82141(62) & 0.80886(65) & 0.81148(69)\\
\bottomrule
  \centering
  \end{tabular}

  \caption{GF coupling data for $L/a=12$.}
  \label{tab:la12}
\end{table}

\begin{table}[pht!]
  \centering		
    \begin{tabular}{llllll}
    \toprule
    $\beta$&$N_{\rm
             msm}$&Zeuthen/mag&Wilson/mag&Zeuthen/ele&Wilson/ele\\
    \midrule
6.0500 & 6000(89) & 5.6477(75) & 5.6817(74) & 5.5341(81) & 5.5437(83)\\
6.0662 & 10000(125) & 5.5383(59) & 5.5741(59) & 5.4189(60) & 5.4305(62)\\
6.1000 & 8000(64) & 5.2911(59) & 5.3278(59) & 5.1751(61) & 5.1894(62)\\
6.1365 & 20000(72) & 5.0562(35) & 5.0940(36) & 4.9470(36) & 4.9627(37)\\
6.1370 & 8000(19) & 5.0551(56) & 5.084(11) & 4.9433(57) & 4.949(11)\\
6.1500 & 8000(21) & 4.9756(54) & 5.000(16) & 4.8670(54) & 4.870(15)\\
6.1700 & 10000(25) & 4.8639(46) & 4.9023(46) & 4.7576(48) & 4.7735(50)\\
6.2000 & 8000(16) & 4.7104(52) & 4.679(67) & 4.6016(51) & 4.552(66)\\
6.2160 & 20000(32) & 4.6283(31) & 4.6669(31) & 4.5320(31) & 4.5500(33)\\
6.2280 & 20000(25) & 4.5814(31) & 4.6201(32) & 4.4797(32) & 4.4981(33)\\
6.2556 & 10000(8) & 4.4561(42) & 4.4939(43) & 4.3571(42) & 4.3759(44)\\
6.4200 & 10000(0) & 3.8494(34) & 3.8854(35) & 3.7677(35) & 3.7870(37)\\
6.4589 & 40000(3) & 3.7357(16) & 3.7711(17) & 3.6584(17) & 3.6779(18)\\
6.5081 & 20000(1) & 3.5991(22) & 3.6337(23) & 3.5227(23) & 3.5416(24)\\
6.5197 & 20000(0) & 3.5698(22) & 3.6040(23) & 3.4986(23) & 3.5173(24)\\
6.5209 & 20000(1) & 3.5686(22) & 3.6031(23) & 3.4932(23) & 3.5122(24)\\
6.5573 & 40000(0) & 3.4789(15) & 3.5127(15) & 3.4056(16) & 3.4246(16)\\
6.5619 & 10000(0) & 3.4562(31) & 3.4900(31) & 3.3830(31) & 3.4010(32)\\
6.5871 & 10000(0) & 3.4039(30) & 3.4369(31) & 3.3359(31) & 3.3549(33)\\
6.6669 & 10000(0) & 3.2287(28) & 3.2604(28) & 3.1602(28) & 3.1784(30)\\
6.7543 & 26000(0) & 3.0621(16) & 3.0922(16) & 3.0015(16) & 3.0189(17)\\
6.7859 & 10000(0) & 3.0048(26) & 3.0344(27) & 2.9416(26) & 2.9581(27)\\
6.8000 & 10000(0) & 2.9811(25) & 3.0107(26) & 2.9205(25) & 2.9378(27)\\
6.8036 & 26000(0) & 2.9740(16) & 3.0033(16) & 2.9153(16) & 2.9320(17)\\
6.8528 & 26000(0) & 2.8940(15) & 2.9226(16) & 2.8355(15) & 2.8521(16)\\
6.8637 & 20000(0) & 2.8723(17) & 2.9007(17) & 2.8172(17) & 2.8341(18)\\
6.9595 & 10000(0) & 2.7283(23) & 2.7551(23) & 2.6776(23) & 2.6934(24)\\
7.1146 & 10000(0) & 2.5282(21) & 2.5529(21) & 2.4804(21) & 2.4954(22)\\
7.2000 & 10000(0) & 2.4279(20) & 2.4516(20) & 2.3842(20) & 2.3987(21)\\
7.6000 & 10000(0) & 2.0640(16) & 2.0835(17) & 2.0277(17) & 2.0393(18)\\
8.0000 & 15000(0) & 1.7960(11) & 1.8122(12) & 1.7661(12) & 1.7761(12)\\
8.5000 & 15000(0) & 1.54889(97) & 1.56216(100) & 1.52559(100) & 1.5339(11)\\
9.0000 & 20000(0) & 1.36404(72) & 1.37499(74) & 1.34555(75) & 1.35247(81)\\
9.5000 & 20000(0) & 1.22019(67) & 1.22942(69) & 1.20446(69) & 1.21022(74)\\
10.0000 & 20000(0) & 1.10469(58) & 1.11268(60) & 1.09035(60) & 1.09538(64)\\
10.2155 & 20000(0) & 1.06058(55) & 1.06805(56) & 1.04732(57) & 1.05210(62)\\
10.2682 & 20000(0) & 1.05109(57) & 1.05847(58) & 1.03900(56) & 1.04377(61)\\
10.2830 & 20000(0) & 1.04805(54) & 1.05547(56) & 1.03521(57) & 1.04001(61)\\
10.2862 & 40000(0) & 1.04684(38) & 1.05418(39) & 1.03428(41) & 1.03895(44)\\
10.3209 & 10000(0) & 1.04153(76) & 1.04881(78) & 1.02809(80) & 1.03265(85)\\
10.5000 & 50000(0) & 1.00839(33) & 1.01530(34) & 0.99639(34) & 1.00078(37)\\
11.0000 & 30000(0) & 0.92827(39) & 0.93434(40) & 0.91824(41) & 0.92211(44)\\
\bottomrule
  \centering
  \end{tabular}

  \caption{GF coupling data for $L/a=10$.}
  \label{tab:la10}
\end{table}

\begin{table}[pht!]
  \centering	
    \begin{tabular}{llllll}
    \toprule
    $\beta$&$N_{\rm
             msm}$&Zeuthen/mag&Wilson/mag&Zeuthen/ele&Wilson/ele\\
    \midrule
5.9600 & 10000(52) & 5.2580(55) & 5.2711(54) & 5.1504(54) & 5.1243(56) \\
5.9900 & 10000(40) & 5.0579(52) & 5.0761(52) & 4.9452(52) & 4.9253(54) \\
6.0662 & 15000(21) & 4.6269(36) & 4.6546(37) & 4.5251(35) & 4.5168(38) \\
6.1287 & 10000(3) & 4.3399(41) & 4.3707(42) & 4.2508(40) & 4.2493(43) \\
6.1700 & 15000(3) & 4.1717(31) & 4.2039(32) & 4.0853(31) & 4.0870(33) \\
6.2556 & 15000(4) & 3.8730(29) & 3.9061(30) & 3.7952(29) & 3.7998(31) \\
6.3597 & 10000(0) & 3.5727(31) & 3.6066(32) & 3.5005(31) & 3.5078(33) \\
6.3604 & 10000(0) & 3.5669(31) & 3.6004(32) & 3.4915(31) & 3.4996(33) \\
6.4147 & 10000(0) & 3.4283(30) & 3.4612(32) & 3.3624(29) & 3.3697(32) \\
6.4198 & 10000(0) & 3.4135(29) & 3.4465(30) & 3.3457(29) & 3.3535(31) \\
6.4200 & 15000(0) & 3.4194(24) & 3.4526(25) & 3.3492(23) & 3.3568(26) \\
6.5619 & 15000(0) & 3.1157(22) & 3.1476(23) & 3.0525(21) & 3.0619(23) \\
6.7859 & 15000(0) & 2.7432(19) & 2.7731(20) & 2.6887(18) & 2.6985(20) \\
6.8000 & 15000(0) & 2.7208(18) & 2.7502(19) & 2.6715(18) & 2.6816(20) \\
6.8637 & 15000(0) & 2.6403(17) & 2.6689(18) & 2.5885(17) & 2.5989(19) \\
7.0425 & 10000(0) & 2.4210(19) & 2.4476(20) & 2.3789(19) & 2.3889(21) \\
7.1146 & 15000(0) & 2.3450(15) & 2.3707(16) & 2.3044(15) & 2.3145(17) \\
7.2000 & 15000(0) & 2.2605(15) & 2.2851(15) & 2.2221(14) & 2.2320(16) \\
7.6000 & 15000(0) & 1.9404(12) & 1.9609(13) & 1.9103(12) & 1.9196(13) \\
8.0000 & 15000(0) & 1.7038(10) & 1.7215(11) & 1.6769(11) & 1.6848(12) \\
8.5000 & 15000(0) & 1.47990(90) & 1.49457(96) & 1.45938(89) & 1.4659(10)  \\
9.0000 & 35000(0) & 1.31185(52) & 1.32419(55) & 1.29494(52) & 1.30076(58)  \\
9.5000 & 35000(0) & 1.17746(46) & 1.18801(48) & 1.16411(45) & 1.16930(51)  \\
9.8663 & 20000(0) & 1.09624(55) & 1.10574(58) & 1.08359(58) & 1.08807(65)  \\
10.0000 & 35000(0) & 1.06989(41) & 1.07906(44) & 1.05719(42) & 1.06163(47)  \\
10.1102 & 1000(0) & 1.0440(25) & 1.009(44) & 1.0347(25) & 0.994(44) \\
10.1127 & 50000(0) & 1.04689(34) & 1.05573(36) & 1.03593(34) & 1.04027(38)  \\
10.1184 & 40000(0) & 1.04653(38) & 1.05531(40) & 1.03508(38) & 1.03936(42)  \\
10.5000 & 35000(0) & 0.97920(38) & 0.98716(40) & 0.96872(38) & 0.97257(42)  \\
11.0000 & 35000(0) & 0.90360(35) & 0.91066(37) & 0.89447(35) & 0.89798(39)  \\
12.0000 & 10000(0) & 0.78447(54) & 0.79008(57) & 0.77644(56) & 0.77901(64)  \\
12.6814 & 20000(0) & 0.71847(36) & 0.72333(38) & 0.71220(36) & 0.71457(41)  \\
13.0000 & 10000(0) & 0.69193(49) & 0.69649(51) & 0.68663(51) & 0.68868(57)  \\
\bottomrule
  \centering
  \end{tabular}

  \caption{GF coupling data for  $L/a=8$.}
  \label{tab:la8}
\end{table}

\begin{table}[ht!]
	\centering
    \begin{tabular}{llllll}
	\toprule
	$L/a$ & $\beta$ & $N_{\rm msm}$ & $\bar{g}^2_{{\rm SF},\nu=0}$ &  $\bar{g}^2_{{\rm SF},\nu=0.3}$ & $\bar{g}^2_{{\rm SF},\nu=-0.3}$ \\ 
	\midrule
	$ 6$ & $7.6000$ & $450000$ & $ 1.7804(12)$ & $ 1.8434(15)$ & $ 1.7216(13)$ \\
	$ 6$ & $8.0000$ & $450000$ & $ 1.57426(96)$ & $ 1.6256(12)$ & $ 1.5260(10)$ \\
	$ 6$ & $8.5000$ & $450000$ & $ 1.37634(76)$ & $ 1.41721(94)$ & $ 1.33776(83)$ \\
	$ 6$ & $9.0000$ & $450000$ & $ 1.22595(62)$ & $ 1.25857(77)$ & $ 1.19498(69)$ \\
	$ 6$ & $9.5000$ & $450000$ & $ 1.10542(52)$ & $ 1.13184(65)$ & $ 1.08020(58)$ \\
	$ 6$ & $10.0000$ & $450000$ & $ 1.00866(45)$ & $ 1.03106(55)$ & $ 0.98722(50)$ \\
	$ 6$ & $10.4250$ & $450000$ & $ 0.93785(40)$ & $ 0.95775(49)$ & $ 0.91876(45)$ \\
	$ 6$ & $10.4262$ & $450000$ & $ 0.93791(40)$ & $ 0.95684(49)$ & $ 0.91971(44)$ \\
	$ 6$ & $10.5000$ & $450000$ & $ 0.92709(39)$ & $ 0.94634(48)$ & $ 0.90860(44)$ \\
	$ 6$ & $11.0000$ & $450000$ & $ 0.85873(34)$ & $ 0.87512(42)$ & $ 0.84293(38)$ \\
	\midrule
	$ 8$ & $7.6000$ & $500000$ & $ 1.9309(17)$ & $ 2.0017(20)$ & $ 1.8649(18)$ \\
	$ 8$ & $8.0000$ & $500000$ & $ 1.6887(13)$ & $ 1.7451(16)$ & $ 1.6358(14)$ \\
	$ 8$ & $8.5000$ & $500000$ & $ 1.4604(10)$ & $ 1.5050(12)$ & $ 1.4185(11)$ \\
	$ 8$ & $9.0000$ & $600000$ & $ 1.29090(76)$ & $ 1.32602(92)$ & $ 1.25759(83)$ \\
	$ 8$ & $9.5000$ & $600000$ & $ 1.15884(63)$ & $ 1.18762(77)$ & $ 1.13143(70)$ \\
	$ 8$ & $10.0000$ & $600000$ & $ 1.05174(53)$ & $ 1.07564(65)$ & $ 1.02887(59)$ \\
	$ 8$ & $10.1106$ & $600000$ & $ 1.03063(52)$ & $ 1.05355(63)$ & $ 1.00869(58)$ \\
	$ 8$ & $10.4382$ & $600000$ & $ 0.97373(47)$ & $ 0.99467(58)$ & $ 0.95366(52)$ \\
	$ 8$ & $10.5000$ & $600000$ & $ 0.96384(46)$ & $ 0.98406(56)$ & $ 0.94444(52)$ \\
	$ 8$ & $10.6581$ & $600000$ & $ 0.93870(44)$ & $ 0.95733(54)$ & $ 0.92078(49)$ \\
	$ 8$ & $10.9270$ & $600000$ & $ 0.89861(42)$ & $ 0.91589(50)$ & $ 0.88197(46)$ \\
	$ 8$ & $11.0000$ & $600000$ & $ 0.88910(41)$ & $ 0.90618(49)$ & $ 0.87265(45)$ \\
	\midrule
	$10$ & $8.0000$ & $900000$ & $ 1.7927(13)$ & $ 1.8556(16)$ & $ 1.7339(14)$ \\
	$10$ & $8.5000$ & $900000$ & $ 1.53723(100)$ & $ 1.5855(12)$ & $ 1.4918(11)$ \\
	$10$ & $9.0000$ & $900000$ & $ 1.34989(80)$ & $ 1.38745(96)$ & $ 1.31431(87)$ \\
	$10$ & $9.5000$ & $900000$ & $ 1.20415(66)$ & $ 1.23490(80)$ & $ 1.17490(72)$ \\
	$10$ & $10.0000$ & $900000$ & $ 1.08927(56)$ & $ 1.11425(67)$ & $ 1.06538(61)$ \\
	$10$ & $10.5000$ & $900000$ & $ 0.99465(48)$ & $ 1.01561(58)$ & $ 0.97454(53)$ \\
	$10$ & $11.0000$ & $900000$ & $ 0.91528(42)$ & $ 0.93277(51)$ & $ 0.89844(47)$ \\
	\midrule
	$12$ & $8.0000$ & $1200000$ & $ 1.8880(14)$ & $ 1.9544(17)$ & $ 1.8259(15)$ \\
	$12$ & $8.5000$ & $1200000$ & $ 1.6070(11)$ & $ 1.6568(13)$ & $ 1.5601(12)$ \\
	$12$ & $9.0000$ & $1200000$ & $ 1.40143(85)$ & $ 1.4413(10)$ & $ 1.36372(92)$ \\
	$12$ & $9.5000$ & $1200000$ & $ 1.24490(69)$ & $ 1.27658(83)$ & $ 1.21474(75)$ \\
	$12$ & $10.0000$ & $1200000$ & $ 1.12205(59)$ & $ 1.14851(70)$ & $ 1.09679(64)$ \\
	$12$ & $10.4258$ & $1200000$ & $ 1.03544(51)$ & $ 1.05795(61)$ & $ 1.01386(56)$ \\
	$12$ & $10.5000$ & $1200000$ & $ 1.02105(50)$ & $ 1.04306(60)$ & $ 0.99996(55)$ \\
	$12$ & $11.0000$ & $1200000$ & $ 0.93837(43)$ & $ 0.95710(52)$ & $ 0.92036(48)$ \\
	\midrule
	$16$ & $8.5000$ & $1600000$ & $ 1.7293(13)$ & $ 1.7877(16)$ & $ 1.6745(14)$ \\
	$16$ & $9.0000$ & $1600000$ & $ 1.4940(10)$ & $ 1.5382(12)$ & $ 1.4524(11)$ \\
	$16$ & $9.5000$ & $1600000$ & $ 1.31496(81)$ & $ 1.34929(96)$ & $ 1.28233(87)$ \\
	$16$ & $10.0000$ & $1600000$ & $ 1.17831(68)$ & $ 1.20691(80)$ & $ 1.15104(73)$ \\
	$16$ & $10.4258$ & $1800000$ & $ 1.08260(55)$ & $ 1.10666(65)$ & $ 1.05956(60)$ \\
	$16$ & $10.5000$ & $1800000$ & $ 1.06889(54)$ & $ 1.09233(64)$ & $ 1.04644(59)$ \\
	$16$ & $10.6581$ & $1800000$ & $ 1.03762(52)$ & $ 1.06036(61)$ & $ 1.01584(56)$ \\
	$16$ & $11.0000$ & $2000000$ & $ 0.97567(44)$ & $ 0.99547(52)$ & $ 0.95664(48)$ \\
	\bottomrule
\end{tabular}
    \caption{SF coupling data.}
    \label{tab:SF}
\end{table}

%%% Local Variables:
%%% mode: latex
%%% TeX-master: "paper"
%%% End:

\clearpage
%\section*{References}
\addcontentsline{toc}{section}{References}
\bibliography{/home/alberto/docs/bib/math,/home/alberto/docs/bib/campos,/home/alberto/docs/bib/fisica,/home/alberto/docs/bib/computing}

\end{document}